\documentclass[pra, 
aps,
superscriptaddress, 
twocolumn, 
amsmath,
amssymb, 
amsfonts,
noeprint,				% hides arxiv citations
a4paper
]{revtex4-1}

\usepackage{braket}                % bra/ket notation
\usepackage{graphicx}  			   % needed for figures
\usepackage{color}     			   % use if color is used in text
\usepackage{hyperref}   		   % use for hypertext links, including those to external documents and URLs
\usepackage[utf8]{inputenc}        % Umlaute
\usepackage{multirow}              % multiple rows in table
\usepackage[caption=false,position=top]{subfig} % package for subfigures
\usepackage{bbold}
\usepackage[normalem]{ulem}

\newcommand{\sx}{\sigma_x}
\newcommand{\sy}{\sigma_y}
\newcommand{\sz}{\sigma_z}

\newcommand{\up}{\ket{\uparrow}}
\newcommand{\down}{\ket{\downarrow}}

\renewcommand{\Re}{\text{Re}}

\newcommand{\MHz}{\text{\,MHz}}
\newcommand{\GHz}{\text{\,GHz}}
\newcommand{\THz}{\text{\,THz}}
\newcommand{\kHz}{\text{\,kHz}}
\newcommand{\mus}{\,\mu\text{s}}
\newcommand{\ns}{\text{\,ns}}
\newcommand{\T}{\text{\,T}} 
\newcommand{\mT}{\text{\,mT}}

\newcommand{\nm}{\text{\,nm}}

\begin{document}
		
%%%%%%%%%%%%%%%% title %%%%%%%%%%%%%%%%%%%%%%
\title{Universal quantum computing using electro-nuclear wavefunctions of rare-earth ions}

%%%%%%%%%%%%%%%% authors %%%%%%%%%%%%%%%%%%%%%%
\author{Manuel Grimm}
\email[]{manuel.grimm@psi.ch}
\affiliation{Condensed Matter Theory Group, LSM, NES, Paul Scherrer Institut, CH-5232 Villigen PSI, Switzerland}
\affiliation{Laboratory for Solid State Physics, ETH Zurich, Zurich, CH-8093, Switzerland.}

\author{Adrian Beckert}
\affiliation{Laboratory for Solid State Physics, ETH Zurich, Zurich, CH-8093, Switzerland.}
\affiliation{Photon Science Division, Paul Scherrer Institut, CH-5232 Villigen PSI, Switzerland.}

\author{Gabriel Aeppli}
\affiliation{Laboratory for Solid State Physics, ETH Zurich, Zurich, CH-8093, Switzerland.}
\affiliation{Photon Science Division, Paul Scherrer Institut, CH-5232 Villigen PSI, Switzerland.}
\affiliation{Institute of Physics, EPF Lausanne, Lausanne, CH-1015, Switzerland.}

\author{Markus M\"uller}
\affiliation{Condensed Matter Theory Group, LSM, NES, Paul Scherrer Institut, CH-5232 Villigen PSI, Switzerland}

%%%%%%%%%%%%%%%% date %%%%%%%%%%%%%%%%%%%%%%
\date{\today}

%%%%%%%%%%%%%%%% abstract %%%%%%%%%%%%%%%%%%%%%%
\begin{abstract}
	We propose a scheme for universal quantum computing based on Kramers rare-earth ions. Their nuclear spins in the presence of a Zeeman-split electronic crystal field ground state act as 'passive' qubits which store quantum information. The qubits can be activated optically by fast coherent transitions to excited crystal field states with a magnetic moment. The dipole interaction between these states is used to implement CNOT gates. We compare our proposal with a similar one based on phosphorus donor atoms in silicon and discuss the significantly improved CNOT gate time as compared to rare-earth implementations via the slower dipole blockade.
\end{abstract}

\maketitle

%%%%%%%%%%%%%%%% sections %%%%%%%%%%%%%%%%%%%%%%
\section{Introduction}
	% RE life and coherence times
	Rare-earth (RE) ions embedded in an insulating solid-state matrix provide an interesting platform for quantum computing and quantum information processing. The nuclear spins and the electronic crystal field (CF) levels of RE ions can be used to store and manipulate quantum states. Due to the long coherence times of the quantum states of RE ions, they are well-suited for the implementation of qubits. Dephasing times ranging from $100 \mus$ for electronic transitions between CF states~\cite{Kukharchyk2018}, to 1.3~s for nuclear transitions~\cite{Rancic2016}, and even up to six hours by employing dynamical decoupling~\cite{Zhong2015} have recently been demonstrated. 
	
	% readout and coupling 
	Furthermore, the possibility to read out single spin states has been demonstrated using detection of photons emitted from yttrium aluminum garnet (YAG)~\cite{Kolesov2012,Siyushev2014}, yttrium orthovanadate (YVO)~\cite{Kindem2020} and  yttrium orthosilicate (YSO)~\cite{Utikal2014,Chen2020,Raha2020}, which makes such RE systems promising platforms for quantum technology. Some RE ions exhibit CF transitions in the frequency range used in telecommunications, which makes them well-suited as quantum repeaters~\cite{Sangouard2011,Saglamyurek2011}.

	% earlier proposals
	Previous schemes for quantum computing with RE ions proposed to use electric dipolar interactions of CF states, suggesting to realize a CNOT gate via an indirect dipole blockade effect~\cite{Lukin2000,Ohlsson2002,Wesenberg2007}. In that scheme, the dipole field from the control qubit shifts the transition frequencies of the target qubit. This is exploited to implement a CNOT gate with a sequence of pulses, which is effective only if the control bit is in the logical 1-state.
	
\begin{figure}[ht] 
			\includegraphics[width=1\columnwidth]{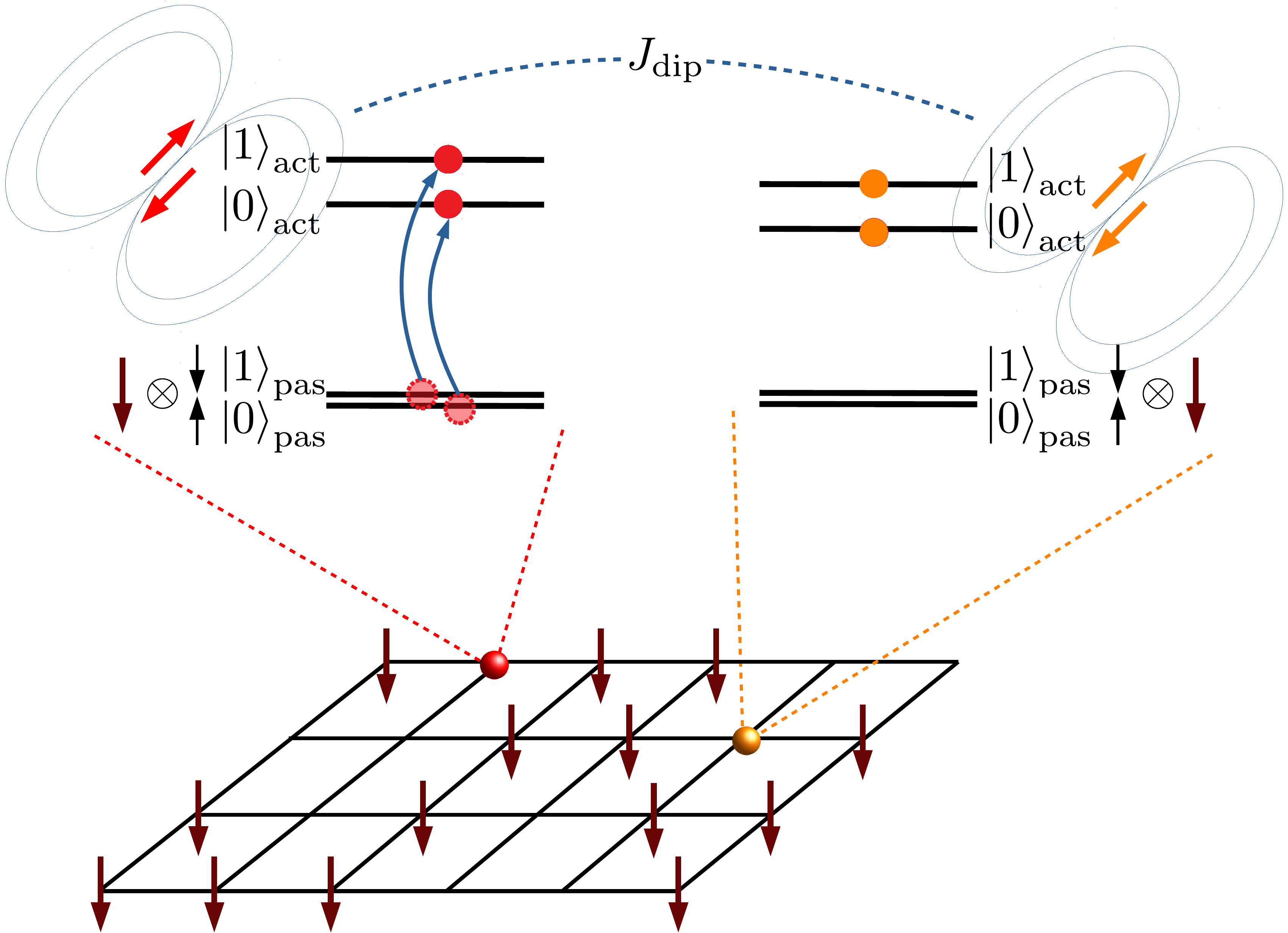} 
			\caption{\label{fig: pair selection} Basic principle of the quantum computing scheme. The passive (memory) qubit consists of nuclear degrees of freedom in the presence of polarized electronic RE spins. Even though the electronic spins create substantial internal fields, they are much smaller than the external field, such that the electronic system remains in its gapped, fully polarized ground state and spin-flips are suppressed. The nuclear qubits can be selectively activated by optical excitation to a (locally tunable) electronic CF doublet with a large magnetic moment. Two activated qubits communicate via the electronic (magnetic) dipolar interaction, allowing for fast two-qubit gates over distances of several 10 nm, as long as the dipolar interaction is sufficiently strong as compared to $1/T_2$ of the active qubits. This enables gates between non-nearest neighbors. 
			}
		\end{figure}

	% our work
	Here instead we propose a faster two-qubit gate based on the magnetic dipole interaction, which is inspired by two-qubit gates implemented using phosphorus donors in silicon in Ref.~\cite{Hill2015} and is similar to hybrid electron and nuclear spin schemes in nitrogen vacancy centers in diamond~\cite{Hegde2020}. We show the basic principle in Fig.~\ref{fig: pair selection} and the underpinning hierarchy of relevant energy scales in Fig.~\ref{fig: qubit level scheme}. In our set-up, the quantum information is stored in nuclear spin states in the presence of the non-degenerate electronic ground state of Kramers ions (having an odd number of electrons) that are subjected to an external field. Those hyperfine states are ideal as quantum memories due to their weak interactions with the environment. To implement single-qubit and two-qubit gates, we exploit hyperfine interactions and optical pulses to transfer the wavefunctions of the nuclear spins to 'active qubits' consisting of excited electronic doublet states that carry a large magnetic moment. Two such active qubits interact at relatively long distances via magnetic dipolar interactions. By making use of the non-trivial unitary evolution of the active qubits under this Hamiltonian, the full interaction strength can be exploited to execute the gate. This reduces the gate time by up to two orders of magnitude as compared to the dipole blockade scheme.

		%
	% comparison to Silicon
	Our scheme differs significantly from proposals based on exchange coupled donors, for example in silicon~\cite{Kane1998,Kalra2014,Stoneham2003}. Those entangle spins indirectly via an effective interaction mediated by an electronic wavefunction. The radius of this wavefunction is manipulated optically or with electrical gates and tunes the hyperfine interaction strength between nuclei and electrons. Thereby the effective interaction between qubits is switched on and off. Our scheme instead entangles nuclear spins by use of the dipolar interaction between electron spins. The latter offers great freedom in choosing the pairs of qubits to be entangled - in contrast to the spatial constraints inherent to schemes where the qubit interaction is due to controlling the spatial extent of electronic wavefunctions and their overlaps.
	
	% comparison to Hill 2015	
	We note that our quantum computing scheme shares several similarities with the one proposed by Hill \textit{et al}.~\cite{Hill2015} for phosphorus donors in silicon. Similarly as in that proposal, we use nuclear spins as passive qubits, dipole-coupled active qubits and essentially the same CNOT implementation. However, our scheme features several additional advantages which we will discuss in Sec.~\ref{sec: Comparison to implementation with phosphorus donors in silicon}. There are also schemes that use electric dipolar interactions, e.g.\ between phosphorus donors in silicon, cf. Ref.~\cite{Tosi2015}. While similar concepts could be implemented with rare-earth ions, we do not pursue this route further here.

	% outline
	The paper is organized as follows: In Sec.~\ref{sec: Set-up and Hamiltonian} we propose a quantum computing scheme based on a RE system and discuss necessary and desirable properties of its CF levels. We describe how to carry out fast single-qubit gates in Sec.~\ref{sec: Single-qubit gates}. In Sec.~\ref{sec: Two-qubit operations} we present the implementation of a CNOT gate via magnetic dipolar interactions. The selective addressability of individual qubits is discussed in Sec.~\ref{sec: Addressability of selected qubits}. A discussion of all other DiVincenzo criteria~\cite{DiVincenzo2000} for quantum computing, in particular the stability of the qubits that are not directly addressed in one- and two-qubit operations, can be found in the Appendix~\ref{app: DiVincenzo criteria and robustness}. We compare our scheme to the CNOT implementation via the dipole blockade and to a phosphorus donor-based set-up by Hill~\textit{et al}.~\cite{Hill2015} in Sec.~\ref{sec: Comparison to similar schemes}. We present a case study of erbium-doped Y$_2$SiO$_5$ as a promising material. We summarize our main results in Sec.~\ref{sec: Conclusion}.

%%%%%%%%%%%%%%%%%%%%%%%%%%%%%%%%%%%%%%%%%%%%%%%%%%%%%%%%%%%%
\section{Set-up and Hamiltonian \label{sec: Set-up and Hamiltonian}}	

	\subsection{Hamiltonian}
\begin{figure}[ht!] 
			\includegraphics[width=1\columnwidth]{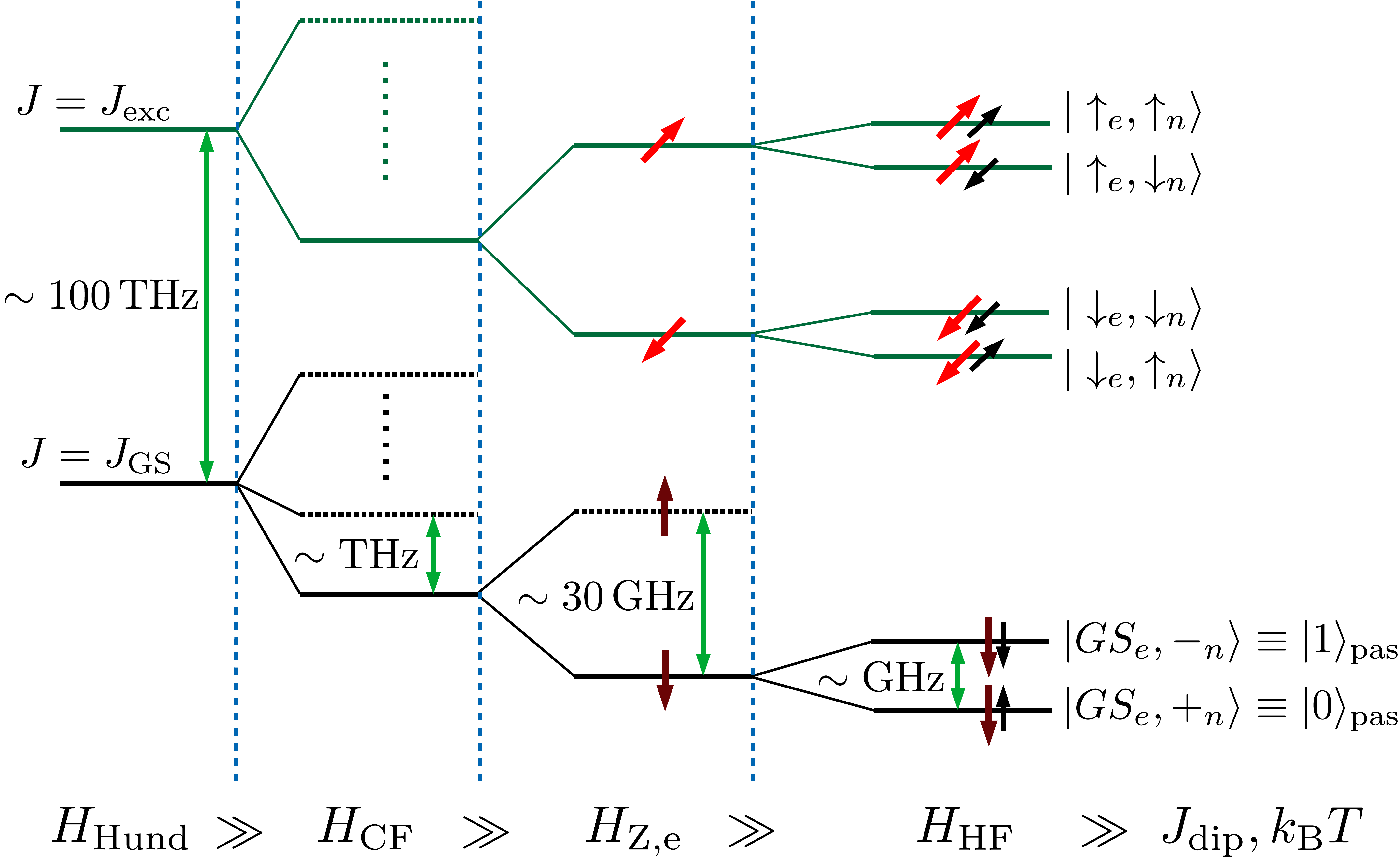} 
			\caption{\label{fig: qubit level scheme} The optimal level scheme of a Kramers RE ion in a CF field, showing the energy scales governing the Hamiltonian~(\ref{eq: H_single}). We use the ground state and one excited spin-orbit manifold, requiring the lowest lying states of both manifolds to be CF doublets with different anisotropy directions, which we then Zeeman-split by an external field. From the ground state doublet we only need the lower Zeeman state. The coupling to a nuclear spin with $I=1/2$ splits the electronic states into pairs of hyperfine states, where the nuclear spins are either aligned or anti-aligned with the electronic spin. The spatial density of RE ions is adjusted such that the dipolar interactions $J_\mathrm{dip}$ are by far the smallest energy scale (apart from the nuclear Zeeman energy), $J_\mathrm{dip} \ll \Delta E_\mathrm{HF}$. An analogous bound $k_\mathrm{B} T\ll  \Delta E_\mathrm{HF}$ applies to the operation temperature.
			}
		\end{figure}

		% CF Hamiltonian
		To realize qubits, we make use of the hyperfine states of single Kramers RE ions in a solid-state matrix and subjected to an external magnetic field. The effect of the crystal environment on the RE ions is captured by the CF potential~\cite{Stevens1952,Hutchings1964}. It breaks the spherical symmetry of the isolated ion and splits its (2$J$+1)-fold degenerate $J$-manifold into CF levels, whose degeneracy and magnetic properties depend on the point symmetry of the RE site in the crystal. The splitting between different $J$-manifolds is governed by Hund's rules and is typically very large, $\Delta_J/h \sim 100 \THz$. Within a given $J$-manifold the single-ion Hamiltonian consists of the CF potential ($H_\mathrm{CF}$), Zeeman terms ($H_\mathrm{Z, e} + H_\mathrm{Z, n}$) in the presence of an external magnetic field $\vec{B}$, and the hyperfine interaction ($H_\mathrm{HF}$) between the electronic spin, $\vec{J}$, and the nuclear spin, $\vec{I}$,
		\begin{align}
			H_\mathrm{single}(\vec{J},\vec{I}) =& H_\mathrm{CF} + H_\mathrm{Z, e} + H_\mathrm{Z, n} + H_\mathrm{HF} \nonumber \\
			=& V_\mathrm{CF}(\vec{J}) + g_J \mu_\mathrm{B} \vec{B} \cdot \vec{J} \nonumber\\*
			&- g_\mathrm{N} \mu_\mathrm{N} \vec{B} \cdot \vec{I} + A_J \vec{J} \cdot \vec{I}.
		\label{eq: H_single}
		\end{align}
		%
		% hierarch and explanation of terms
		Figure~\ref{fig: qubit level scheme} illustrates the hierarchy of the terms in the Hamiltonian~(\ref{eq: H_single}) of the ideal qubit setup. The CF potential $V_\mathrm{CF}(\vec{J})$ induces the largest splitting within a given $J$-manifold, typically of the order of hundreds of GHz up to several THz. The electronic Zeeman term with the Land\'e factor $g_J$ lifts the degeneracy of time-reversal-symmetry protected Kramers doublets. It induces a splitting linear in the field, of the order of $\sim 2 \mu_\mathrm{B}/h \approx 30 \GHz/\text{T}$. These two terms determine the electronic level structure, which we will use to coherently manipulate the electronic wavefunctions of the RE ions. 
		
		The hyperfine interactions with the nuclear spin $I$ are usually well captured by a contact interaction of the form $\vec{J} \cdot \vec{I}$ with the hyperfine constant $A_J$. This interaction is used to transfer quantum information from nuclear to electronic degrees of freedom, and back. Since the hyperfine interaction is much smaller than the electronic Zeeman splitting, its main effect is to (anti-)align the nuclear spins with the polarized magnetic moments of the electronic doublet states. The hyperfine splitting between different nuclear spin orientations (in a manifold with non-zero $\vec{J}$ and $\vec{L}$) is typically of the order of $A_J/h \sim \text{GHz}$. In this paper we assume the nuclear spin to be $I= 1/2$, but similar reasonings and results apply to larger nuclear spins.
		
		The last relevant energy scale is the (magnetic) dipole interaction between RE ions. We will make use of the interaction between excited electronic states to build a gate between two qubits located at a finite spatial distance. A system of two active qubits is governed by two single-ion Hamiltonians as in Eq.~(\ref{eq: H_single}), and by the magnetic dipolar interaction between the RE electrons
    	\begin{align}
    		H =& \sum_{i=1,2} H_\mathrm{single}^{(i)} + \sum_{\alpha,\beta} J_\mathrm{dip}^{\alpha,\beta} J_\alpha^{(1)} J_\beta^{(2)},
    	\label{eq: H two qubit}
    	\end{align}
    	with the dipolar coupling
    	\begin{align}
    		J_\mathrm{dip}^{\alpha,\beta} =& \frac{\mu_0 (\mu_\mathrm{B} g_J)^2}{4\pi r_{12}^3}  \left(\delta_{\alpha, \beta} - \frac{3r_{12,\alpha}r_{12,\beta}}{r_{12}^2} \right).
    	\end{align}
    	Here the superscripts $(1)$ and $(2)$ label the two RE ions, while $r_{12,\alpha}$ ($\alpha = x,y,z$) are the Cartesian components of the vector $\vec{r}_{12}$ connecting the locations of the two qubits in the lattice. In the following, we always assume the dipole interaction to be much smaller than the hyperfine splitting, which is the case for qubit distances exceeding $r\gtrsim \left(\mu_0 \mu_\mathrm{B}^2/(4 \pi A_J) \right)^{1/3}\sim 2  \text{\,\AA}$. 
		
		Due to the tiny nuclear moment $\mu_\mathrm{N}$, the nuclear Zeeman term and dipolar interactions involving nuclear spins are yet much smaller energy scales. For our scheme the Zeeman term only plays an indirect role, insofar as it increases the fidelity of our two-qubit gate by tuning off resonance a class of undesired higher-order spin-flip processes that could occur while carrying out the gate operation. The effects of dipolar interactions between electronic and nuclear spins is analyzed in Appendix~\ref{app: robustness} where we find that they are negligible. The purely nuclear dipole interactions create entanglement between the nuclear spin qubits on a timescale corresponding to $(\mu_\mathrm{B}/\mu_\mathrm{N})^2 \sim 10^6$ two-qubit operations, which constitutes an intrinsic decoherence time of the qubits.
	
	\subsection{Active and passive qubits: an ideal RE system \label{sec: Active and passive qubits}}
	
	    The optimal CF level structure for our qubit scheme is shown in Fig.~\ref{fig: qubit level scheme}. The level and qubit structure is chosen such as to maximize life- and coherence times, while minimizing single-qubit and two-qubit gate times.
	    
		% passive qubit
		To achieve long coherence times, we store the quantum information in the nuclear spin states ($\ket{+_n}$, $\ket{-_n}$) on RE ions, whose electron shell is in its ground state. These nuclear states serve as memory or 'passive' qubits. We denote its states as $\ket{0}_\mathrm{pas}$ and $\ket{1}_\mathrm{pas}$, respectively. Given the small magnetic moments of the nuclear spins, they hardly interact with each other. Moreover, they couple to phonons only very weakly (indirectly via the electrons), so their coherence time is much bigger than that of electronic states. Coherence times of such hyperfine states of up to six hours have been measured in RE systems~\cite{Zhong2015}.
		
		% switch off electronic interaction --> non-degenerate GS
		For the electronic dipolar interaction~(\ref{eq: H two qubit}) not to entangle the passive qubits, we require a non-degenerate electronic~\footnote{
		We note that the perfect degeneracy between the different passive qubit levels allows for resonant excitation hopping between RE ions, i.e.\ the transfer to a different RE of the $\ket{1}_\mathrm{pas}$ state due to virtual transitions involving excited electronic states. The effective hopping scales as $J_\mathrm{hop} \sim J_\mathrm{dip} \left(A_J/E_\mathrm{Z} \right)^2$ which amounts to about $J_\mathrm{hop}/h \sim 10 \text{\,Hz}$ for a qubit spacing of $r=10 \nm$. This weak hopping is typically rendered off-resonant by small residual inhomogeneities among the RE sites.
		}
		ground state $\ket{GS_e}$, which is achieved by polarizing electronic doublets with a magnetic field. We further assume that all RE ions can be cooled efficiently (e.g.\ by pumping schemes involving higher excited CF states~\cite{Rancic2016,Kindem2020,Lauritzen2008,Cruzeiro2018}) to occupy that state. Due to the relative smallness of the dipolar interaction compared to the Zeeman splitting (provided that the RE's are far enough apart in the host medium, $J_\mathrm{dip}/(\mu_\mathrm{B} B) \sim 10^{-6}$ for distances $r=10 \nm$ and $B = 1 \T$), we can neglect dipole-induced spin-flip terms in the electronic ground state. This effectively 'switches off' the electronic interaction between the passive qubits~\footnote{The longitudinal part of the dipolar interaction is kept. These static dipolar fields can be taken into account as an effective static field that adds to the externally applied field.}. Note that while a non-degenerate electronic ground state could also be achieved with non-Kramers ions, our fast single-qubit gate implementation would fail as we will see below.  
		
\begin{figure}[ht]
			\subfloat[]{
				\includegraphics[width=0.85\columnwidth]{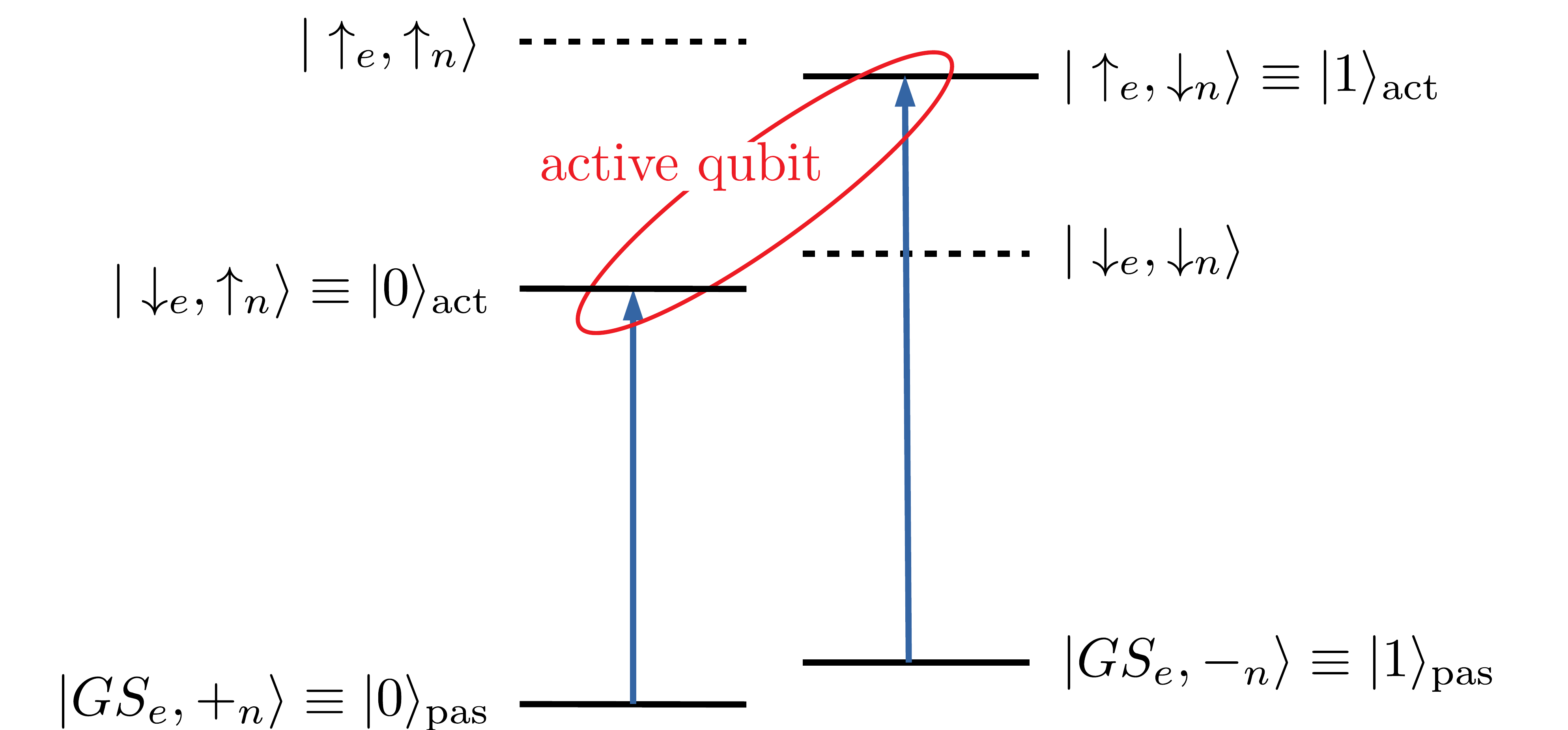}
				\label{Fig: swap_a}
			}
			\vspace{0.2cm}
			\subfloat[]{
				\includegraphics[width=0.85\columnwidth]{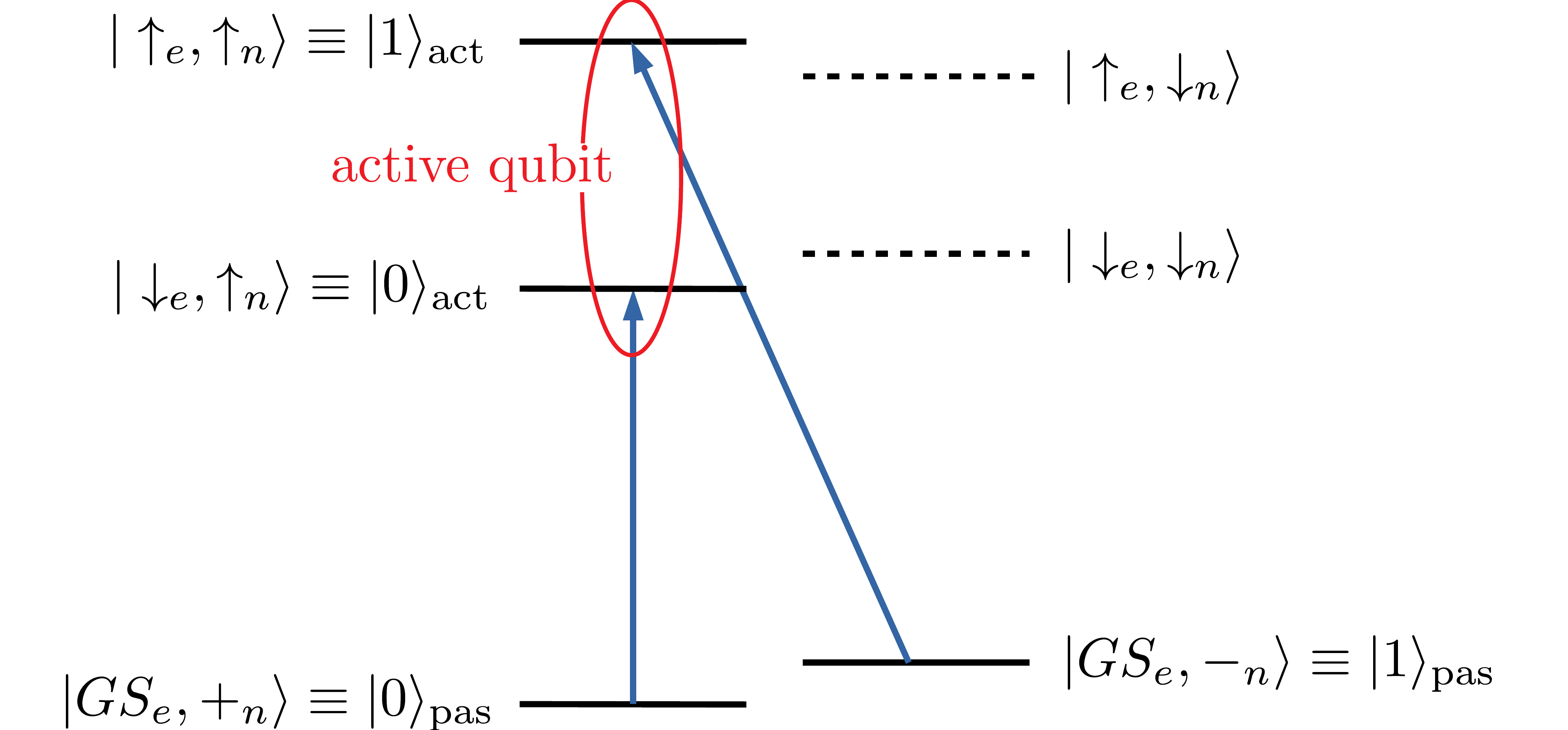}
				\label{Fig: swap_b}
			}
			\caption{\label{fig: swap} Two possible choices for the mapping between passive and active qubits: \protect\subref{Fig: swap_a} electro-nuclear qubit and \protect\subref{Fig: swap_b} purely electronic qubit. Activation of a qubit is achieved by two $\pi$-pulses as discussed in the main text. Each pulse leaves one hyperfine state of the passive qubit untouched and takes the other state into one of two target hyperfine states within the excited electronic doublet. Those two states constitute the active qubit. The two target states have opposite (electronic) magnetic moments, and thus the active qubits can interact over a long distance.} 
		\end{figure}

		% two type of active qubits & activation
		Since the passive qubits are well protected (on the timescales discussed above), their quantum information needs to be transferred to 'active' qubits before one can implement efficient quantum gates on them. A natural candidate for an active qubit is a doublet of excited CF states, which we denote $\up_e$,$\down_e$. In general those have sizable magnetic moments of a few $\mu_\mathrm{B}$, which allow them to interact with other active qubits through dipolar coupling. The hyperfine interaction with the nuclear spin splits each of these CF states into a pair of hyperfine states, similarly as in the passive ground state. One level of either hyperfine pair is selected to represent one of the two active qubit states $\ket{0}_\mathrm{act}$ and $\ket{1}_\mathrm{act}$. To limit decay processes, it is best to choose the qubit state $\ket{0}_\mathrm{act}$ as the lowest of the four hyperfine states. This leaves us with two choices of how to assign the other qubit state $\ket{1}_\mathrm{act}$ to either of the two hyperfine levels of the higher lying Zeeman state. Accordingly we distinguish between an electro-nuclear qubit and a purely electronic qubit, respectively, referring to the degrees of freedom in which the two qubit states differ, cf. Fig.~\ref{fig: swap}. As we will see, in cases where the perpendicular $g$-factor of the doublet is non-zero~\footnote{
				The $g$-matrix governs how an external magnetic field $\vec{B}$ couples to a doublet of CF states $H_\mathrm{doublet} = \frac{\mu_B}{2} \sum_{\alpha,\beta} B_\alpha g_{\alpha \beta} \sigma_\beta$. In many point symmetry groups the basis of the doublet can be chosen such that the $g$-matrix reduces to a diagonal matrix with a longitudinal and transverse component $g_\parallel, g_\perp$, i.e.\ 
				\begin{equation*}
					H_\mathrm{doublet} =  \frac{\mu_B}{2} [g_\parallel B_z \sz + g_\perp (B_x \sx + B_y \sy)].
				\end{equation*}
				For simplicity, we restrict to point symmetries where the g-matrix takes this form.
		}, the electro-nuclear qubit turns out to be the best choice to suppress spin-flip errors during a two-qubit gate. This is because the opposite orientation of the two nuclear spins impedes the two selected electronic spins from flipping simultaneously, which we refer to as an electronic spin flip-flop. However, in special point symmetries the perpendicular $g$-factor ($g_\perp$) can vanish exactly. Under that condition the electronic qubit is the optimal choice.
		
		% our set-up in short: anisotropy & excited doublet
		It is advantageous to further require that the doublets of electronic ground and excited states come with different $g$-factor anisotropy and therefore favor different magnetization directions. This excludes implementations with cubic point symmetries at the RE site. The different anisotropies help to reduce the single-qubit gate time by allowing for an efficient implementation of nuclear spin flips in the ground state by driving optical transitions via the excited doublet states since there is significant overlap between all nuclear spin states in the ground and excited doublet. Vastly different anisotropies are unusual, and a vanishing perpendicular $g$-factor requires special symmetries at the RE sites: tetragonal symmetry with $J=3/2$, or trigonal/hexagonal symmetry with $J>1/2$, cf.\ Table~\ref{tab: Kramers} in the Appendix~\ref{app: Crystal Symmetry} for more details on the symmetry constraints. Nonetheless, as long as the anisotropies of the ground and excited state doublets differ sufficiently, one can carry out fast single-qubit gates. Note that finite perpendicular $g$-factors cannot occur in non-Kramers doublets. Thus, it is impossible to obtain non-parallel magnetization directions in different doublets in these systems~\footnote{
			This symmetry consideration does not take into account the possibility of mixing two different, but energetically close CF levels by a static magnetic field. If two such states are connected by a finite $J_{x/y}$ matrix element, a magnetic field hybridizes them to create two polarized states with a magnetic moment in the transverse direction, even for non-Kramers ions. However, significantly hybridized states with non-negligible transverse magnetization occur only for fine-tuned CF Hamiltonians.
		}.
        We therefore restrict ourselves to Kramers ions.
			
	% activation of qubits
		To activate qubits (i.e., to transfer the quantum wavefunction from a nuclear spin to electronic degrees of freedom) one can use laser pulses that are resonant with specific transitions in the level scheme shown in Fig.~\ref{fig: swap}. By coupling to the transition dipole moments between CF states, one can coherently manipulate their populations via Rabi oscillations, as for example demonstrated in Refs.~\cite{Kukharchyk2018,Rancic2016,Ortu2018,Bottger2009}. 

		To achieve long lifetimes, the active excited doublet should be the lowest CF level of a higher-lying $J$-manifold, whose gap to lower $J$-manifolds is significantly bigger than the Debye frequency. Under those conditions the active qubit state {$\ket{0}_\mathrm{act}$} can only decay via multi-phonon emission or slow photon emission~\footnote{For RE ions the photon emission rate is strongly suppressed, because the electric dipole operator only couples states with opposite parity.}. The decay rate of the excited Zeeman state due to phonon emission is generally very small as well, since the  Debye density of states at the frequency of the Zeeman splitting is tiny. Indeed, lifetimes as long as 60\,s have been reported e.g.\ in (the ground state doublet of) Er-doped Y$_2$SiO$_5$~\cite{Hastings-Simon2008}.
	
%%%%%%%%%%%%%%%%%%%%%%%%%%%%%%%%%%%%%%%%%%%%%%%%%%%%%%%%%%%%	
\section{Single-qubit gates \label{sec: Single-qubit gates}}

\subsection{Implementation of single-qubit gates}
		
	% slow single-qubit gate in ground state
	Single-qubit gates on passive qubits could be implemented by coupling directly to the nuclear spins with microwaves, which drive Rabi oscillations within the pair of hyperfine states constituting the passive qubit. However, this has the severe drawback that the coupling to the nuclear moment is very weak, which implies very long gate times of the order of $t\sim \hbar/(\mu_N B_\mathrm{ac}) = 21 \mus$ (for a amplitude of the driving field of $B_\mathrm{ac} = 1\mT$). Similar qubit proposals based on phosphorus donors in silicon as in Refs.~\cite{Kalra2014,Hill2015} suffer from this problem~\footnote{%
    	Working instead in the regime where the magnetic field is of the order of the hyperfine interaction allows one to exploit states at avoided hyperfine crossings,where the relevant eigenstates are superpositions of both electronic polarization states, that are entangled with the nuclear spin. This structure enables fast transitions within the (passive) qubit due to purely electronic matrix elements~\cite{Morley2010,Morley2013,Wolfowicz2013}. This regime, however, has the drawback that it is unclear how such qubits could be individually addressed and how magnetic dipolar interactions could be ’switched on’ in a fast manner so as to entangle qubits. Furthermore, the (passive) qubit lifetimes are drastically reduced compared to our proposed qubits, since two passive qubits can swap their states directly via the magnetic dipolar interaction. 
    	}.
    
\begin{figure}[ht] 
		\centering 
		\includegraphics[width=1\columnwidth]{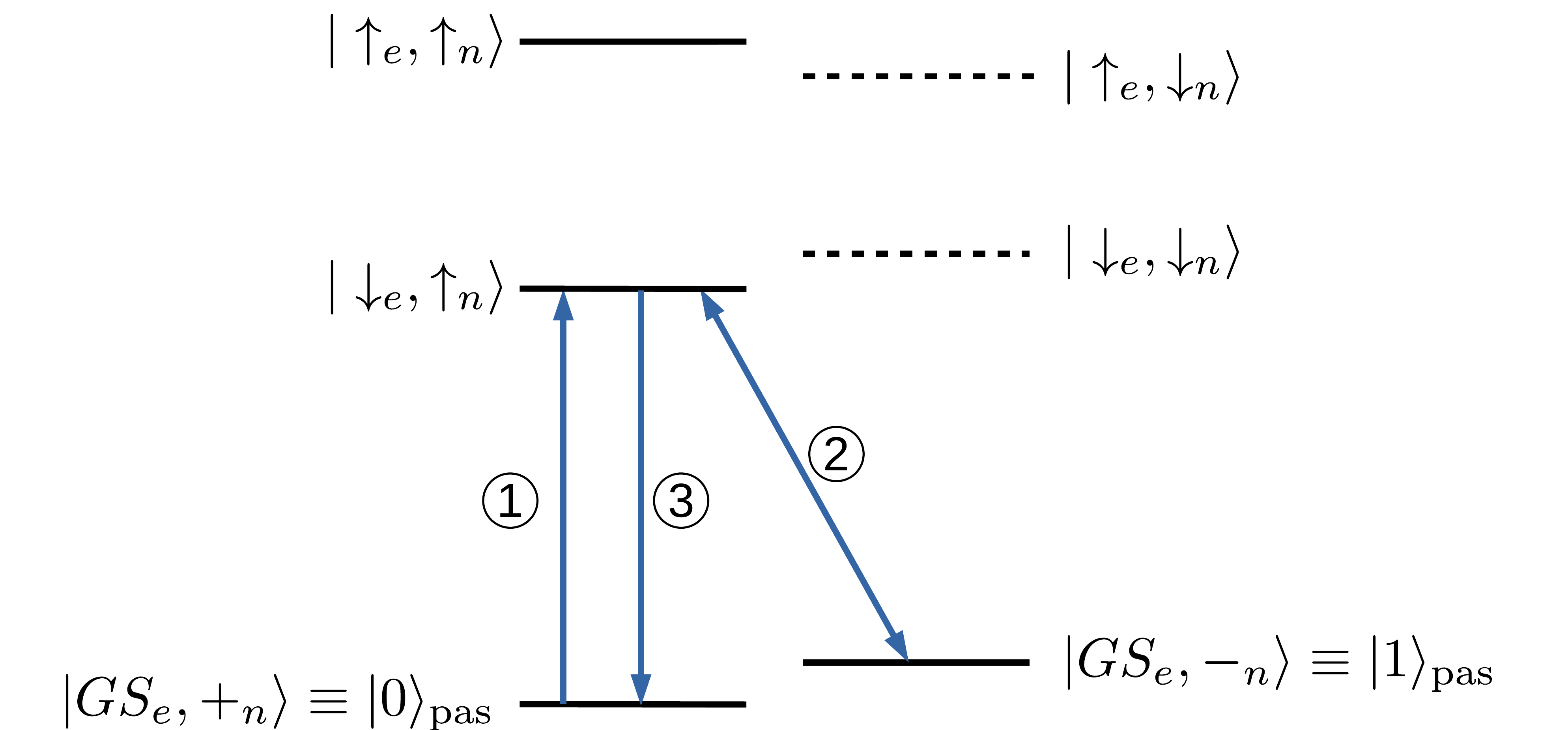}
		\caption{\label{fig: single qubit gate} Implementation of a fast $X$-gate via an excited state. The blue arrows denote $\pi$-pulses which switch the population of the states they connect. The pulses are applied in the order of the indicated numbers. The differing $g$-factor anisotropy in the ground and excited doublet ensure sizable matrix elements for all three transitions.}
	\end{figure}

	% address time of gate with new proposal (different polarizations)
	Here we present a much faster alternative, which takes advantage of the non-parallel magnetic polarizations in ground and excited electronic doublet states. 
	The latter implies that the nuclear spin is not conserved upon switching between ground and excited electronic states. This allows to drive transitions between the passive qubit levels by passing via excited doublet states, which merely requires coupling to the electronic degrees of freedom. For example an $X$-gate, i.e.\ a logical spin-flip, is achieved with three consecutive laser pulses, as sketched in Fig.~\ref{fig: single qubit gate}. In this scheme, the gate time is limited by the inverse of the electronic matrix elements, which are typically much larger than the Zeeman coupling to the nuclear moments. 
	%

	% speed-up for magnetic transitions
	The speed-up of this method depends on selection rules, which involve the value of $J$ in the ground and excited doublet states. The largest speed-up is achieved if magnetic dipole transitions are allowed between the ground and excited $J$-manifold, i.e.\ if $\Delta J = 0,\, \pm1$~\footnote{		
		% speed-up for electric transitions
		In some cases it may nevertheless be favorable to choose an excited manifold which cannot be reached from the ground state by a magnetic dipole transition. In such cases the electric dipole transition from the ground state to the excited doublet can be used. The associated electric dipole moments are typically in the range $\mu_\mathrm{e} = 10^{-34}-10^{-32} \, \text{C\,m}$ (values for LiYF$_4$:Ho$^{3+}$~\cite{Matmon2016}). The speed-up as compared to direct nuclear spin transitions is still of the order of a factor $(c \mu_\mathrm{e})/(3\mu_\mathrm{N}) = 2-200$. 
	}.
	As compared to the direct manipulation of the nuclear spin one achieves a speed-up by a factor of the order of $\mu_\mathrm{B}/(3\mu_\mathrm{N})= 600$~\footnote{%
	To minimize the single-qubit gate time, the overlap of the nuclear spin states in the passive and active qubits should be maximized. For $I=1/2$, the optimal angle between the polarization directions of the active and passive doublet is $\pi/2$, whereas for larger nuclear spins $I>1/2$ it is less than $\pi/2$.
	}. The corresponding gate time is of order $t\sim 3 \hbar/ (\mu_\mathrm{B} B_\mathrm{ac}) = 35 \ns$ for an amplitude of the laser field of $B_\mathrm{ac} = 1\mT$. 
	
\subsection{Fidelity of the single-qubit gate}
    
    Gate fidelities $\mathcal{F}$ or, equivalently, the gate errors $1-\mathcal{F}$ are important because the inverse of the latter yield an estimate of how many gate operations can be carried out until quantum error correction is required. The fidelity of the proposed single-qubit gate is intrinsically limited by $1-\mathcal{F} \gtrsim t_\mathrm{gate}/T_2$ due to the excited state's decoherence time $T_2$ and the finite single-qubit gate time $t_\mathrm{gate}$. Due to the long $T_2$ times of the excited states (up to several milliseconds) such errors are negligible for our fast single-qubit gates that last typically $t_\mathrm{gate}\sim 35\ns$. 
    
    A further source of gate errors is the population transfer of the addressed qubit to (or from) an untargeted hyperfine state on the same ion. In order to ensure that the single-qubit operations have a high fidelity, one needs that (i) the inverse timescale for these processes, i.e.\ the single-qubit Rabi frequency $\Omega$, is significantly smaller than the hyperfine splitting of the addressed level set and (ii) smooth, selective pulses such as (truncated) Gaussian pulses should be chosen so as to limit the range of frequencies they contain. Since the hyperfine splittings are large ($\sim \GHz$), they do not seriously limit the experimental Rabi frequencies as we show in Appendix~\ref{app: Single-qubit fidelity}. There we also discuss further negligible sources of errors due to nuclear and electronic spin-flips of nearby passive qubits during an activation pulse.
    
    The probably most important source of single-qubit errors in our scheme is the unintended activation of non-targeted qubits. This can be prevented by a large detuning of the levels of nearby ions, which can be achieved e.g.\ by locally inducing Stark shifts via electrical gating. Since such detunings are limited to moderate shifts, it is paramount to use selective pulses to minimize the errors on non-targeted qubits, while maintaining a sensible gate speed. The required detunings and electrical fields are discussed in more detail in Sec.~\ref{sec: Addressability of selected qubits}. In conclusion, we expect that sufficiently large Stark shifts can be applied, such that the single-qubit gate fidelity will not be a limiting factor for quantum computing.

%%%%%%%%%%%%%%%%%%%%%%%%%%%%%%%%%%%%%%%%%%%%%%%%%%%%%%%%%%%%%
\section{Two-qubit operations \label{sec: Two-qubit operations}}

    \subsection{Implementation of a CNOT gate \label{sec: CNOT gate implementation}}

    	% universal QC needs two-qubit gate;
    	To achieve universal quantum computing, one needs gates that entangle two qubits~\cite{DiVincenzo2000, Dodd2002}. We  show here an implementation of a logical CNOT gate which follows a similar scheme proposed in Refs.~\cite{Kalra2014,Hill2015}. The CNOT gate flips the target qubit if and only if the control qubit is in state $\ket{1}$.
	
		% transfer of the quantum information; CNOT: Ising-like interaction
		Before carrying out a two-qubit gate operation, we need to activate the two passive qubits by two $\pi$-pulses as illustrated in Fig.~\ref{fig: swap}. Since the pair of active moments is no longer fully polarized (unlike in the passive ground state) but in a superposition of different polarization states, the dipolar interaction entangles the qubits by inducing a configuration-dependent phase. In order to avoid resonances due to flip-flops, i.e.\ the simultaneous flipping of (electronic and/or nuclear) spins on both RE ions, both active qubits are chosen of the same type ('electro-nuclear', or, for $g_\perp=0$, 'electronic'), cf.\ Fig.~\ref{fig: CNOT initialization} and the next subsection.
		
		Upon projection onto the qubit subspace, the dipole interaction between two active qubits reduces to an Ising-like interaction
		\begin{align}
			H_\mathrm{dip} &\approx \frac{\mu_0 (\mu_\mathrm{B} g_{\parallel}/2)^2}{4\pi r_{12}^3} \left(1 - \frac{3(\vec{r}_{12}\cdot {\hat{m}})^2}{r_{12}^2} \right) \sigma_z^{(1)} \sigma_z^{(2)} \nonumber \\
			&\equiv J_\mathrm{dip} \sigma_z^{(1)} \sigma_z^{(2)},
		\label{eq: Ising}
		\end{align}
		where $\hat{m}$ denotes the direction of the magnetic moment in the excited doublet, and $\sigma_z$ is the Pauli matrix acting in the eigenbasis of an active qubit. The small corrections due to transverse interaction terms are addressed in the next subsection.
	    
		% CNOT implementation
		Using the Ising interaction~(\ref{eq: Ising}) in conjunction with single-qubit gates, one can implement a CNOT gate on two active qubits as ~\cite{Chuang2005} 
		\begin{equation}
			\mathrm{CNOT} = \sqrt{Z}_1 \sqrt{Z}_2^\dagger \sqrt{X}_2 C(\pi/2) \sqrt{Y}_2.
			\label{eq: CNOT}
		\end{equation}
		Here $\sqrt{\alpha}_i$ ($\alpha = X,Y,Z)$ denotes a $\pi/4$ rotation of qubit~$i$ around the axis $\alpha$, i.e.\ $\sqrt{\alpha}_i = e^{-i \frac{\pi}{4} \sigma_\alpha^{(i)}}$, and $C(\pi/2) = e^{-i\frac{\pi}{4} \sz^1 \sz^2}$ is implemented by unitary time-evolution under the Ising interaction for a time $t_\mathrm{CNOT} = \pi/(4J_\mathrm{dip})$, which is approximately the gate time of the two-qubit gate. 
		
		% How to generate Ising-like interaction: effective Hamiltonian
		However, the full Hamiltonian~(\ref{eq: H two qubit}) does not only consist of this Ising interaction, but contains additional terms, namely the single-ion Zeeman and hyperfine interactions in Eq.~(\ref{eq: H_single}) as well as those two-qubit interactions that are non-Ising. Even with such terms present, we can reproduce an effective $C(\phi)$ operation 
        % spin-echo
		by eliminating their effect with a spin echo, namely
		\begin{equation}
			C(\pi/2) = X_1 X_2 e^{-\frac{i}{2 \hbar} H_\mathrm{dip} t_\mathrm{CNOT}} X_2 X_1 e^{-\frac{i}{2 \hbar} H_\mathrm{dip} t_\mathrm{CNOT}}.
		\label{eq: spin echo}
		\end{equation}
		Here it is assumed that the $X$-gates are much faster than $t_\mathrm{CNOT}$ and thus can be approximated as instantaneous. 	
		% SWAP back to passive qubits
		After completing a CNOT operation, the quantum information is transferred from active back to passive qubits, analogous to the activation.
		
		Note that the interaction of active qubits with all other (passive) qubits is small: Due to the applied spin-echo in the scheme above, the longitudinal part of the dipolar field (with respect to $\hat{m}$) created by passive qubits cancels out for the active qubit. The dephasing effect of the transverse part is negligible since the dipolar interaction is much smaller than the Zeeman energy, as discussed in more detail in Appendix~\ref{app: robustness}. As a consequence, the activated qubits effectively only talk to each other.

	\subsection{Fidelity of the CNOT gate\label{sec: CNOT gate fidelity}}	
	    % errors by spin-flip terms
		Let us consider the fidelity of our CNOT implementation. So far we have neglected the transverse (non-Ising) parts of the hyperfine and the dipolar interaction, that is, those terms that flip one or both (electronic or nuclear) spins within the excited doublet states. This was motivated by the fact that if we choose to use active electro-nuclear-qubits (and also for electronic qubits, if we assume $g_\perp = 0$) spin-flip terms appear only in higher order perturbation theory in the transverse terms in Hamiltonian~(\ref{eq: H two qubit}). Nevertheless, such transverse perturbations can become important if they connect states that are close in energy.
		\begin{figure}[ht] 
			\subfloat[]{
				\includegraphics[width=1\columnwidth] {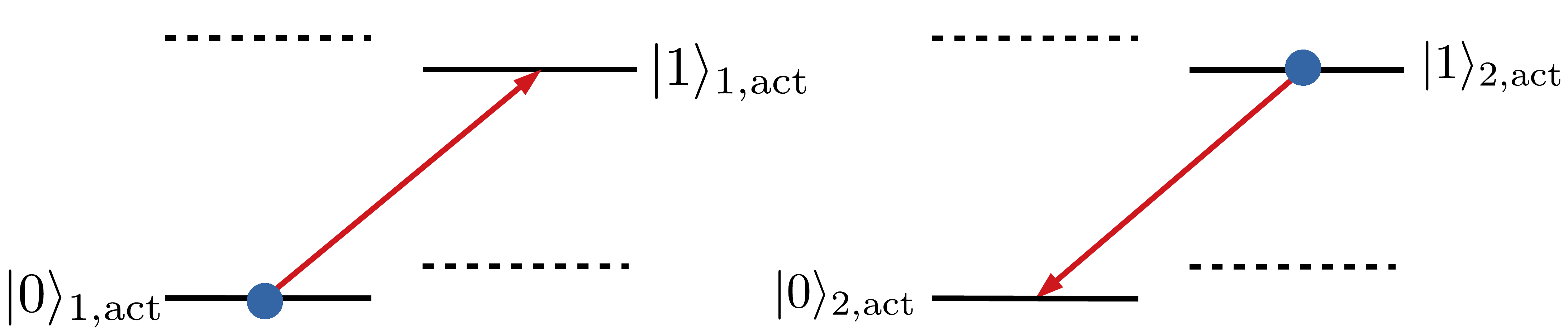}
				\label{Fig: CNOT initialization_a}
			}
			\vspace{0.5cm}
			\subfloat[]{
				\includegraphics[width=1\columnwidth] {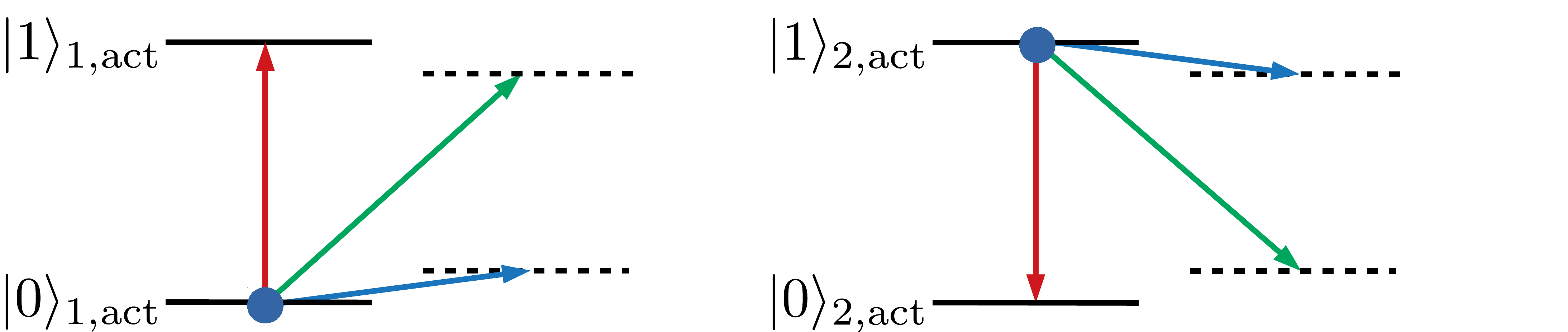}
				\label{Fig: CNOT initialization_b}
			}
			\caption{\label{fig: CNOT initialization} Two possible active qubit choices which are found by minimizing the number of resonant channels and maximizing the lifetime of the active qubits: \protect\subref{Fig: CNOT initialization_a} the electro-nuclear qubit and \protect\subref{Fig: CNOT initialization_b} the electronic qubit. We use here the same level structure as for the excited doublet in Fig.~\ref{fig: swap}, with the indices $1$ and $2$ labeling the two ions on which the CNOT gate acts. The red arrows denote the electro-nuclear and electronic 'spin flip-flop' matrix elements for \protect\subref{Fig: CNOT initialization_a} and \protect\subref{Fig: CNOT initialization_b}, respectively. Such a simultaneous swap of both RE qubits is exactly resonant. The blue and green arrows in \protect\subref{Fig: CNOT initialization_b} denote the nuclear and electro-nuclear flip-flops, respectively, which are nearly resonant up to the nuclear Zeeman energy. These processes limit the fidelity of the CNOT gate in the most favorable case of active qubits with $g_\perp=0$, for which the electronic qubit \protect\subref{Fig: CNOT initialization_b} minimizes the gate error. In most host materials $g_\perp$ is non-zero and the electro-nuclear qubit \protect\subref{Fig: CNOT initialization_a} should be used. 
			}
		\end{figure} 

		In order to minimize the number of such resonant matrix elements, one should realize the active qubits of the same type ('electronic' or 'electro-nuclear') on the two concerned ions, cf.\ Fig.~\ref{fig: CNOT initialization}. Indeed, this leaves only one residual resonant process that has an initial state in the two-qubit subspace. The optimal choice, as well as the resulting fidelity depend on whether or not the transverse $g$-factor of the excited doublet vanishes. If $g_\perp = 0$, there are no residual interactions that act directly within the excited doublet. Any non-Ising matrix elements are thus due to tunneling through intermediate, higher lying CF states. This reduces the magnitude of residual matrix elements and therefore leads to a high fidelity of our CNOT implementation. Typically the electronic active qubit is the better choice in this case. The intrinsic gate error due to the non-Ising terms is then limited by the off-resonant nuclear or electro-nuclear 'spin flip-flops', cf.\ Fig.~\ref{Fig: CNOT initialization_b}. It can be estimated to be negligibly small, $1-\mathcal{F}_\mathrm{min} \sim 10^{-14}$ for a distance of $r=10 \nm$ between the qubits, as we show in detail in Appendix~\ref{app: Fidelity of the CNOT gate}. The fidelity of our CNOT gate implementation will thus rather be limited by the finite coherence time of the active qubits.

		In the case of a finite $g_\perp$, the purely electronic active qubit is ruled out, since the transverse part of the dipole interaction mediates a resonant electronic flip-flop process with matrix element $V \propto \frac{g_\perp}{g_{\parallel}} J_\mathrm{dip}$, which leads to a substantial gate error $1-\mathcal{F}_\mathrm{min} = \mathcal{O}\left((g_\perp/g_\parallel)^2\right)$. However, the electro-nuclear qubit is still viable. The gate error due to the residual interactions can then be estimated to be $1-\mathcal{F}_\mathrm{min} \sim 10^{-5}$ for a distance of $r=10 \nm$ between the qubits, see Appendix~~\ref{app: Fidelity of the CNOT gate}. While this is substantially smaller than the intrinsic fidelity for the case $g_\perp=0$, it is still reasonably small.
		
		The two most important error sources in an experimental setting are most likely (i) the coherence time ($T_2$) of active qubits,  and (ii) the finite time ($t_\mathrm{act}$) to activate the qubits and for the spin-echo pulse. These impose a lower bound on the typical error 
		\begin{equation}
    		1-\mathcal{F} \gtrsim \max \left[\frac{t_\mathrm{CNOT}}{T_2}, \left( \frac{t_\mathrm{act}}{t_\mathrm{CNOT}} \right)^2\right],
    	\label{eq: 1-F bounds}
		\end{equation}
		with the two-qubit gate time $t_\mathrm{CNOT}$. Both $t_\mathrm{act}$ and $t_\mathrm{CNOT}$, and thus both of these lower bounds, can be tuned within some range by the angle of the applied magnetic field. On one hand, the orientation of the field determines the overlap between nuclear spin states in the active and passive qubits and thereby the single-qubit gate time. On the other hand, the field direction also determines the magnetic moment $\mu_m$ of the Zeeman split active qubit states, $2\mu_\mathrm{m}/\mu_\mathrm{B} = \sqrt{(g_{\parallel} \sin(\theta))^2 + (g_{\perp} \cos(\theta))^2}$, and thereby affects the dipolar interaction strength and thus the two-qubit gate time. The orientation of the magnetic field should be chosen such as to minimize these errors.
		
%%%%%%%%%%%%%%%%%%%%%%%%%%%%%%%%%%%%%%%%%%%%%%%%%%%
\section{A scalable quantum computing scheme: addressability of selected qubits \label{sec: Addressability of selected qubits}}

	% scalability/addressability: diluted sample
	A key feature of any viable programmable computer, quantum or classical, is the ability to address only the desired (qu)bits during the fundamental one- and two-(qu)bit gate operations, even as we scale up the processor to handle  many qubits. Below we address this issue  explicitly. All other DiVincenzo criteria~\cite{DiVincenzo2000} for scalable quantum computing are discussed in Appendix~\ref{app: DiVincenzo criteria}.

	Our scheme implements single-qubit manipulations by resonant electro-magnetic pulses. Those should address one and only one qubit at a time. In general one will spatially separate the qubits and work with dilute magnetic RE ions, using materials in which most RE sites are occupied by non-magnetic ions, e.g.\ trivalent Yttrium. Nevertheless, the distance between qubits will still be much smaller than the wavelength of the laser light used for the various gate operations, such that a laser cannot be focused on a single qubit~\footnote{Near field optical excitation could potentially go beyond this limit for samples with a thin, near-surface layer of rare-earth impurities}. Therefore moderate dilution by non-magnetic ions is by itself insufficient to ensure the selective addressability of single qubits. This problem can be overcome by locally altering the gap between the electronic ground state doublet and the excited state doublet of the RE ion to be addressed. 
	
	% magnetic adatom
	\textit{Addressability by static modulation:} If scalability is not an immediate concern, but the aim is merely to selectively address and entangle a few qubits (which is sufficient, e.g., for quantum repeaters), one can achieve a spatially restricted, static modulation of qubit parameters with polarized magnetic adatoms, most easily realized by another species of RE which will be subject to the same (polarizing) external field already imposed on the qubit atoms. Similar ideas to address qubits by statically shifting transition frequencies have been proposed previously, see for example Refs.~\cite{Wesenberg2007,Ahlefeldt2013,Longdell2004,Ahlefeldt2020}.
	 
\begin{figure}[ht] 
		\includegraphics[width=1\columnwidth]{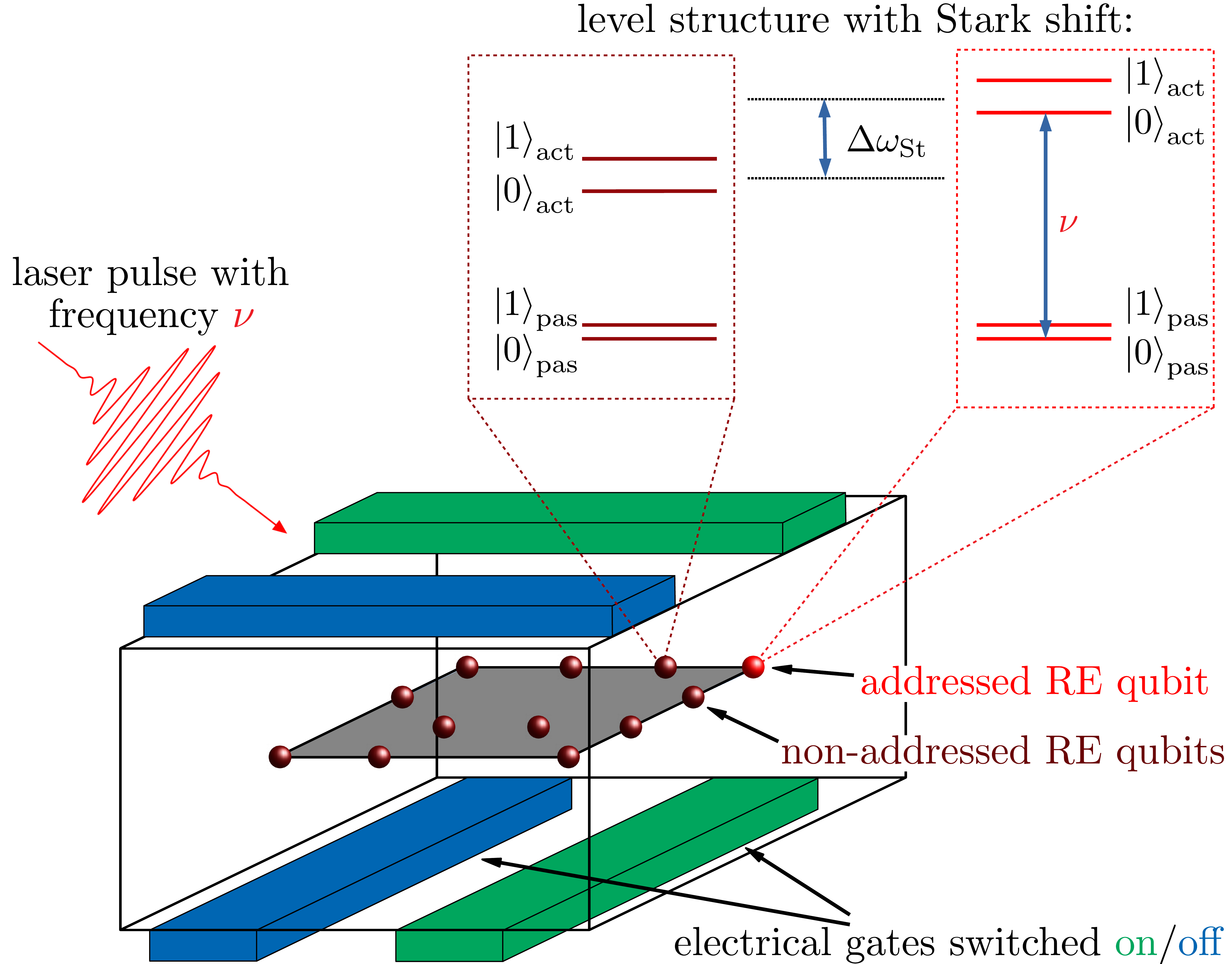} 
		\caption{\label{fig: gating} Sketch of our RE qubit setup with a qubit spacing of order 10\,nm. The qubits are activated with laser pulses. A grid of electrical gates, which can be switched on and off, induces a Stark shift on selected qubits. The difference $\Delta \omega_\mathrm{St}$ in the Stark shift between neighboring RE ions, which is proportional to the electric field gradient, allows one to address single qubits. Note that the sketch is not true to scale, in particular the thickness of the wires and their distance to the qubits are scaled down to make the figure readable.}
	\end{figure}

	\textit{Addressability by dynamic modulation:} For a quantum computer, it is more convenient to generate the spatial modulation of transition frequencies dynamically. To achieve this, we propose to apply local electric fields by local gating. An electric field induces a Stark shift in all CF levels that have electric dipole moments. An array of gates similar to a set-up proposed in Ref.~\cite{Hill2015} for phosphorus donors in silicon or in Ref.~\cite{Wesenberg2007} for Eu ions doped into a Y$_2$SiO$_5$ matrix allows to tune selected RE ions into resonance with a particular laser resonance line, so as to either activate a qubit or perform selective single qubit operations, as sketched in Fig.~\ref{fig: gating}. 
	
	This dynamic modulation requires the relevant CF states to have a non-vanishing electric dipole moment. This is the case for Kramers doublets for almost every crystallographic point symmetry group that does not contain the inversion symmetry~\footnote{The only exceptions occur in the groups $C_{3h}$, $D_{3h}$ and $T_d$: For the double groups of $C_{3h}$ and $D_{3h}$ only the doublets of the $\bar{E}_3$ representation have an electric dipole moment (using the notation of Ref.~\cite{Bradley2010}). The group $T_d$ does not contain an inversion symmetry, but nevertheless does not allow for electric dipole moments in Kramers doublets.}. The point groups for which the Kramers doublets always have electric dipole moments ~- and therefore allow for electrical gating~- are highlighted in bold in Table~\ref{tab: Kramers} in Appendix~\ref{app: Crystal Symmetry}.	

    We estimate the required electric field differences assuming that the single-qubit fidelity is limited by the activation error of the non-targeted qubits and the CNOT gate is limited by the finite activation time. We use a truncated Gaussian pulse with the activation (or single-qubit gate) time $t_\mathrm{act} \approx 2\sqrt{\pi}/\Omega_\mathrm{act}$, where $\Omega_\mathrm{act}$ is the Rabi frequency. For simplicity we assume the same Stark detuning for all $N$ non-targeted qubits within the laser spot size. In order to achieve an activation/single-qubit fidelity $\mathcal{F}_\mathrm{act}$, the Rabi frequency needs to be smaller than $\Omega_\mathrm{act} \lesssim \Delta \omega_\mathrm{St}  \sqrt{\pi/(2 \log(N \pi^2/ (4(1-\mathcal{F}_\mathrm{act}))}$, as we show in Eq.~(\ref{eq: activation detuning}) of the Appendix, and thus the activation time is larger than $t_\mathrm{act} \gtrsim 2\sqrt{\pi}/\Omega_\mathrm{act}$. On the other hand, in order to achieve a CNOT fidelity $\mathcal{F}_\mathrm{CNOT}$, the activation has to be faster than the bound given by the second term in Eq.~(\ref{eq: 1-F bounds}). Combining these two restrictions, the required Stark detuning between targeted and non-targeted qubits can be estimated as
    \begin{equation}
        \Delta \omega_\mathrm{St} \gtrsim \frac{2 \sqrt{2} \mu_0 \mu_B^2}{\pi^2 \hbar r^3} \sqrt{\log \left( \frac{N \pi^2}{4 (1-\mathcal{F}_\mathrm{act})} \right) \frac{1}{1-\mathcal{F}_\mathrm{CNOT}}}.
    \end{equation}

    We consider two levels of desired fidelities  $\mathcal{F}_\mathrm{act} = \mathcal{F}_\mathrm{CNOT}=99\%$ (a lower bound required for topological error correcting codes~\cite{Raussendorf2007}) and $\mathcal{F}_\mathrm{act} = \mathcal{F}_\mathrm{CNOT}=99.99\%$. Assuming a qubit spacing of $r=10\nm$ and a laser spot size of $\sim 1\,\mu\text{m}^2$ (i.e.\ $N\sim 10^4$ affected qubits) with a Stark coefficient $\partial \nu_\mathrm{St}/ \partial \mathcal{E} = 35 \text{\,kHz/(V/cm)}$ as for Eu doped Y$_2$SiO$_5$, this requires a difference of electric fields acting on targeted and non-targeted qubits of order $\Delta \mathcal{E} = \frac{\Delta \omega_\mathrm{St}}{2 \pi \partial \nu_\mathrm{St} / \partial \mathcal{E}} \gtrsim 50-600\frac{\mathrm{V}}{\mathrm{cm}}$. Such field strengths are indeed realistic to achieve in an experimental setting.
		
	% 2d-array
	To be able to electrically gate any selected qubit, one needs to arrange the qubits in a 2d-array. This can be achieved, e.g., by growing a single layer containing dilute magnetic ions on top of bulk material that contains no RE qubits, and covering it with a thin layer of magnetically inert bulk material. This could be realized, e.g., using liquid phase epitaxy~\cite{Rogin1997, Douysset-Bloch1998, Starecki2013} or pulsed laser deposition~\cite{Camposeo2004, Anwar-ul-Haq2009, Secu2017} with first results reviewed in Ref.~\cite{Zhong2019}. Another route is to implant RE ions in silicon~\cite{Tang1989,Lourenco2016,Zhang2019,England2019}, which would allow us to profit from the advanced technology developed for silicon electronics~\cite{Zwanenburg2013}. Single-ion implantation with secondary electron detection for feedback enables the high-fidelity fabrication of 2d-arrays of implanted ions~\cite{England2019}. If using methods that incorporate RE's with considerable positional randomness (unlike in Ref.~\cite{England2019}), the RE layer in Fig.~\ref{fig: gating} will need to be surveyed to identify those atoms usable for qubits and quantum gates. Approaches could include superresolution or near-field microscopy where we take advantage of the strong fluorescence of the rare earths. Voltages applied to the gating grids could  tune in and tune out the fluorescence response and so locate the RE atoms with respect to the gating grids.
	%

%%%%%%%%%%%%%%%%%%%%%%%%%%%%%%%%%%%%%%%%%%%%%%%%%%%
\section{Comparison to similar schemes \label{sec: Comparison to similar schemes}}
	\subsection{Comparison to CNOT implementation via dipole blockade\label{sec: Comparison to dipole blockade implementation}}

		% dipole blockade scheme
		We now compare our scheme to alternative implementations of CNOT gates with dipole-coupled RE ions. Refs.~\cite{Lukin2000,Ohlsson2002,Wesenberg2007} proposed to use the so-called dipole blockade. Here the qubits consist either of two electronic CF states or of two hyperfine levels of a single electronic level. An auxiliary state, to which one qubit state can be excited, is needed as well. All these states interact via their magnetic or electronic dipole moments. For RE ions, magnetic and electric dipolar interactions are usually of similar strength, since the electric dipole moments are typically very small, while typical magnetic moments are fairly large. 
		
\begin{figure}[ht] 
			\centering 
			\includegraphics[width=1\columnwidth]{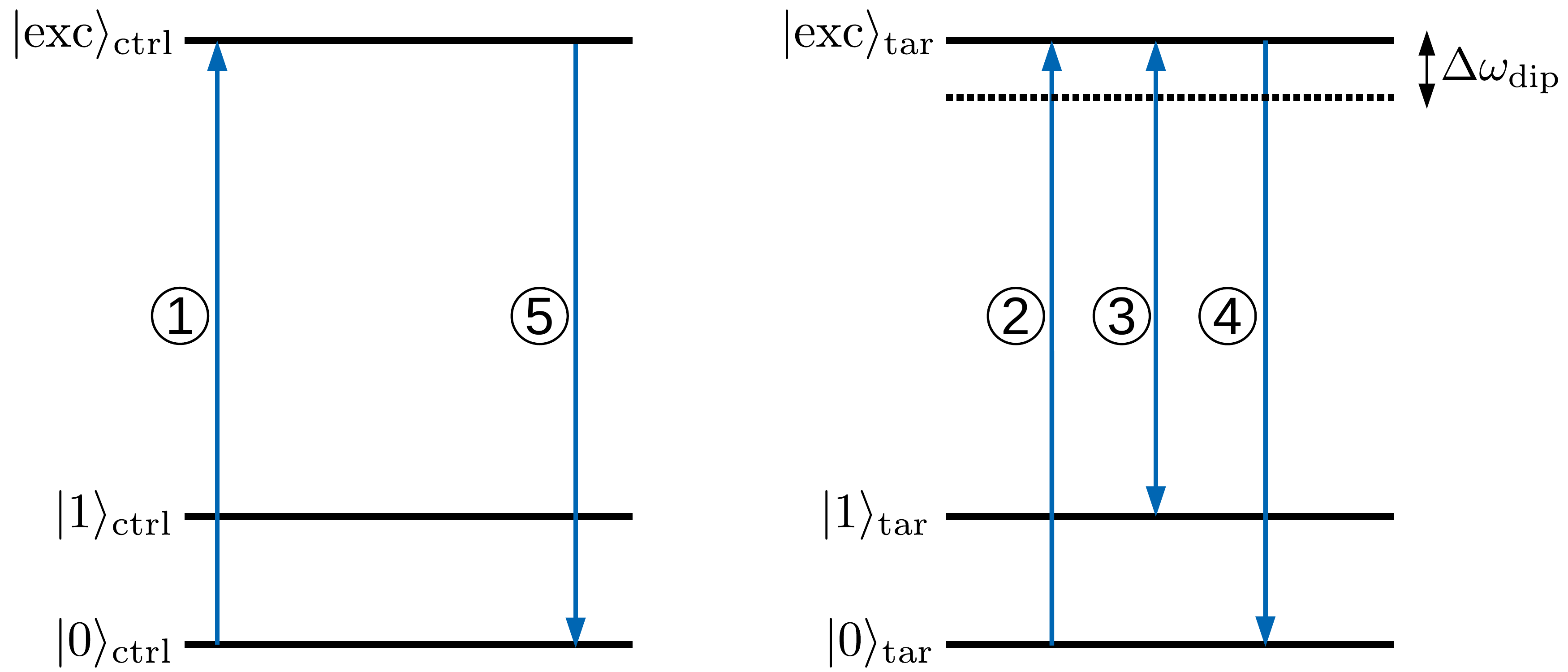} 
			\caption{Pulse sequence for the dipole blockade scheme. If the control qubit is in state $\ket{0}$, it is excited, which shifts the levels of the target qubit out of resonance with the swapping pulses. A swap is thus only carried out if the control qubit was in state $\ket{1}$. The indices 'ctrl' and 'tar' denote the control and target qubit, respectively.}
			\label{fig: dipole blockade} 
		\end{figure}

		The dipole blockade works as follows: If the control qubit is in the $\ket{0}$ state it is transferred to an excited state with different dipole moment. Through its dipolar interaction with the target qubit, the level structure of the latter is changed. Now a sequence of three pulses, whose frequencies match the unmodified level structure of the target qubit, is applied. This swaps the states $\ket{0}$ and $\ket{1}$ of the target qubit, but only if the control qubit was left in the state $\ket{1}$, since the pulse sequence is off-resonant if the control qubit was excited, cf. Fig.~\ref{fig: dipole blockade}. After this conditional swap, the control qubit is brought back to its initial state.
		
\begin{figure}[ht] 
			\includegraphics[width=1\columnwidth]{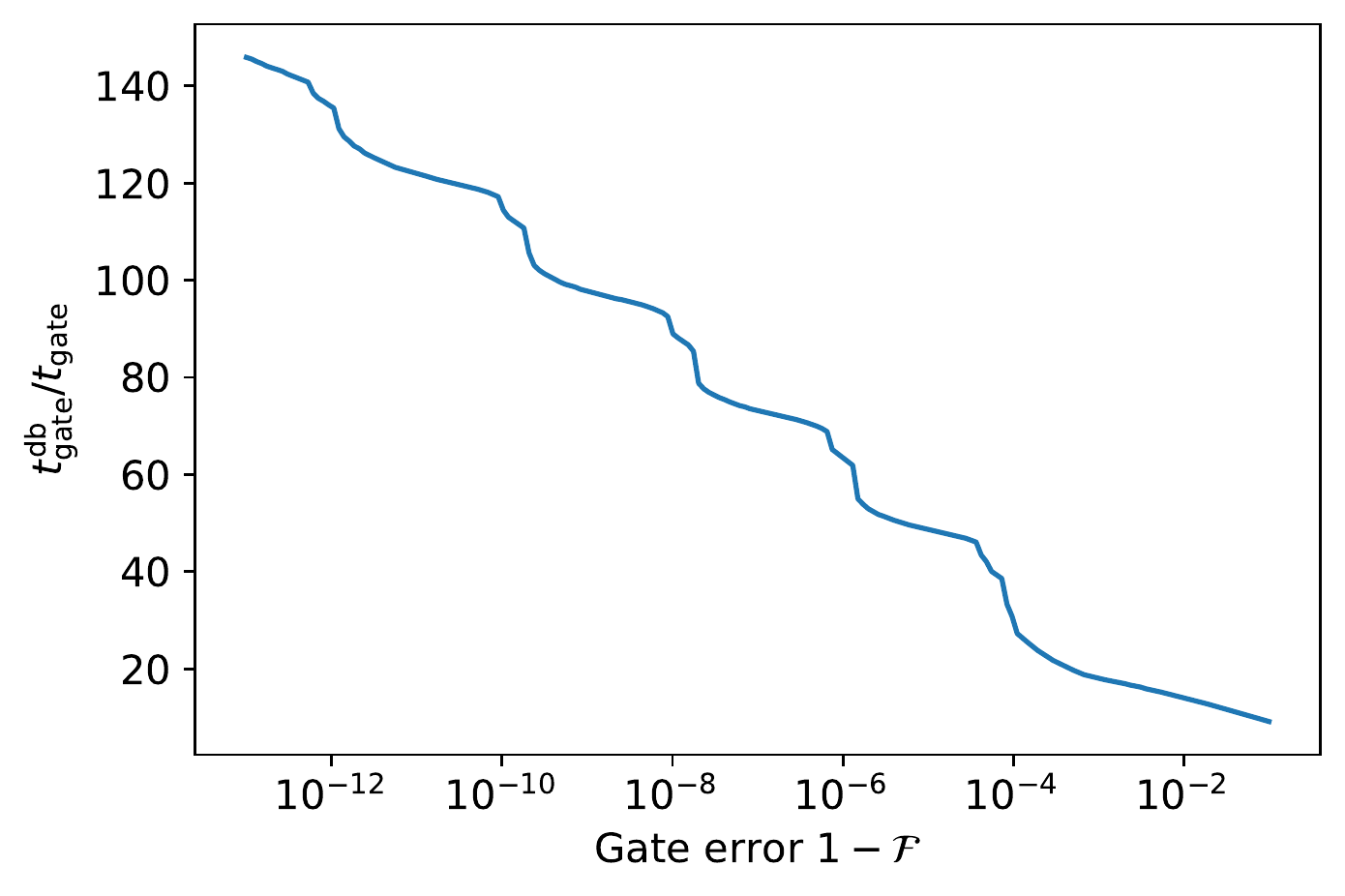} 
			\caption{\label{fig: gate fidelity} The speed-up factor $t_{\rm gate}^{\rm db}/t_{\rm gate}$ by which the gate time of our CNOT implementation is reduced as compared to that of a dipole blockade scheme carried out with Gaussian pulses (assuming that the dipolar interaction is smaller than the hyperfine splitting and the largest achievable Rabi frequencies). The speed-up increases logarithmically with the maximally admissible gate error $1-\mathcal{F}$ which we assume to be given by the probability of a non-resonant excitation (rather than by the decoherence time of the qubits). The oscillatory behavior traces back to Rabi oscillations in the unwanted transition probability.
			}
		\end{figure}

		% our scheme 
		Such a dipole blockade could also be implemented with our magnetic states, and it is thus interesting to compare it to our proposed scheme. We consider the regime where the dipolar interaction is much smaller than the experimentally achievable Rabi frequencies, such that our scheme can be applied (while in the opposite limit only the dipole blockade can be used). This is the regime of qubit distances beyond $r\gtrsim \left( \mu_0 \mu_\mathrm{B}/(4\pi B_\mathrm{ac} \right)^{1/3} \approx 1 \nm$ (with $B_\mathrm{ac} = 1\mT$). Here our CNOT implementation is significantly faster, because it makes use of the full interaction strength between two RE ions, achieving a gate time equal to  $t_\mathrm{CNOT} \approx \hbar \pi/(4 J_\mathrm{dip})$. The dipole blockade instead relies on the ability to resolve the frequency shift induced by the dipole interaction. This requires Rabi frequencies smaller than the interaction strength and consequently gate times that are substantially longer than in our scheme. We estimate the gate time of the dipole blockade in Appendix~\ref{app: Comparison to dipole blockade implementation} by minimizing the $\pi$-pulse duration for a fixed fidelity goal. We numerically evaluated the unitary time-evolution in the rotating-wave approximation and optimized with respect to both the pulse cut-off $T_\mathrm{cut}$ and the Gaussian pulse width $T$, such that for all detunings larger than $J_\mathrm{dip}$ the gate error is below a threshold value $1-\mathcal{F}$. Like for our direct gate, the gate time of the dipole blockade is proportional to $\hbar/J_\mathrm{dip}$, but in addition, it grows logarithmically with the desired gate fidelity, cf.\ Fig.~\ref{fig: gate fidelity}. Therefore, if one requires gate errors to be less than $10^{-4}$, the indirect dipole blockade is by a factor of roughly 50 slower than our gate. 
		
		Fig.~\ref{fig: gate fidelity} shows that the minimal gate time required in the dipole blockade increases monotonically with the required fidelity, but does so in an oscillatory manner. This originates from Rabi-like oscillations of the (unwanted) transition probability, which in turn result from the non-adiabatic switching-on and -off of the pulses at $t=\pm T_\mathrm{cut}$, cf. Appendix~\ref{app: Comparison to dipole blockade implementation} for a detailed calculation. Viewing the fidelity as a function of the $\pi$-pulse duration $T_\pi= 2T_\mathrm{cut}$, the local maxima of the fidelity are spaced by $\Delta T_\pi = 4\pi/\Delta \omega_\mathrm{dip}$.

		We note that a CNOT gate speed-up implies that the gate error due to the finite decoherence time in Eq.~(\ref{eq: 1-F bounds}) is smaller by the same factor in our scheme as compared with the dipole blockade. This allows for the entanglement of qubits at longer distances with better (coherence limited) fidelity in our scheme. On the other hand, the dipole blockade has the advantage that it can still be used in the regime where the dipolar interaction is larger than the hyperfine splitting. In fact for very close-by qubits ($r \lesssim 1 \nm$), the dipole blockade is the better choice of two-qubit gate. If such neighbors could be addressed individually (by magnetic adatoms or gating) in an experimental setting, a combination of dipole blockade and the scheme presented here might be useful to entangle both very close and distant qubits.
	
	\subsection{Comparison to implementation with phosphorus donors in silicon \label{sec: Comparison to implementation with phosphorus donors in silicon}}
		Our scheme resembles in several aspects the one proposed for phosphorus donors in Ref.~\cite{Hill2015}, where the nuclear spins of $^{31}$P$^+$ are used to store quantum information. In contrast to our scheme, the donors have no electronic spin. Instead they are tunnel-coupled to single-electron transistor structures. By gating those, an extra electron is loaded into a spin-down state onto the donor. This operation is the equivalent of our gate-tuning the RE qubits by a local Stark shift, which singles out the qubits which are later selectively addressed. In the proposal of Ref.~\cite{Hill2015}, resonant microwave pulses then transfer the stored quantum information from the nuclear spins to the electronic states. This is the analog of the activation step of our $e$-qubits. Two active qubits interact by dipolar interactions, and the CNOT gate is realized in a similar way as we discuss in this paper. In comparison to donors in silicon, our scheme has several advantages: (i) The magnetic moment of (effective) RE spins are much larger (typically by a factor $5-7$) than those of single electrons. This allows for a much shorter two-qubit gate time, which is inversely proportional to the square of the magnetic moment entering the dipole interaction. (ii) Our single-qubit gate time is up to a factor $\sim 600$ faster since we can make use of the hyperfine interaction and manipulate the nuclear spin states via electronic transitions instead of coupling electromagnetic pulses directly to the nuclear spins. (iii) The qubit activation in our scheme is simpler and does not involve a charge transfer over inter-atomic distances (with associated dipole moments and the danger of exciting the lattice). This is likely to reduce the activation error. 
	
\section{Case study of Y$_2$SiO$_5$:Er$^{3+}$ \label{sec: Case study}}

		To illustrate our scheme and the main ideas of this paper, we discuss a specific RE material: $^{167}$Er-doped Y$_2$SiO$_5$, $^{167}$Er$^{3+}$ being a Kramers ion with nuclear spin $I=7/2$. There are two inequivalent Er sites each with $C_1$ symmetry. The absence of inversion symmetry allows for electrical gating of the qubits. For simplicity we discuss only the Er site~1 (using the notation of Ref.~\cite{Sun2008}). Similar results hold for site~2 as well. A detailed evaluation with all the experimentally measured parameters can be found in Appendix~\ref{app: Case study}. Here we summarize the main results.
		
		We choose the active qubit to be of electron-nuclear type, consisting of the lowest and second highest hyperfine level of the excited doublet, i.e.\ the states with nuclear spin projection $-7/2$ and $-5/2$ with respect to the polarization axis of the excited doublet. The latter us the lowest CF level in the first excited manifold $J=13/2$, accessible with the convenient optical transition wavelength $1536.5 \nm$. The finite coherence time $T_2$ of the excited states has been measured to be as long as $T_2= 4.4 \text{\,ms}$~\cite{Bottger2009}.
		
		A moderate magnetic field of $B \gtrsim 0.26\T$ suffices to achieve the energy scales needed for our scheme to work ($B \gg I A_J/\mu_\mathrm{B} = 26 \mT$). The direction of the externally applied magnetic field is still a further parameter to tune the single- and two-qubit gate times as discussed in Eq.~(\ref{eq: 1-F bounds}). We optimize it numerically in order to achieve the best two-qubit fidelities for qubit spacings $r=10 \nm$. This leads to a coherence-time limited gate fidelity of the CNOT implementation,  with $\mathcal{F}_\mathrm{min} \approx 99.9 \%$ and a total gate time of $4.2 \mus$. A fidelity of above $99\%$ can be achieved for a range of qubit distances $r$ that allow to directly entangle qubits beyond the nearest neighbor distance.
 
		This theoretical fidelity is already above the threshold for quantum error correction. If even higher fidelities are desired, faster single-qubit gates (stronger pulses) and/or longer decoherence times are needed. To improve the latter, one could try to decouple the electronic spins from the surrounding nuclear spins sitting on other (non-RE) atoms. This can be achieved via dynamical decoupling and/or by choosing a host-material where the non-RE ions have smaller or even vanishing magnetic moments. 
		
        The above fidelity estimation shows that Y$_2$SiO$_5$ is a very good host system. It might be an interesting candidate also when doped with other RE ions such as Nd$^{3+}$~\cite{Wolfowicz2015} or Yb$^{3+}$~\cite{Welinski2016}. Another interesting host system to explore is isotopically enriched silicon, where decoherence due to nuclear spins is minimized. However, for that material we currently do not have precise $g$-tensor values and coherence time measurements.

%%%%%%%%%%%%%%%%%%%%%%%%%%%%%%%%%%%%%%%%%%%%%%
\section{Conclusion \label{sec: Conclusion}}
	
	% summarize scheme & necessary ingredients
	In this paper we have proposed an implementation of qubits using RE ions subject to a CF and hyperfine coupling. Our scheme constitutes a promising platform for universal quantum computing. We use nearly decoupled passive (memory) qubits and transfer them to less protected active qubits only when they are involved in a two qubit gate. The passive qubits are well-protected nuclear spin states serving as quantum memory. The active qubit states are magnetized electronic states that can interact over relatively long distances. 
	
	The level scheme of a single ion (illustrated in Fig.~\ref{fig: qubit level scheme}) has the following essential ingredients: (i) a Zeeman-split magnetic ground state; (ii) a nuclear spin and strong hyperfine interactions; (iii) a doublet state at the bottom of an excited $J$-manifold, having a magnetic anisotropy differing from that of the ground state. The last requirement can only be met with Kramers RE ions with non-cubic point symmetry groups. Our estimation of gate errors due to residual (non-Ising) interactions suggest that typically the two-qubit gate fidelity is limited by the coherence time $T_2$ of the active qubit. Using a crystal where the RE site has no inversion symmetry generally allows for electrical gating of the qubits and thereby assures that individual qubits can be addressed selectively. 
	
	As a concrete example and illustration of our scheme, we have discussed the case of Er-doped Y$_2$SiO$_5$. However, the scheme proposed here equally applies to molecular magnets and crystals hosting transition metal ions, as long as they generate a CF level structure that matches the above three criteria. Given the possibilities offered by isotope enrichment and semiconductor processing, we believe silicon doped with RE ions~\cite{Yin2013,Weiss2020} to be a very promising platform.
	
	% summarize CNOT gate
	Using the above ingredients, it is possible to implement a CNOT gate using the Ising-like dipole interaction between two active qubits. This achieves a speed-up by up to two orders of magnitude as compared to the standard dipole blockade, in the regime where dipolar interactions are weak as compared to typical optical Rabi frequencies and the hyperfine splitting.
	
	A decisive advantage of our scheme as compared to many others lies in the fact that the selective activation and the slow power law decay of the dipole interaction allows gates between relatively distant qubits, even if many non-activated qubits are located between the two qubits that are being coupled. 
	In contrast to other similar schemes which also use nuclear spins as passive qubits, we can reach a speed-up for single-qubit gates by a factor of order 600. This is achieved by coupling to electronic instead of nuclear matrix elements which is possible due to different $g$-factors in the ground and excited doublets. Furthermore, the use of the effective spin of a RE shell (as opposed to spins of single electrons as in phosphorus doped silicon) allows for larger magnetic moments of several $\mu_\mathrm{B}$. At equal distances of qubits, the dipolar interaction is enhanced by the square of the larger magnetic moment and the two-qubit gate time is accordingly reduced by an order of magnitude. Thus, we are proposing the implementation of RE qubits with coupled electro-nuclear degrees of freedom and fast single-qubit and CNOT gates that can entangle not only nearest neighbors, but also relatively distant qubits in an array. Combined with the readout capability of single ions and the possibility of coupling RE ions to optical modes, networks of such ions are promising candidates for scalable quantum computing.

\section*{Acknowledgments}
	The authors would like to thank A.~Grimm for useful comments on the manuscript. This work was supported by the Swiss National Science Foundation (SNSF) under grant No. 200021\_166271.
	
%%%%%%%%%%%%%%%%%%%%%%%%%%%%%%%%%%%%%%%%%%%%%%%
%%%%%%%%%%%%%%%%%%%%%%%%%%%%%%%%%%%%%%%%%%%%%%%
%%%%%%%%%%%%%%%%%%%%%%%%%%%%%%%%%%%%%%%%%%%%%%%

\appendix

%%%%%%%%%%%%%%%%%%%%%
\section{Comparing direct dipole interaction vs dipole blockade as CNOT implementations \label{app: Comparison to dipole blockade implementation}}

	% comparison electric and magnetic dipole interaction
	We compare the CNOT gate time of our implementation with the gate time of the dipole blockade as explained in Sec.~\ref{sec: Comparison to dipole blockade implementation}, cf.\ Fig.~\ref{fig: dipole blockade}.
	
	First let us discuss the magnitudes of the magnetic and electric dipolar interactions. As mentioned in the main text, the $4f$-configurations of RE ions have fairly small electric dipole moments. On the other hand, one can have large magnetic moments which compensate for the weaker magnetic interaction. As a consequence, it turns out that the magnetic and electric dipole interactions are typically comparable in magnitude. For example in the material Y$_2$SiO$_5$:Eu$^{3+}$, proposed in Ref.~\cite{Wesenberg2007} as a candidate material for quantum computing, the electric dipole moments of ground and excited manifold differ by $\Delta \mu_\mathrm{e} = 7.7 \times 10^{-32} \,\text{C\,m}$. Typical dipole moments are smaller but of similar order of magnitude. A magnetic moment of only $\mu_\mathrm{m}= 2.5 \mu_\mathrm{B}$ of the doublet states leads to the same interaction strength between the two active qubits assuming vacuum permittivity, i.e.\ $\Delta \mu_\mathrm{e}^2/\epsilon_0 \approx \mu_\mathrm{m}^2/\mu_0$.
	
	% gate time for dipole blockade
	For a distance $r_{12}=10 \nm$ between qubits $1$ and $2$, the electric dipole shift between RE ions is $\Delta \omega_\mathrm{dip} \sim (\Delta \mu_\mathrm{e})^2/(4 \epsilon_0 \hbar \pi r_{12}^3)$. This evaluates to $\Delta \omega_\mathrm{dip} = 2\pi \times 80 \kHz /((r_{12}/10 \nm)^3 )$, which is much smaller than both the typical hyperfine splittings of RE ions. It is also much smaller than the highest experimentally achievable Rabi frequencies for electric dipole transitions. This implies that it is the smallness of the dipolar shift in the spectrum which limits the gating time. Indeed, if the swapping operation on the target qubit is to take place only if the control qubit is in state $\ket{1}$, the Rabi frequency of the three pulses used to swap the target qubit must be significantly smaller than the dipolar shift. In our implementation, which exploits the dipolar interaction directly, the gating time is instead limited by the time evolution under the magnetic dipole interaction, leading to a gate time of $t_\mathrm{CNOT}\approx \hbar \pi/(4 J_\mathrm{dip})$, where the dipolar interaction strength is comparable to the dipole energy shift above $\Delta \omega_\mathrm{dip} \sim J_\mathrm{dip}/\hbar$. The advantage of our scheme is that the system gets entangled directly by the dynamics under this Hamiltonian. Below we estimate the speed-up of our implementation compared to the dipole blockade scheme with Gaussian pulses.

	\subsection{Dipole blockade with Gaussian pulses}
	
		To estimate the gate time of the dipole blockade scheme, we use Gaussian pulses (with cutoffs), since they drive the desired transition more selectively than other pulse shapes~\cite{Chuang2005}. The temporal envelope of the pulse is described by $\Omega(t) = \Omega_0 e^{-t^2/T^2} \theta(T_\mathrm{cut}-|t|)$, where $\Omega_0$ is the Rabi frequency corresponding to the maximal amplitude at time $t=0$ and $T$ is the pulse width. The latter is not to be confused with the pulse duration equal to $T_\pi = 2T_\mathrm{cut}$. Our goal is to optimize the two parameters $T_\mathrm{cut}$ and $\Omega_0$ such as to minimize the $\pi$-pulse duration for a given desired CNOT gate fidelity $\mathcal{F}$ and dipolar frequency shift $\Delta \omega_\mathrm{dip} = J_\mathrm{dip}/\hbar$. Note that since the pulse should carry out a $\pi$-rotation on the resonant state ($\Delta \omega = 0$), the pulse width $T$ is not a free parameter, but depends on $T_\mathrm{cut}$ and $\Omega_0$, as shown below.
		
		For simplicity we consider a two-level system, with one state being the addressed qubit state and the other one being the excited ancillary state. The non-addressed qubit state as well as all other CF states are neglected, as they are assumed to be far off-resonance. The Hamiltonian of this two-level system in the rotating frame of the pulse frequency and with the rotating-wave approximation is~\cite{Vasilev2004}
		\begin{align}
			H^\mathrm{RWA} &= \frac{\hbar \Delta\omega}{2}\sz + \frac{\hbar \Omega_0}{2} e^{-(t/T)^2} \theta(T_\mathrm{cut}-|t|) \sx \nonumber\\
			&\equiv H_0 + V(t),
		\label{eq: HRWA}
		\end{align}
		where $\Delta \omega$ is the detuning between the transition and pulse frequencies. In the dipole blockade scheme $\Delta \omega$ corresponds to the dipolar interaction $\Delta \omega/2 = J_\mathrm{dip}/\hbar$. 
		
		We first derive the pulse width $T$ as a function of $\Omega_0$ and $T_\mathrm{cut}$. The Gaussian pulse should carry out a $\pi$-rotation on the resonant state ($\Delta \omega=0$). Because the Hamiltonian~(\ref{eq: HRWA}) commutes with itself at different times for $\Delta \omega = 0$, the pulse width is determined from the condition on the action integral $\int_{-T_\mathrm{cut}}^{+T_\mathrm{cut}} \mathrm{d}t \, V(t)=\frac{\pi}{2}\sigma_x$. We find
		\begin{equation}
			T=\frac{\sqrt{\pi}}{\Omega_0 \mathrm{erf}(T_\mathrm{cut}/T)},
		\label{eq: T}
		\end{equation}
		where $\mathrm{erf}(x)= (\pi)^{-1/2} \int_{-x}^{x}\mathrm{d}t\, e^{-t^2}$ denotes the error function. 
		
\begin{figure}[t] 
            \includegraphics[width=1\columnwidth]{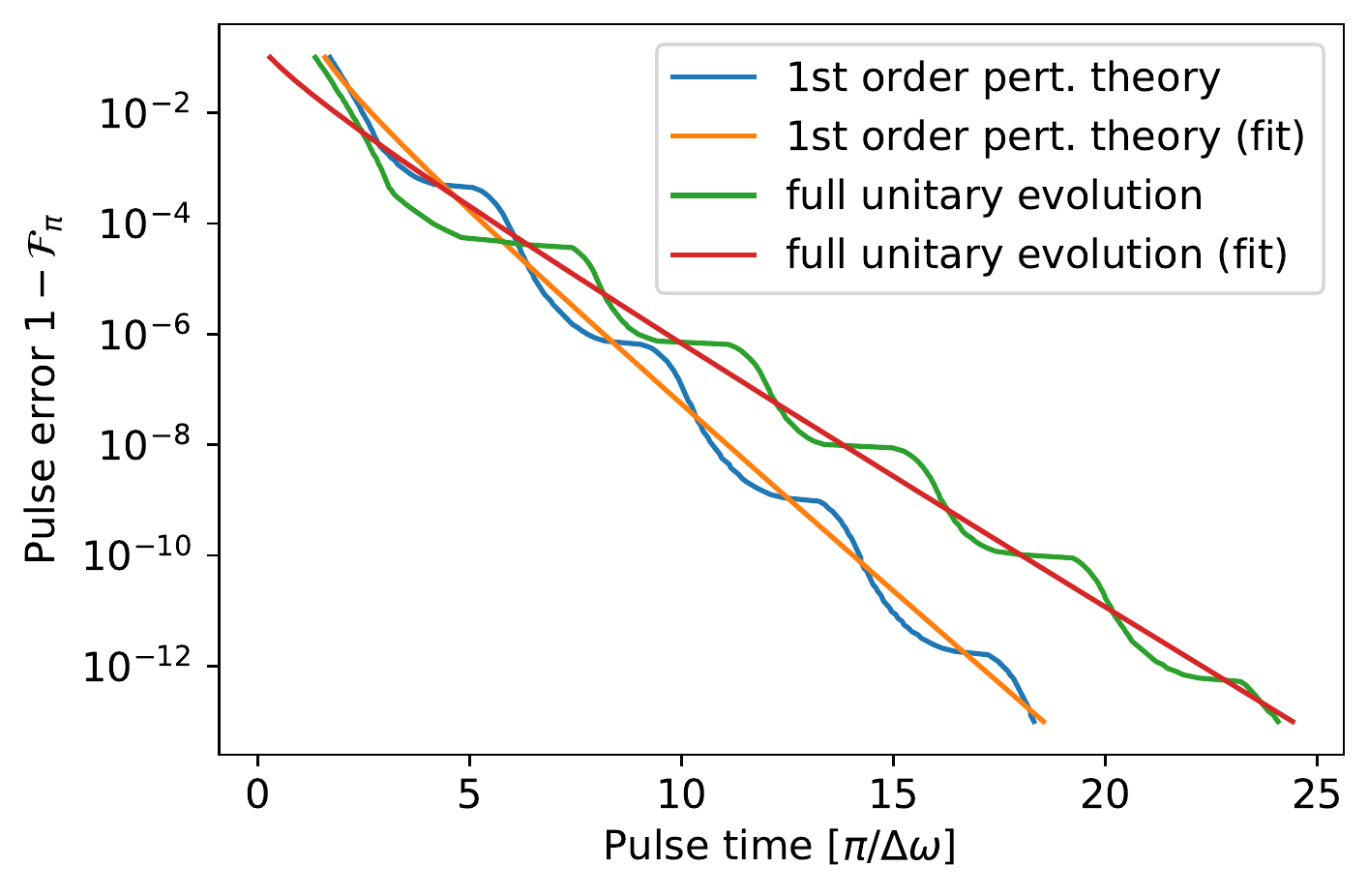} \caption{\label{fig: pi pulse} The maximal error of an optimized $\pi$-pulse with truncated Gaussian envelope as a function of the pulse time $T_\pi = 2 T_\mathrm{cut}$ is plotted for a fixed dipolar coupling $\Delta \omega$. The error decreases exponentially with the pulse time on top of which there are Rabi-like oscillations with period $4\pi/\Delta \omega$. The fits to the numerically optimized pulse times are of the form~(\ref{eq: T pi}).}
        \end{figure}

		We minimize the $\pi$-pulse duration $T_\pi=2T_\mathrm{cut}$ numerically by simulating the unitary time-evolution with the Hamiltonian~(\ref{eq: HRWA}) for fixed $\Omega_0$ and varying $T_\mathrm{cut}$ and $\Delta \omega$. We then determine the minimal pulse duration $T_\pi$ as follows: For each $T_\mathrm{cut}$ we find the smallest detuning $\Delta \omega$ such that for all larger detunings the gate error is below some desired threshold. From this we deduce the minimal $T_\mathrm{cut}$ in units of $1/\Delta \omega$. The result is shown in Fig.~\ref{fig: pi pulse}, where we plot the maximal error $1-\mathcal{F}_\pi$ of the optimized pulse as a function of its duration. For truncated Gaussian pulses one finds $T_\pi$ to grow logarithmically with $(1-\mathcal{F}_\pi)^{-1}$ and exhibiting $4\pi/\Delta \omega$ periodic oscillations in $\log(1-\mathcal{F}_\pi)$.
		
		To understand this result, we use first-order time-dependent perturbation theory and estimate the small spin-flip probability on a state with given detuning $\Delta \omega$. For the following calculations we assume a large detuning $\Delta \omega \gg \Omega_0$, as otherwise the spin-flip probability is non-negligible. Using the interaction picture formalism, the time-evolution operator is expressed as
		\begin{align}
			&U(T_\mathrm{cut},-T_\mathrm{cut}) \nonumber\\
			&= \mathcal{T} \exp\left( -\frac{i}{\hbar} \int_{-T_\mathrm{cut}}^{T_\mathrm{cut}} \mathrm{d}t  \,  e^{\frac{i}{\hbar}H_0t}V(t)e^{-\frac{i}{\hbar}H_0t}\right) \nonumber\\
			&= \mathbb{1}-\frac{i}{\hbar} \int_{-T_\mathrm{cut}}^{T_\mathrm{cut}} \mathrm{d}t  \, e^{\frac{i}{\hbar}H_0t} V(t) e^{-\frac{i}{\hbar}H_0t} + \mathcal{O}(V^2) \nonumber\\
			 &\approx \mathbb{1}- i \sqrt{\pi}\frac{\Omega_0 T}{2} e^{-\left(\frac{\Delta\omega\, T}{2}\right)^2} \Re \left[\text{erf}\left(\frac{T_\mathrm{cut}}{T}-\frac{i\Delta\omega\, T}{2} \right)\right]\sx,
		\label{eq: U approx}
		\end{align}
		where $\mathcal{T}$ is the time-ordering operator. The second term term in the last line (as well as the neglected higher order terms) are expected to be small for $\Delta \omega\, T/2 \gg 1$. In this limit, we can expand the error function for large absolute values of the argument and find the fidelity $\mathcal{F}_\pi$ of a single $\pi$-pulse as
		\begin{equation}
		\begin{split}
			1-&\mathcal{F}_\pi = \lVert \mathbb{1}- U \rVert^2 \approx \left( \frac{\pi}{2 \text{erf}(x)}\right)^2  \\
			&\times \left[ e^{-y^2}- \frac{e^{-x^2}}{\sqrt{\pi}(x^2+y^2)}  \left( x \cos(2xy)-y \sin(2xy) \right) \right]^2,
		\end{split}
		\label{eq: 1-F erf expansion}
		\end{equation}
		where $x\equiv T_\mathrm{cut}/T$ and $y\equiv \Delta \omega\, T/2$ are the real and imaginary parts of the argument of the error function, respectively. Our aim is to minimize the  pulse time $T_{\pi}=2 \, T_\mathrm{cut}= 4 xy$ under the constraint of a fixed error rate $1-\mathcal{F}_\pi$. For the optimal pulse, $x$ and $y$ will both be large and relatively close to each other. Thus due to the exponential factors in Eq.~(\ref{eq: 1-F erf expansion}), $x$ and $y$ scale to leading order with the large parameter $S \equiv \sqrt{\log[1/(1-\mathcal{F}_\pi)]/2}$. Expanding both in $S$, i.e.\ making the ansatz $x= S-\log(S)/(2S) + \alpha/S$ and $y = S+ \beta/S$, $\alpha$ and $\beta$ have to satisfy
		\begin{equation}
        \begin{split}
            &\frac{\pi}{16} e^{-4 (\alpha+ \beta)}  \bigg( e^{2\beta} \big[ \cos(\log(1-\mathcal{F}_\pi)) + \sin(\log(1-\mathcal{F}_\pi)) \big]\\
            &-2 \sqrt{\pi} e^{2\alpha} \bigg)^2 -1 \approx 0,
        \end{split}
        \label{eq: oscillations}
        \end{equation}
        to leading order in $S$. The gate time follows as 
		\begin{equation}
		\begin{split}
		    T_\pi \Delta \omega = 4xy  &\approx 2 \log \left( \frac{1}{1-\mathcal{F}_\pi} \right) \\
		    &- \log \left(\frac{1}{2} \log\left( \frac{1}{1-\mathcal{F}_\pi} \right) \right) +  4 (\alpha+ \beta) .
		\end{split}
		\label{eq: T pi analytical}
		\end{equation}
		With this we reproduce the aforementioned scaling $T_\pi \sim 2 \log[1/(1-\mathcal{F}_\pi)]$. ($\alpha+ \beta$) now has to be minimized subject to the constraint~(\ref{eq: oscillations}) which yields a term that periodically oscillates with $\log(1/(1-\mathcal{F}_\pi))$. This reflects Rabi-like oscillations of the transition probability as $T_\mathrm{cut} \Delta \omega$ increases by multiples of $2\pi$, which results in oscillations with $4\pi/\Delta \omega$ periodicity in $T_\pi$.
        The numerically evaluated maximal gate error as a function of the pulse time is shown in Fig.~\ref{fig: pi pulse} both for the first order expression~(\ref{eq: U approx}), as well as for the full unitary time-evolution. The first order result exhibits the expected slope $T_\pi \Delta \omega \approx 2 \log[1/(1-\mathcal{F})]$. We demonstrate this by fitting a curve of the following form to the numerical result
        \begin{equation}
        \begin{split}
            T_{\pi} \Delta \omega =& a \bigg[ 2\log\left( \frac{1}{1-\mathcal{F}_\pi}\right) \\
            &-  \log \left(\frac{1}{2} \log\left( \frac{1}{1-\mathcal{F}_\pi} \right) \right) \bigg] + b.
        \end{split}
        \label{eq: T pi}
        \end{equation}
        We find $a = 1.06$, which agrees very well with the analytical result~(\ref{eq: T pi analytical}). The result of the full time-evolution shows the same functional behavior with $1-\mathcal{F}_\pi$, albeit with $a = 1.51$ and a different value of $b$. An oscillatory behaviour with $4\pi/\Delta \omega$ is still present. 

		Finally, we estimate the total gate time of the dipole blockade scheme for the pulse sequence in Fig.~\ref{fig: dipole blockade}. Only the pulses on the target qubit are taken into account since their durations are limited by the small dipole interaction strength. Using our approximation of the single $\pi$-pulses in Eq.~(\ref{eq: T pi}), the minimal gate time $t_\mathrm{CNOT}^\mathrm{db}$ of the dipole blockade scheme divided by the gate time $t_\mathrm{CNOT}= \hbar\pi/(4 J_\mathrm{dip}) = \pi/(2 \Delta \omega)$ of our CNOT implementation is
		\begin{equation}
			\frac{t_\mathrm{CNOT}^\mathrm{db}}{t_\mathrm{CNOT}} \gtrsim \frac{6}{\pi} \times  \left( 2a \log \left(  \frac{1}{1-\mathcal{F}}\right) + b \right) .
		\label{eq: speed-up}
		\end{equation}
		In conclusion, we find that the CNOT implementation via dipole blockade is a factor of 25 slower than our implementation for a desired fidelity of $ \mathcal{F}= 99.9\%$, and even a factor of 75 slower for $1-\mathcal{F}=10^{-7}$ (the fidelity we estimated for our scheme in the ideal case $g_\perp=0$ in the main text). The resulting speed-up calculated using the full numerical optimization is presented in Fig.~\ref{fig: gate fidelity} in the main text. Note that for the numerical evaluation of the total gate time as a function of the targeted error rate we take into account that the duration of pulses 2 and 4 is determined by imposing half of the desired gate error, i.e.\ $(1-\mathcal{F})/2$, whereas the duration of pulse 3 is adjusted such that a maximal error of $(1-\mathcal{F})$ results.
		
\section{Fidelities}
\subsection{Single-qubit fidelity \label{app: Single-qubit fidelity}}
    The minimal gate fidelity is defined as~\cite{Gilchrist2005,Nielsen2010},
	\begin{equation}
		\mathcal{F}_\mathrm{min} = \min_{\psi} |\braket{\psi|U_\mathrm{ideal}^\dagger U_\mathrm{exp} |\psi}|^2,
	\label{eq: def F_min}
	\end{equation}
	where $U_\mathrm{ideal}$ is the unitary operator of the ideal quantum gate to be implemented, while $U_\mathrm{exp}$ is its actual implementation. 

    To ensure that the single-qubit operations have a high fidelity, the inverse timescale for these processes (the single-qubit Rabi frequency $\Omega$) has to be  significantly smaller than the hyperfine splitting of the addressed qubit. For simple square pulses, the error due to 'nearly resonant' states with detuning $\Delta \omega$ is estimated from the  theory of Rabi oscillations as $\sim (\Omega/\Delta \omega)^2$. The weak algebraic suppression in $\Omega/\Delta \omega$ is due to the sharp cutoff of a square pulse which entails a broad frequency spectrum. The selectivity of the pulses can be drastically increased (i.e., substantially suppressing errors due to detuned transitions) by choosing smoother pulses such as (cut-off) Gaussian pulses. The error of a single Gaussian $\pi$-pulse is exponentially suppressed for large detuning $\Delta \omega \gg \Omega$
    \begin{equation}
        1-\mathcal{F}_\pi \approx \left(\frac{\pi}{2}\right)^2 \exp \left(-\frac{\pi}{2} \left(\frac{\Delta \omega}{\Omega} \right)^2 \right),
    \label{eq: activation fidelity}
    \end{equation}
    as follows from first order perturbation theory, cf.\ Eq.~(\ref{eq: 1-F erf expansion}) in the Appendix~\ref{app: Comparison to dipole blockade implementation}. Note that in an experimental setting one should optimize the pulse cut-off so as to minimize the pulse duration for a given fidelity, in a similar way as discussed in the Appendix~\ref{app: Comparison to dipole blockade implementation} on the CNOT gate based on the dipole blockade. From Eq.~(\ref{eq: activation fidelity}) it follows that for a desired fidelity of a single $\pi$-pulse, $\mathcal{F}_\pi$, and a given detuning $\Delta \omega$, the admissible Rabi frequency $\Omega$ is bounded from above by
    \begin{equation}
        \frac{\Omega}{\Delta \omega} \lesssim \sqrt{\frac{\pi}{2 \log \left(\frac{ \pi^2}{4(1-\mathcal{F}_\pi)} \right)} }.
    \label{eq: activation detuning}
    \end{equation}
    It follows that for a desired single-pulse  fidelity of e.g.\ $1-\mathcal{F}_\pi=10^{-6}$ the maximally admissible Rabi frequency is $\Omega/\Delta \omega \approx 0.33$ (numerically one finds $\Omega/\Delta \omega \approx 0.26$). This does not impose a serious restriction on the Rabi pulses. Indeed, for experimental pulses with magnetic field strengths of the order of  $B_\mathrm{ac} \sim 1 \mT$ the Rabi frequencies are of order $\mu_B B_\mathrm{ac}/\hbar \sim 90 \MHz$, which is already much smaller than the hyperfine energy scale $\Delta \omega \sim \GHz$.
    
    Note that the finite decoherence time of the excited state also leads to an effective Lorentzian distribution of its energy and thus to a (weak) quadratic asymptotic tail of the pulse fidelity (even for Gaussian pulses). We checked that the error due to excitations from this contribution is negligible for the Rabi frequencies, fidelities and coherence times ($T_2 \sim \text{ms}$)) considered in this paper.
    
    Above we have only discussed gate errors due to undesired excitations. The laser will, however, also induce an AC Stark effect~\cite{Drake2006}, which amounts to a coherent phase accumulation of the qubits. These possibly non-negligible effects have to be taken into account when carrying out multi-qubit operations.

\subsection{Fidelity of the CNOT gate \label{app: Fidelity of the CNOT gate}}	

	% error estimation
	The error of the CNOT implementation due to the off-diagonal matrix elements is quantified by the minimal gate fidelity Eq.~(\ref{eq: def F_min}) and can be estimated as follows. A residual interaction connects two states with energy difference $\Delta$ by a matrix element $V$, which we assume to be sufficiently small: For states with mismatch $\Delta$ larger than the inverse gate time $1/t_\mathrm{CNOT} \sim J_\mathrm{dip}$, we assume the matrix element to be much smaller than $\Delta$, $V\ll \Delta$. Otherwise, we assume that $V\ll J_\mathrm{dip}$. In the latter case, the maximal gate error due to a single residual interaction is estimated as
	\begin{equation}
		\mathcal{F}_\mathrm{min} \approx \cos^2\left({V t_\mathrm{CNOT}/\hbar} \right) \approx 1- \left( \frac{\pi V}{4J_\mathrm{dip}} \right)^2.
	\label{eq: F min resonant}
	\end{equation}
	This decrease in fidelity is due to the resonant matrix element between the (nearly) degenerate configurations which leads to oscillations of the state population between the degenerate states during the gate time.
	
	% non-resonant
	On the other hand, if a gate error is dominated by a non-resonant process ($\Delta \gg J_\mathrm{dip}, V$), the minimal gate fidelity is approximated as
	\begin{equation}
		\mathcal{F}_\mathrm{min} \gtrsim 1- \left(\frac{ 2V}{\Delta}\right)^2.
	\label{eq: F min non-resonant}
	\end{equation}
	This error is due to a finite admixing of a non-resonant state, which leads to an incorrect rotation within the qubit space. The decrease in gate fidelity due to any residual interaction, resonant or not, can be summarized as
	\begin{equation}
		1-\mathcal{F}_\mathrm{min} \propto \left( \frac{V}{\max(J_\mathrm{dip},\Delta)} \right)^2.
	\label{eq: F min}
	\end{equation}

	To optimize the gate fidelity of our CNOT implementation, we have to look at all possible spin-flip processes with the initial state in the active qubit subspace. Here we consider only processes to a final state within the excited electronic doublet space, since processes connecting to other states are detuned by the large CF splitting. Note also, that only spin-flip terms mediated by the dipolar interaction lead to gate errors, whereas single-qubit terms by themselves only lead to a renormalization of the energy levels. {In a given experimental set-up, one should then choose the active qubit realization that minimizes the gate error 
	\begin{align}
		1-\mathcal{F}_\mathrm{min} &\approx \sum_{\alpha} \left(\frac{V_\alpha}{\max(\Delta_\alpha, J_\mathrm{dip})}\right)^2 \nonumber\\
		 &\approx \max_{\alpha}\left(\frac{V_\alpha}{\max(\Delta_\alpha, J_\mathrm{dip})}\right)^2,
	\end{align}
	where the $\{V_\alpha\}$ are the residual matrix elements and $\{\Delta_\alpha\}$ the corresponding energy mismatches.} The latter typically involve the Zeeman energy or the hyperfine coupling and are thus always bigger than the dipolar interaction. The only exceptions are processes that are exactly resonant, $\Delta_\alpha=0$, and, potentially, those that have a mismatch of order of the nuclear Zeeman energy $E_\mathrm{Z,n}$.

	The active qubits with the best decoherence properties are realized in an excited doublet with $g_\perp = 0$. In this case there are no residual interactions that act directly within the excited doublet. Any non-Ising matrix elements are thus due to tunneling through intermediate, higher lying CF states. This reduces the magnitude of residual matrix elements and therefore leads to a high fidelity of our CNOT implementation. The scaling of the matrix elements and therefore the dominant gate errors (assuming the hierarchy of energy scales $J_\mathrm{dip} \ll A_J \ll E_\mathrm{Z} \ll \Delta_\mathrm{CF}$) can be deduced from perturbation theory for the electro-nuclear ($en$) and purely electronic ($e$) qubits as
	%(assuming $g_\perp/g_\parallel = \mathcal{O}(1)$)
	\begin{align}
		en: & \quad 1-\mathcal{F}_\mathrm{min} \approx \max\left[ \left(\frac{J_\mathrm{dip}}{\Delta_\mathrm{CF}}\right)^2, \left(\frac{A_J}{\Delta_\mathrm{CF}}\right)^4 \right], \\
		e: &\quad 1-\mathcal{F}_\mathrm{min} \approx \nonumber\\
		\max&\left[ \left(\frac{J_\mathrm{dip}}{\Delta_\mathrm{CF}}\right)^2, \left(\frac{A_J}{\Delta_\mathrm{CF}}\right)^4 \left( \frac{J_\mathrm{dip}}{\max(J_\mathrm{dip}, E_\mathrm{Z,n})}\right)^2 \right],
	\end{align}
	where $\Delta_\mathrm{CF}$ is the gap to the dominant admixed CF state. For strong dipole interactions $J_\mathrm{dip} > \left(A_J^2/\Delta_\mathrm{CF}\right)$, both qubit choices lead to the same gate error. On the other hand, for weaker dipole interactions $J_\mathrm{dip} < \left(A_J^2/\Delta_\mathrm{CF}\right)$, 
	and if in addition $A_J^2/\Delta_{\rm CF}<E_\mathrm{Z,n}$, the purely electronic qubit is the better choice. For typical values $A_J/h\sim\text{GHz}$, $\Delta_\mathrm{CF}/h\sim 100\GHz$ and an applied field of $B =1 \T$, this is the case if the qubits are separated by distances $r\gtrsim 1.2 \nm$. The gate error is then limited by the off-resonant nuclear or electro-nuclear 'spin flip-flops'. It can be estimated to be as small as $1-\mathcal{F}_\mathrm{min} \sim 10^{-14}$ for a distance of $r=10 \nm$ between the qubits. 
		
	Let us also briefly comment on the case where the transverse $g$-factor of the excited doublet is finite, $g_\perp >0$. We have to choose the electro-nuclear active qubit due to the resonant electronic spin flip-flop matrix element. Assuming a sizable CF splitting, the gate error will be dominated by a process which does not involve higher lying CF states. It scales as
	\begin{equation}
		1-\mathcal{F}_\mathrm{min} \approx \max\left[ \left(\frac{A_J}{E_\mathrm{Z}}\right)^4 , \left(\frac{J_\mathrm{dip}}{A_J}\right)^2 \right].
	\end{equation}
	For typical values of the hyperfine interaction and an applied field of $B=1\T$, the gate error is dominated by the resonant electro-nuclear flip-flop as soon as the qubit spacing exceeds $r\gtrsim 1.3 \nm$ (assuming $g_\perp/g_\parallel = \mathcal{O}(1)$). The gate error due to the residual interactions can then be estimated to be $1-\mathcal{F}_\mathrm{min} \sim 10^{-5}$ for a distance of $r=10 \nm$ between the qubits.

\section{DiVincenzo criteria robustness of non-targeted passive qubits during gate operations \label{app: DiVincenzo criteria and robustness}}

\subsection{DiVincenzo criteria \label{app: DiVincenzo criteria}}

	% DiVincenzo criteria
	For a system to qualify as a platform for quantum computing the DiVincenzo criteria~\cite{DiVincenzo2000} require: \\
	
	\textit{1. A scalable physical system with well characterized qubits:} In our case the well characterized single RE ions serve as qubits as described in the previous sections. Scalability {and the addressability of single qubits} will be discussed in the next subsection.\\
	
	\textit{2. Ability to initialize the state of the qubits to a simple fiducial state:} Qubits can be initialized in their ground state, $\ket{0}_\mathrm{pas}$, by cooling to a temperature significantly below the gap to the first excited state (the upper state of the passive qubit), which is set by the hyperfine interaction. Cooling via a cold finger (heat bath) is usually not sufficient, but dynamic protocols are required. Several experiments have demonstrated such cooling by pumping hyperfine states via excited CF states~\cite{Rancic2016,Kindem2020,Lauritzen2008,Cruzeiro2018}. \\
	
	\textit{3. Long decoherence times as compared to gate times:} We achieve long life and decoherence times by working with long-lived passive memory qubits, which are only transferred into active qubit states right before a multi-qubit operation is carried out, cf. Sec.~\ref{sec: Active and passive qubits}.\\
	
	\textit{4. A universal set of quantum gates:} Single-qubit gates can be implemented efficiently, as described in Sec.~\ref{sec: Single-qubit gates}. To realize a universal set of quantum gates, one needs only one entangling two-qubit gate~\cite{Dodd2002}. We chose to implement the CNOT gate, cf. Sec.~\ref{sec: CNOT gate implementation}. Our implementation is about two orders of magnitude faster than previously proposed implementations based on the dipole blockade. \\
	
	\textit{5. A qubit-specific measurement capability:} In principle, one could measure single RE spins by detecting the emission of a photon after excitation to a high-energy manifold~\cite{Kolesov2012, Siyushev2014}. However, since single-photon measurements have low fidelities, repeated measurements would have to be made; and since the emission process destroys the quantum information, this approach is not viable for our RE qubits. Another way to measure RE spins was suggested in Ref.~\cite{Wesenberg2007}: A different RE ion nearby may serve as the readout ion. One then uses the electric (or magnetic) dipole blockade to shift the transition frequencies of the readout ion. If the qubit to be measured is in state $\ket{0}$, an external laser excites the readout ion and emission can be detected. The excitation and the emission measurement can be repeated without altering the qubit information (apart from the initial wavefunction collapse which projects onto the eigenstate basis of the qubit), assuming that the excited readout ion decays back to its ground state sufficiently rapidly and that there is no component of its wavefunction that becomes entangled with the qubit ions. Alternatively, we could also perform a CNOT operation analogous to the two-qubit gate, with the qubit as the control bit and the readout ion (in its ground state) as the target bit. This allows for a faster readout than the dipole blockade scheme as discussed below, but it necessitates more control over the readout ion.  The requirements for the readout ion are discussed in Ref.~\cite{Wesenberg2007}. A further possibility to read out the qubits consists in measuring the RE charge state after an optical excitation. Such single-ion measurements have been demonstrated for example with erbium defects in silicon~\cite{Yin2013}.
	
\subsection{Robustness of non-targeted passive qubits during gate operations \label{app: robustness}}
	
	While the unintended activation of non-targeted qubits can be suppressed with local detunings as discussed in the main text, we also need to consider the interactions between targeted and non-targeted qubits during (single- and two-qubit) gate operations. These induce errors which cannot be eliminated by the Stark shift above and are therefore intrinsic to our scheme. We argue below that these errors are extremely small and most likely negligible compared to other error sources.
	
	During (de-)activation pulses, one potential source of errors are flips of the electronic or nuclear spins of non-targeted RE ions induced by the time-varying dipolar field of the addressed RE ion. However, the dipolar fields that drive those transitions are much weaker than the fields associated with activation pulses, and thus are a subdominant source of dephasing~\footnote{
    The Rabi frequency associated with the activation of a qubit and a simultaneous spin-flip of a nearby (non-targeted) qubit is suppressed by a factor $J_\mathrm{dip}/E_\mathrm{Z}$ and the transition is detuned by the hyperfine energy as compared to the activation with no spin-flip. Since one already has to minimize activation errors due to undesired hyperfine transitions of the targeted qubit (which are detuned by the hyperfine energy as well) to achieve high single-qubit gate fidelities, simultaneous spin-flips of passive qubits are inherently suppressed. They scale with an additional factor $\sim \left(\frac{J_\mathrm{dip}}{E_\mathrm{Z}}\right)^2 \sim \left( \frac{\mu_0 \mu_B}{4 \pi r^3 B} \right)^2 \sim 10^{-12}$ for qubit spacings of $r=10\nm$ and magnetic fields $B=1 \text{\,T}$, and are therefore negligible.
    }.
    In addition to spin-flip errors, the dipolar field during the pulses results in accumulated phases on the surrounding passive qubits. This error is also negligible since the phase amounts to at most $\delta \phi \sim J_\mathrm{dip} \frac{\mu_\mathrm{N}}{\mu_\mathrm{B}} \frac{t_\mathrm{act}}{\hbar} \sim \frac{\mu_0 \mu_\mathrm{N}}{4 \pi r^3  B_\mathrm{ac}} \sim  5 \times 10^{-7} $ per single-qubit operation for experimental pulses $B_\mathrm{ac} = 1 \mT$ and a qubit spacing of $r=10 \nm$. These essentially random phases accumulate to an $O(1)$ error only after $\sim 1/\delta\phi^2 \sim 10^{12}$  operations. 
    
    The last error source that we have to consider is due to interactions between active qubits and nearby passive qubits during two-qubit gates where the targeted RE ions stay activated for an extended period. These interactions can be described in terms of a dipolar field that depends on the active qubit's state and acts on the passive qubits. The field can be decomposed into components parallel and transverse to the existing average local field on the passive qubit. The effect of the longitudinal part cancels out once a spin-echo (within the active qubit) is applied, as is indeed the case for our CNOT scheme. However, the  transverse part of the field induces a second order correction to the level splitting, which does not reverse under the spin-echo. This results in a deterministic phase accumulation in the passive qubits. The (distance-dependent) phase factor $\delta \phi$  for nearby ions is of order $\delta \phi \sim A_J \left(\frac{J_\mathrm{dip}}{E_\mathrm{Z}} \right)^2 t_\mathrm{CNOT} \sim \frac{A_J J_\mathrm{dip}}{(\mu_\mathrm{B} B)^2} \sim 10^{-8}$ (with $r=10 \nm$, $B=1 \text{\,T}$) and can theoretically be taken into account when multi-qubit operations are carried out. However, it is most likely negligible compared to other error sources.
    
    The considerations above show the robustness of the passive qubits during gate operations in our proposed scheme.

\section{Requirements on the crystal symmetry \label{app: Crystal Symmetry}}
	
	% requirements for doublet states 
	Our scheme relies on the existence of magnetic doublet states with different magnetic anisotropy within the same CF environment. This enables fast single-qubit operations and qubit activation/deactivation processes. We have also shown that $g_\perp=0$ in the excited doublet is a desirable (but not necessary) property since it significantly enhances the intrinsic fidelity of our CNOT gate implementation. These requirements impose constraints on the point symmetry group of the RE ion. 
	
	% Kramers 
	\textit{Kramers ions:} For Kramers ions, (at least) doubly-degenerate states are ensured by time-reversal symmetry whose longitudinal $g$-factor $g_\parallel$ is generally non-vanishing independent of crystal symmetry. Furthermore, due to time-reversal symmetry, CF doublets that contain wavefunctions of the $J$ manifold with $J_z$ eigenvalues $\ket{m_z =\pm 1/2}$ always have a finite transverse $g$-factor $g_\perp \neq 0$. Thus, doublets with different anisotropy directions are expected for Kramers ions independent of the crystal symmetry. This does not hold for the cubic point groups, where the $g$-factors are isotropic. Thus cubic symmetries are not suitable for this scheme. In Table~\ref{tab: Kramers}, we indicate which point symmetry groups also allow for doublets having $g_\perp = 0$, and which ones allow for Stark shifts of the doublets, enabling the electrical gating of qubits.
	\begin{table}
		\caption{Kramers ions will always have some doublets with non-vanishing $g_\perp \neq 0$, irrespective of the point group. Here we tabulate the possibility that also doublets with $g_\perp=0$ occur in the various crystallographic point groups. Point symmetry groups with Kramers doublets that also allow for electric dipole moments are highlighted in bold. \label{tab: Kramers}} 
		\begin{ruledtabular}
			\begin{tabular}{c|c|c}
				Crystal structure & symmetry group & doublets with $g_\perp = 0$ \\
				\hline
				Triclinic & $\boldsymbol{C_1}, S_2$ & no \\
				\hline
				Monoclinic & $\boldsymbol{C_2}, \boldsymbol{C_S}, C_{2h}$ & no \\
				\hline
				Orthorombic & $\boldsymbol{D_2}, \boldsymbol{C_{2v}}, D_{2h}$ & no\\
				\hline
				\multirow{2}{*}{Tetragonal} & $\boldsymbol{C_4}, \boldsymbol{S_4}, C_{4h}$ & \multirow{2}{*}{yes ($J=3/2$)} \\
				\cline{2-2}
				& $\boldsymbol{D_4}, \boldsymbol{C_{4v}}, \boldsymbol{D_{2d}}, D_{4h}$ & \\
				\hline
				\multirow{2}{*}{Trigonal} & $\boldsymbol{C_3}, S_6$ & \multirow{ 2}{*}{yes ($J>1/2$)}\\
				\cline{2-2}
				& $\boldsymbol{D_3}, \boldsymbol{C_{3v}}, D_{3d}$ & \\
				\hline 
				\multirow{2}{*}{Hexagonal} & $\boldsymbol{C_6}, \boldsymbol{C_{3h}}, C_{6h}$ & \multirow{2}{*}{yes ($J>1/2$)} \\
				\cline{2-2}
				& $\boldsymbol{D_6}, \boldsymbol{C_{6v}}, D_{3h}, D_{6h}$ & \\
				\hline
				\multirow{ 2}{*}{Cubic} & $\boldsymbol{T}, T_h$ & no \\
				\cline{2-3}
				& $T_d, \boldsymbol{O}, O_h$ & no
			\end{tabular}
		\end{ruledtabular}
	\end{table}
	%
	
	% non-Kramers
	\textit{Non-Kramers ions:} Non-Kramers ions can host magnetic (degenerate) CF states only if the point symmetry group of the RE ion has irreducible representations that are higher-dimensional or complex. This is the case for the tetragonal, trigonal, hexagonal, and cubic point symmetry groups.  However, all these non-Kramers doublets are of Ising type, having vanishing perpendicular $g$-factors. Only the triplets of the cubic groups allow for finite matrix elements of the transverse magnetization operator between degenerate states. However, those come with an isotropic $g$-factor, which is not suitable for our scheme.

\section{Case study of Y$_2$SiO$_5$:Er$^{3+}$ \label{app: Case study}}
				
	The ground state manifold has angular momentum $J_{\rm GS}=15/2$. Due to the low symmetry at the RE site, the $g$-tensor has three inequivalent eigenvalues $g_x \neq g_y \neq g_z$ (in contrast to the discussion in the main part of the paper with $g_x=g_y \equiv g_\parallel$). The full $g$-tensor and its principal components are listed in Table~\ref{tab: g-tensor}. Each doublet state splits into an octuplet of hyperfine states upon coupling to the nuclear spin $I=7/2$. The passive qubit is encoded in the two states of lowest energy, i.e.\ with nuclear spin projection $-7/2$ and $-5/2$, respectively, with respect to the polarization axis of the electronic ground state. The hyperfine interaction tensor was measured in Ref.~\cite{Guillot-Noel2006}, which seems to be close to the isotropic form in Eq.~(\ref{eq: H_single}). We deduce the hyperfine strength from the largest principal component as $A_{15/2}/h= 103.6 \MHz$. The nuclear $g$-factor is $g_\mathrm{N}=-0.16$~\cite{Stone2005}. 
	\begin{table}
		\caption{ $g$-tensors of the ground state doublets in the $J=15/2$ and $J=13/2$ manifolds of $^{167}$Er:Y$_2$SiO$_5$~\cite{Bottger2009}. We show here only site~1 in orientation~I, orientation II is related by a symmetry operation. \label{tab: g-tensor}} 
		\begin{ruledtabular}
			\begin{tabular}{c c}
			     %& \multicolumn{2}{c}{manifold} \\
			     $J=15/2$ & $J=13/2$ \\
			     \hline
			    $ \begin{pmatrix}
			        \phantom{-}3.07 & -3.12 & \phantom{-}3.40 \\
			        -3.12 & \phantom{-}8.16 & -5.76 \\
			        \phantom{-}3.40 & -5.76 & \phantom{-}5.79
			    \end{pmatrix}$ & %
			    $ \begin{pmatrix}
			        \phantom{-}1.95 & -2.21 & \phantom{-}3.58 \\
			        -2.21 & \phantom{-}4.23 & -5.00 \\
			        \phantom{-}3.58 & -5.00 & \phantom{-}7.89
			    \end{pmatrix} $ 
			\end{tabular}
		\end{ruledtabular}
	\end{table}

\begin{figure}[ht] 
			\includegraphics[width=1\columnwidth]{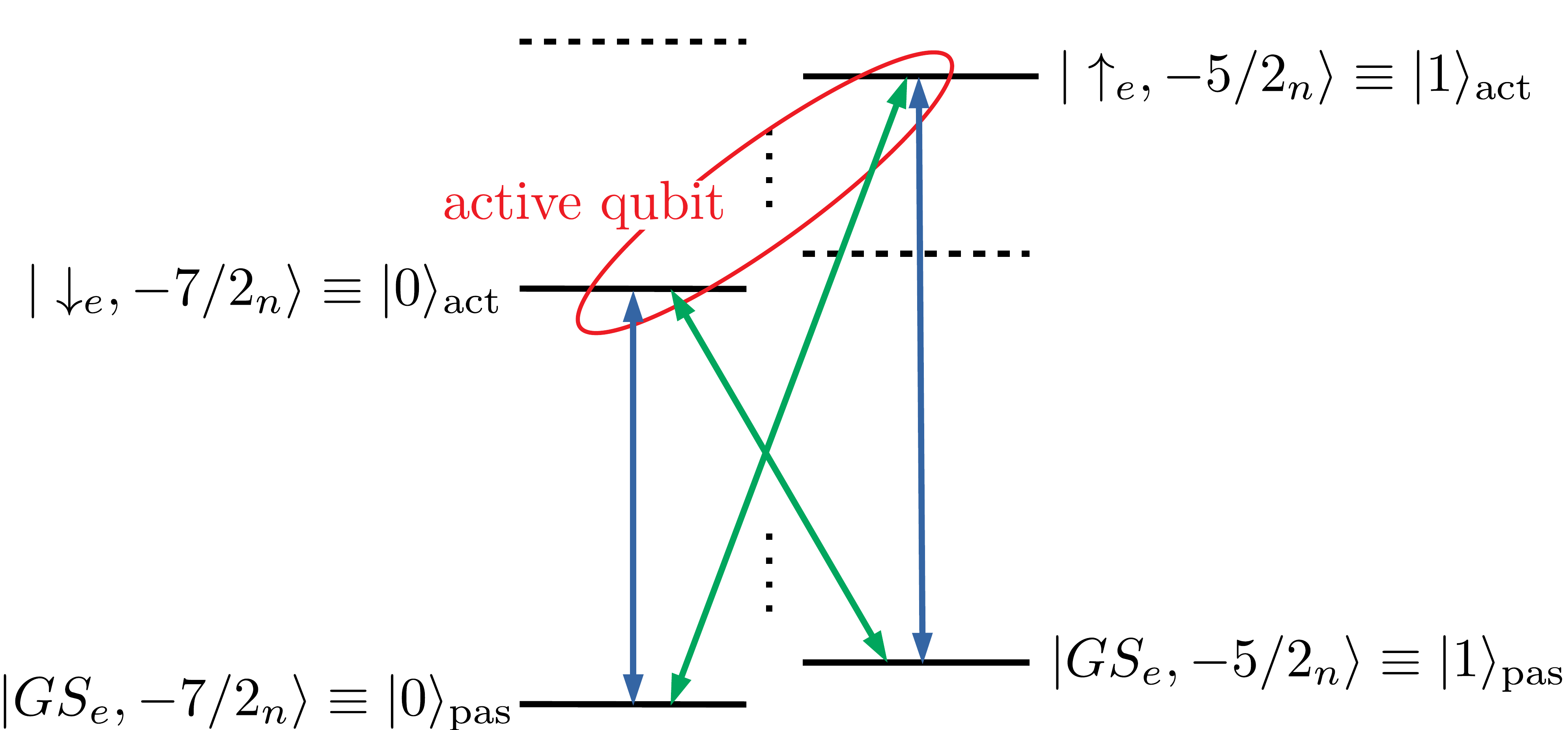} 
			\caption{\label{fig: lierf} Sketch of our choice of passive and active qubits in Y$_2$SiO$_5$:$^{167}$Er$^{3+}$. The nuclear spin states denote the projections in the electronic polarization directions, i.e.\ they are not the same for the ground state and excited doublet. Due to the similar $g$-factor anisotropies in the ground and excited doublet, the transitions indicated with blue arrows are a factor 3.5 faster than the transitions indicated with green arrows.}
		\end{figure}

	The first excited $J$-manifold has total angular momentum $J_{\rm exc}=13/2$ at an energy of $\Delta_J/h = 195 \THz$ corresponding to  wavelength of $1536.5\nm$. The $g$-tensor of the lowest doublet is given in Table~\ref{tab: g-tensor}. Since we have no data on this manifold's hyperfine constant, which usually does not change dramatically between manifolds, we assume that $A_{13/2} = A_{15/2}$. As the transverse $g$-factors do not vanish, the active qubit must be chosen to be an electro-nuclear qubit. More specifically, we choose the active qubit to consist of the lowest and second highest hyperfine level of the excited doublet, i.e.\ the states with nuclear spin projection $-7/2$ and $-5/2$ with respect to the polarization axis of the excited doublet, cf.\ Fig.~\ref{fig: lierf}.
	
	Due to the similar anisotropies of the $g$-factors in the ground and excited doublet, the activation of the qubits (i.e.\ the transition $\ket{0/1}_\mathrm{pas} \leftrightarrow \ket{0/1}_\mathrm{act}$) is faster by a factor of $\sim 3.5$ than the application of an $X$-gate on the active qubits, which involves the slow transitions $\ket{0/1}_\mathrm{pas} \leftrightarrow \ket{1/0}_\mathrm{act}$, cf.\ Fig.~\ref{fig: lierf}. To maximize the gate fidelity it is therefore better to deactivate one of the two qubits while applying an $X$-gate (for the spin echo) on the other active qubit.

	We now estimate the total gate time and CNOT fidelity by taking into account the finite activation and $X$-gate times. For simplicity we neglect all single-qubit gates in the scheme except for the spin echo which implements the two-qubit operation $C(\pi/2)$, cf.\ Eq.~(\ref{eq: CNOT}). The activation times are estimated with the oscillator strength of the transition between the two doublets $f=1.1 \times 10^{-7}$~\cite{Thiel2011}, which corresponds to an electric transition dipole moment of $\mu_\mathrm{e} = 2.0 \times 10^{-32} \text{\,C\,m}$. Assuming pulses with magnetic field amplitudes $B_\mathrm{ac} = 1 \mT$, we find with the optimal polar and azimuthal angles of the magnetic field $\phi=132^\circ$ and $\theta = 35^\circ$, respectively, for a qubit spacing of $r=10\nm$, leading to an extrinsic gate fidelity of the CNOT implementation of $\mathcal{F}_\mathrm{min} \approx 99.9 \%$ and a total gate time of $4.2 \mus$.


\begin{thebibliography}{74}%
	\makeatletter
	\providecommand \@ifxundefined [1]{%
		\@ifx{#1\undefined}
	}%
	\providecommand \@ifnum [1]{%
		\ifnum #1\expandafter \@firstoftwo
		\else \expandafter \@secondoftwo
		\fi
	}%
	\providecommand \@ifx [1]{%
		\ifx #1\expandafter \@firstoftwo
		\else \expandafter \@secondoftwo
		\fi
	}%
	\providecommand \natexlab [1]{#1}%
	\providecommand \enquote  [1]{``#1''}%
	\providecommand \bibnamefont  [1]{#1}%
	\providecommand \bibfnamefont [1]{#1}%
	\providecommand \citenamefont [1]{#1}%
	\providecommand \href@noop [0]{\@secondoftwo}%
	\providecommand \href [0]{\begingroup \@sanitize@url \@href}%
	\providecommand \@href[1]{\@@startlink{#1}\@@href}%
	\providecommand \@@href[1]{\endgroup#1\@@endlink}%
	\providecommand \@sanitize@url [0]{\catcode `\\12\catcode `\$12\catcode
		`\&12\catcode `\#12\catcode `\^12\catcode `\_12\catcode `\%12\relax}%
	\providecommand \@@startlink[1]{}%
	\providecommand \@@endlink[0]{}%
	\providecommand \url  [0]{\begingroup\@sanitize@url \@url }%
	\providecommand \@url [1]{\endgroup\@href {#1}{\urlprefix }}%
	\providecommand \urlprefix  [0]{URL }%
	\providecommand \Eprint [0]{\href }%
	\providecommand \doibase [0]{https://doi.org/}%
	\providecommand \selectlanguage [0]{\@gobble}%
	\providecommand \bibinfo  [0]{\@secondoftwo}%
	\providecommand \bibfield  [0]{\@secondoftwo}%
	\providecommand \translation [1]{[#1]}%
	\providecommand \BibitemOpen [0]{}%
	\providecommand \bibitemStop [0]{}%
	\providecommand \bibitemNoStop [0]{.\EOS\space}%
	\providecommand \EOS [0]{\spacefactor3000\relax}%
	\providecommand \BibitemShut  [1]{\csname bibitem#1\endcsname}%
	\let\auto@bib@innerbib\@empty
	%</preamble>
	\bibitem [{\citenamefont {Kukharchyk}\ \emph {et~al.}(2018)\citenamefont
		{Kukharchyk}, \citenamefont {Sholokhov}, \citenamefont {Morozov},
		\citenamefont {Korableva}, \citenamefont {Kalachev},\ and\ \citenamefont
		{Bushev}}]{Kukharchyk2018}%
	\BibitemOpen
	\bibfield  {author} {\bibinfo {author} {\bibfnamefont {N.}~\bibnamefont
			{Kukharchyk}}, \bibinfo {author} {\bibfnamefont {D.}~\bibnamefont
			{Sholokhov}}, \bibinfo {author} {\bibfnamefont {O.}~\bibnamefont {Morozov}},
		\bibinfo {author} {\bibfnamefont {S.~L.}\ \bibnamefont {Korableva}}, \bibinfo
		{author} {\bibfnamefont {A.~A.}\ \bibnamefont {Kalachev}},\ and\ \bibinfo
		{author} {\bibfnamefont {P.~A.}\ \bibnamefont {Bushev}},\ }\bibfield  {title}
	{\bibinfo {title} {{Optical coherence of $^{166}$Er:$^7$LiYF$_4$ crystal
				below 1 K}},\ }\href {https://doi.org/10.1088/1367-2630/aaa7e4} {\bibfield
		{journal} {\bibinfo  {journal} {New J. Phys.}\ }\textbf {\bibinfo {volume}
			{20}},\ \bibinfo {pages} {023044} (\bibinfo {year} {2018})}\BibitemShut
	{NoStop}%
	\bibitem [{\citenamefont {Ran{\v{c}}i{\'{c}}}\ \emph
		{et~al.}(2018)\citenamefont {Ran{\v{c}}i{\'{c}}}, \citenamefont {Hedges},
		\citenamefont {Ahlefeldt},\ and\ \citenamefont {Sellars}}]{Rancic2016}%
	\BibitemOpen
	\bibfield  {author} {\bibinfo {author} {\bibfnamefont {M.}~\bibnamefont
			{Ran{\v{c}}i{\'{c}}}}, \bibinfo {author} {\bibfnamefont {M.~P.}\ \bibnamefont
			{Hedges}}, \bibinfo {author} {\bibfnamefont {R.~L.}\ \bibnamefont
			{Ahlefeldt}},\ and\ \bibinfo {author} {\bibfnamefont {M.~J.}\ \bibnamefont
			{Sellars}},\ }\bibfield  {title} {\bibinfo {title} {{Coherence time of over a
				second in a telecom-compatible quantum memory storage material}},\ }\href
	{https://doi.org/10.1038/nphys4254} {\bibfield  {journal} {\bibinfo
			{journal} {Nat. Phys.}\ }\textbf {\bibinfo {volume} {14}},\ \bibinfo {pages}
		{50} (\bibinfo {year} {2018})}\BibitemShut {NoStop}%
	\bibitem [{\citenamefont {Zhong}\ \emph {et~al.}(2015)\citenamefont {Zhong},
		\citenamefont {Hedges}, \citenamefont {Ahlefeldt}, \citenamefont
		{Bartholomew}, \citenamefont {Beavan}, \citenamefont {Wittig}, \citenamefont
		{Longdell},\ and\ \citenamefont {Sellars}}]{Zhong2015}%
	\BibitemOpen
	\bibfield  {author} {\bibinfo {author} {\bibfnamefont {M.}~\bibnamefont
			{Zhong}}, \bibinfo {author} {\bibfnamefont {M.~P.}\ \bibnamefont {Hedges}},
		\bibinfo {author} {\bibfnamefont {R.~L.}\ \bibnamefont {Ahlefeldt}}, \bibinfo
		{author} {\bibfnamefont {J.~G.}\ \bibnamefont {Bartholomew}}, \bibinfo
		{author} {\bibfnamefont {S.~E.}\ \bibnamefont {Beavan}}, \bibinfo {author}
		{\bibfnamefont {S.~M.}\ \bibnamefont {Wittig}}, \bibinfo {author}
		{\bibfnamefont {J.~J.}\ \bibnamefont {Longdell}},\ and\ \bibinfo {author}
		{\bibfnamefont {M.~J.}\ \bibnamefont {Sellars}},\ }\bibfield  {title}
	{\bibinfo {title} {{Optically addressable nuclear spins in a solid with a
				six-hour coherence time}},\ }\href {https://doi.org/10.1038/nature14025}
	{\bibfield  {journal} {\bibinfo  {journal} {Nature}\ }\textbf {\bibinfo
			{volume} {517}},\ \bibinfo {pages} {177} (\bibinfo {year}
		{2015})}\BibitemShut {NoStop}%
	\bibitem [{\citenamefont {Kolesov}\ \emph {et~al.}(2012)\citenamefont
		{Kolesov}, \citenamefont {Xia}, \citenamefont {Reuter}, \citenamefont
		{St{\"{o}}hr}, \citenamefont {Zappe}, \citenamefont {Meijer}, \citenamefont
		{Hemmer},\ and\ \citenamefont {Wrachtrup}}]{Kolesov2012}%
	\BibitemOpen
	\bibfield  {author} {\bibinfo {author} {\bibfnamefont {R.}~\bibnamefont
			{Kolesov}}, \bibinfo {author} {\bibfnamefont {K.}~\bibnamefont {Xia}},
		\bibinfo {author} {\bibfnamefont {R.}~\bibnamefont {Reuter}}, \bibinfo
		{author} {\bibfnamefont {R.}~\bibnamefont {St{\"{o}}hr}}, \bibinfo {author}
		{\bibfnamefont {A.}~\bibnamefont {Zappe}}, \bibinfo {author} {\bibfnamefont
			{J.}~\bibnamefont {Meijer}}, \bibinfo {author} {\bibfnamefont
			{P.}~\bibnamefont {Hemmer}},\ and\ \bibinfo {author} {\bibfnamefont
			{J.}~\bibnamefont {Wrachtrup}},\ }\bibfield  {title} {\bibinfo {title}
		{{Optical detection of a single rare-earth ion in a crystal}},\ }\href
	{https://doi.org/10.1038/ncomms2034} {\bibfield  {journal} {\bibinfo
			{journal} {Nat. Commun.}\ }\textbf {\bibinfo {volume} {3}},\ \bibinfo {pages}
		{1029} (\bibinfo {year} {2012})}\BibitemShut {NoStop}%
	\bibitem [{\citenamefont {Siyushev}\ \emph {et~al.}(2014)\citenamefont
		{Siyushev}, \citenamefont {Xia}, \citenamefont {Reuter}, \citenamefont
		{Jamali}, \citenamefont {Zhao}, \citenamefont {Yang}, \citenamefont {Duan},
		\citenamefont {Kukharchyk}, \citenamefont {Wieck}, \citenamefont {Kolesov},\
		and\ \citenamefont {Wrachtrup}}]{Siyushev2014}%
	\BibitemOpen
	\bibfield  {author} {\bibinfo {author} {\bibfnamefont {P.}~\bibnamefont
			{Siyushev}}, \bibinfo {author} {\bibfnamefont {K.}~\bibnamefont {Xia}},
		\bibinfo {author} {\bibfnamefont {R.}~\bibnamefont {Reuter}}, \bibinfo
		{author} {\bibfnamefont {M.}~\bibnamefont {Jamali}}, \bibinfo {author}
		{\bibfnamefont {N.}~\bibnamefont {Zhao}}, \bibinfo {author} {\bibfnamefont
			{N.}~\bibnamefont {Yang}}, \bibinfo {author} {\bibfnamefont {C.}~\bibnamefont
			{Duan}}, \bibinfo {author} {\bibfnamefont {N.}~\bibnamefont {Kukharchyk}},
		\bibinfo {author} {\bibfnamefont {A.~D.}\ \bibnamefont {Wieck}}, \bibinfo
		{author} {\bibfnamefont {R.}~\bibnamefont {Kolesov}},\ and\ \bibinfo {author}
		{\bibfnamefont {J.}~\bibnamefont {Wrachtrup}},\ }\bibfield  {title} {\bibinfo
		{title} {{Coherent properties of single rare-earth spin qubits}},\ }\href
	{https://doi.org/10.1038/ncomms4895} {\bibfield  {journal} {\bibinfo
			{journal} {Nat. Commun.}\ }\textbf {\bibinfo {volume} {5}},\ \bibinfo {pages}
		{3895} (\bibinfo {year} {2014})}\BibitemShut {NoStop}%
	\bibitem [{\citenamefont {Kindem}\ \emph {et~al.}(2020)\citenamefont {Kindem},
		\citenamefont {Ruskuc}, \citenamefont {Bartholomew}, \citenamefont {Rochman},
		\citenamefont {Huan},\ and\ \citenamefont {Faraon}}]{Kindem2020}%
	\BibitemOpen
	\bibfield  {author} {\bibinfo {author} {\bibfnamefont {J.~M.}\ \bibnamefont
			{Kindem}}, \bibinfo {author} {\bibfnamefont {A.}~\bibnamefont {Ruskuc}},
		\bibinfo {author} {\bibfnamefont {J.~G.}\ \bibnamefont {Bartholomew}},
		\bibinfo {author} {\bibfnamefont {J.}~\bibnamefont {Rochman}}, \bibinfo
		{author} {\bibfnamefont {Y.~Q.}\ \bibnamefont {Huan}},\ and\ \bibinfo
		{author} {\bibfnamefont {A.}~\bibnamefont {Faraon}},\ }\bibfield  {title}
	{\bibinfo {title} {{Control and single-shot readout of an ion embedded in a
				nanophotonic cavity}},\ }\href {https://doi.org/10.1038/s41586-020-2160-9}
	{\bibfield  {journal} {\bibinfo  {journal} {Nature}\ }\textbf {\bibinfo
			{volume} {580}},\ \bibinfo {pages} {201} (\bibinfo {year}
		{2020})}\BibitemShut {NoStop}%
	\bibitem [{\citenamefont {Utikal}\ \emph {et~al.}(2014)\citenamefont {Utikal},
		\citenamefont {Eichhammer}, \citenamefont {Petersen}, \citenamefont {Renn},
		\citenamefont {G{\"{o}}tzinger},\ and\ \citenamefont
		{Sandoghdar}}]{Utikal2014}%
	\BibitemOpen
	\bibfield  {author} {\bibinfo {author} {\bibfnamefont {T.}~\bibnamefont
			{Utikal}}, \bibinfo {author} {\bibfnamefont {E.}~\bibnamefont {Eichhammer}},
		\bibinfo {author} {\bibfnamefont {L.}~\bibnamefont {Petersen}}, \bibinfo
		{author} {\bibfnamefont {A.}~\bibnamefont {Renn}}, \bibinfo {author}
		{\bibfnamefont {S.}~\bibnamefont {G{\"{o}}tzinger}},\ and\ \bibinfo {author}
		{\bibfnamefont {V.}~\bibnamefont {Sandoghdar}},\ }\bibfield  {title}
	{\bibinfo {title} {{Spectroscopic detection and state preparation of a single
				praseodymium ion in a crystal}},\ }\href {https://doi.org/10.1038/ncomms4627}
	{\bibfield  {journal} {\bibinfo  {journal} {Nat. Commun.}\ }\textbf {\bibinfo
			{volume} {5}},\ \bibinfo {pages} {3627} (\bibinfo {year} {2014})}\BibitemShut
	{NoStop}%
	\bibitem [{\citenamefont {Chen}\ \emph {et~al.}(2020)\citenamefont {Chen},
		\citenamefont {Raha}, \citenamefont {Phenicie}, \citenamefont {Ourari},\ and\
		\citenamefont {Thompson}}]{Chen2020}%
	\BibitemOpen
	\bibfield  {author} {\bibinfo {author} {\bibfnamefont {S.}~\bibnamefont
			{Chen}}, \bibinfo {author} {\bibfnamefont {M.}~\bibnamefont {Raha}}, \bibinfo
		{author} {\bibfnamefont {C.}~\bibnamefont {Phenicie}}, \bibinfo {author}
		{\bibfnamefont {S.}~\bibnamefont {Ourari}},\ and\ \bibinfo {author}
		{\bibfnamefont {J.}~\bibnamefont {Thompson}},\ }\bibfield  {title} {\bibinfo
		{title} {{Parallel single-shot measurement and coherent control of
				solid-state spins below the diffraction limit}},\ }\href
	{http://www.nature.com/articles/s41586-020-2160-9
		http://arxiv.org/abs/2006.01823} {\bibfield  {journal} {\bibinfo  {journal}
			{Nature}\ }\textbf {\bibinfo {volume} {580}},\ \bibinfo {pages} {201}
		(\bibinfo {year} {2020})}\BibitemShut {NoStop}%
	\bibitem [{\citenamefont {Raha}\ \emph {et~al.}(2020)\citenamefont {Raha},
		\citenamefont {Chen}, \citenamefont {Phenicie}, \citenamefont {Ourari},
		\citenamefont {Dibos},\ and\ \citenamefont {Thompson}}]{Raha2020}%
	\BibitemOpen
	\bibfield  {author} {\bibinfo {author} {\bibfnamefont {M.}~\bibnamefont
			{Raha}}, \bibinfo {author} {\bibfnamefont {S.}~\bibnamefont {Chen}}, \bibinfo
		{author} {\bibfnamefont {C.~M.}\ \bibnamefont {Phenicie}}, \bibinfo {author}
		{\bibfnamefont {S.}~\bibnamefont {Ourari}}, \bibinfo {author} {\bibfnamefont
			{A.~M.}\ \bibnamefont {Dibos}},\ and\ \bibinfo {author} {\bibfnamefont
			{J.~D.}\ \bibnamefont {Thompson}},\ }\bibfield  {title} {\bibinfo {title}
		{{Optical quantum nondemolition measurement of a single rare earth ion
				qubit}},\ }\href {https://doi.org/10.1038/s41467-020-15138-7} {\bibfield
		{journal} {\bibinfo  {journal} {Nat. Commun.}\ }\textbf {\bibinfo {volume}
			{11}},\ \bibinfo {pages} {1605} (\bibinfo {year} {2020})}\BibitemShut
	{NoStop}%
	\bibitem [{\citenamefont {Sangouard}\ \emph {et~al.}(2011)\citenamefont
		{Sangouard}, \citenamefont {Simon}, \citenamefont {{deRiedmatten}},\ and\
		\citenamefont {Gisin}}]{Sangouard2011}%
	\BibitemOpen
	\bibfield  {author} {\bibinfo {author} {\bibfnamefont {N.}~\bibnamefont
			{Sangouard}}, \bibinfo {author} {\bibfnamefont {C.}~\bibnamefont {Simon}},
		\bibinfo {author} {\bibfnamefont {H.}~\bibnamefont {{deRiedmatten}}},\ and\
		\bibinfo {author} {\bibfnamefont {N.}~\bibnamefont {Gisin}},\ }\bibfield
	{title} {\bibinfo {title} {{Quantum repeaters based on atomic ensembles and
				linear optics}},\ }\href {https://doi.org/10.1103/RevModPhys.83.33}
	{\bibfield  {journal} {\bibinfo  {journal} {Rev. Mod. Phys.}\ }\textbf
		{\bibinfo {volume} {83}},\ \bibinfo {pages} {33} (\bibinfo {year}
		{2011})}\BibitemShut {NoStop}%
	\bibitem [{\citenamefont {Saglamyurek}\ \emph {et~al.}(2011)\citenamefont
		{Saglamyurek}, \citenamefont {Sinclair}, \citenamefont {Jin}, \citenamefont
		{Slater}, \citenamefont {Oblak}, \citenamefont {Bussi{\`{e}}res},
		\citenamefont {George}, \citenamefont {Ricken}, \citenamefont {Sohler},\ and\
		\citenamefont {Tittel}}]{Saglamyurek2011}%
	\BibitemOpen
	\bibfield  {author} {\bibinfo {author} {\bibfnamefont {E.}~\bibnamefont
			{Saglamyurek}}, \bibinfo {author} {\bibfnamefont {N.}~\bibnamefont
			{Sinclair}}, \bibinfo {author} {\bibfnamefont {J.}~\bibnamefont {Jin}},
		\bibinfo {author} {\bibfnamefont {J.~A.}\ \bibnamefont {Slater}}, \bibinfo
		{author} {\bibfnamefont {D.}~\bibnamefont {Oblak}}, \bibinfo {author}
		{\bibfnamefont {F.}~\bibnamefont {Bussi{\`{e}}res}}, \bibinfo {author}
		{\bibfnamefont {M.}~\bibnamefont {George}}, \bibinfo {author} {\bibfnamefont
			{R.}~\bibnamefont {Ricken}}, \bibinfo {author} {\bibfnamefont
			{W.}~\bibnamefont {Sohler}},\ and\ \bibinfo {author} {\bibfnamefont
			{W.}~\bibnamefont {Tittel}},\ }\bibfield  {title} {\bibinfo {title}
		{{Broadband waveguide quantum memory for entangled photons}},\ }\href
	{https://doi.org/10.1038/nature09719} {\bibfield  {journal} {\bibinfo
			{journal} {Nature}\ }\textbf {\bibinfo {volume} {469}},\ \bibinfo {pages}
		{512} (\bibinfo {year} {2011})}\BibitemShut {NoStop}%
	\bibitem [{\citenamefont {Lukin}\ and\ \citenamefont
		{Hemmer}(2000)}]{Lukin2000}%
	\BibitemOpen
	\bibfield  {author} {\bibinfo {author} {\bibfnamefont {M.~D.}\ \bibnamefont
			{Lukin}}\ and\ \bibinfo {author} {\bibfnamefont {P.~R.}\ \bibnamefont
			{Hemmer}},\ }\bibfield  {title} {\bibinfo {title} {{Quantum Entanglement via
				Optical Control of Atom-Atom Interactions}},\ }\href
	{https://doi.org/10.1103/PhysRevLett.84.2818} {\bibfield  {journal} {\bibinfo
			{journal} {Phys. Rev. Lett.}\ }\textbf {\bibinfo {volume} {84}},\ \bibinfo
		{pages} {2818} (\bibinfo {year} {2000})}\BibitemShut {NoStop}%
	\bibitem [{\citenamefont {Ohlsson}\ \emph {et~al.}(2002)\citenamefont
		{Ohlsson}, \citenamefont {{Krishna Mohan}},\ and\ \citenamefont
		{Kr{\"{o}}ll}}]{Ohlsson2002}%
	\BibitemOpen
	\bibfield  {author} {\bibinfo {author} {\bibfnamefont {N.}~\bibnamefont
			{Ohlsson}}, \bibinfo {author} {\bibfnamefont {R.}~\bibnamefont {{Krishna
					Mohan}}},\ and\ \bibinfo {author} {\bibfnamefont {S.}~\bibnamefont
			{Kr{\"{o}}ll}},\ }\bibfield  {title} {\bibinfo {title} {{Quantum computer
				hardware based on rare-earth-ion-doped inorganic crystals}},\ }\href
	{https://doi.org/10.1016/S0030-4018(01)01666-2} {\bibfield  {journal}
		{\bibinfo  {journal} {Opt. Commun.}\ }\textbf {\bibinfo {volume} {201}},\
		\bibinfo {pages} {71} (\bibinfo {year} {2002})}\BibitemShut {NoStop}%
	\bibitem [{\citenamefont {Wesenberg}\ \emph {et~al.}(2007)\citenamefont
		{Wesenberg}, \citenamefont {M{\o}lmer}, \citenamefont {Rippe},\ and\
		\citenamefont {Kr{\"{o}}ll}}]{Wesenberg2007}%
	\BibitemOpen
	\bibfield  {author} {\bibinfo {author} {\bibfnamefont {J.~H.}\ \bibnamefont
			{Wesenberg}}, \bibinfo {author} {\bibfnamefont {K.}~\bibnamefont
			{M{\o}lmer}}, \bibinfo {author} {\bibfnamefont {L.}~\bibnamefont {Rippe}},\
		and\ \bibinfo {author} {\bibfnamefont {S.}~\bibnamefont {Kr{\"{o}}ll}},\
	}\bibfield  {title} {\bibinfo {title} {{Scalable designs for quantum
				computing with rare-earth-ion-doped crystals}},\ }\href
	{https://doi.org/10.1103/PhysRevA.75.012304} {\bibfield  {journal} {\bibinfo
			{journal} {Phys. Rev. A}\ }\textbf {\bibinfo {volume} {75}},\ \bibinfo
		{pages} {012304} (\bibinfo {year} {2007})}\BibitemShut {NoStop}%
	\bibitem [{\citenamefont {Hill}\ \emph {et~al.}(2015)\citenamefont {Hill},
		\citenamefont {Peretz}, \citenamefont {Hile}, \citenamefont {House},
		\citenamefont {Fuechsle}, \citenamefont {Rogge}, \citenamefont {Simmons},\
		and\ \citenamefont {Hollenberg}}]{Hill2015}%
	\BibitemOpen
	\bibfield  {author} {\bibinfo {author} {\bibfnamefont {C.~D.}\ \bibnamefont
			{Hill}}, \bibinfo {author} {\bibfnamefont {E.}~\bibnamefont {Peretz}},
		\bibinfo {author} {\bibfnamefont {S.~J.}\ \bibnamefont {Hile}}, \bibinfo
		{author} {\bibfnamefont {M.~G.}\ \bibnamefont {House}}, \bibinfo {author}
		{\bibfnamefont {M.}~\bibnamefont {Fuechsle}}, \bibinfo {author}
		{\bibfnamefont {S.}~\bibnamefont {Rogge}}, \bibinfo {author} {\bibfnamefont
			{M.~Y.}\ \bibnamefont {Simmons}},\ and\ \bibinfo {author} {\bibfnamefont
			{L.~C.~L.}\ \bibnamefont {Hollenberg}},\ }\bibfield  {title} {\bibinfo
		{title} {{A surface code quantum computer in silicon}},\ }\href
	{https://doi.org/10.1126/sciadv.1500707} {\bibfield  {journal} {\bibinfo
			{journal} {Sci. Adv.}\ }\textbf {\bibinfo {volume} {1}},\ \bibinfo {pages}
		{e1500707} (\bibinfo {year} {2015})}\BibitemShut {NoStop}%
	\bibitem [{\citenamefont {Hegde}\ \emph {et~al.}(2020)\citenamefont {Hegde},
		\citenamefont {Zhang},\ and\ \citenamefont {Suter}}]{Hegde2020}%
	\BibitemOpen
	\bibfield  {author} {\bibinfo {author} {\bibfnamefont {S.~S.}\ \bibnamefont
			{Hegde}}, \bibinfo {author} {\bibfnamefont {J.}~\bibnamefont {Zhang}},\ and\
		\bibinfo {author} {\bibfnamefont {D.}~\bibnamefont {Suter}},\ }\bibfield
	{title} {\bibinfo {title} {{Efficient Quantum Gates for Individual Nuclear
				Spin Qubits by Indirect Control}},\ }\href
	{https://doi.org/10.1103/PhysRevLett.124.220501} {\bibfield  {journal}
		{\bibinfo  {journal} {Phys. Rev. Lett.}\ }\textbf {\bibinfo {volume} {124}},\
		\bibinfo {pages} {220501} (\bibinfo {year} {2020})}\BibitemShut {NoStop}%
	\bibitem [{\citenamefont {Kane}(1998)}]{Kane1998}%
	\BibitemOpen
	\bibfield  {author} {\bibinfo {author} {\bibfnamefont {B.~E.}\ \bibnamefont
			{Kane}},\ }\bibfield  {title} {\bibinfo {title} {{A silicon-based nuclear
				spin quantum computer}},\ }\href {https://doi.org/10.1038/30156} {\bibfield
		{journal} {\bibinfo  {journal} {Nature}\ }\textbf {\bibinfo {volume} {393}},\
		\bibinfo {pages} {133} (\bibinfo {year} {1998})}\BibitemShut {NoStop}%
	\bibitem [{\citenamefont {Kalra}\ \emph {et~al.}(2014)\citenamefont {Kalra},
		\citenamefont {Laucht}, \citenamefont {Hill},\ and\ \citenamefont
		{Morello}}]{Kalra2014}%
	\BibitemOpen
	\bibfield  {author} {\bibinfo {author} {\bibfnamefont {R.}~\bibnamefont
			{Kalra}}, \bibinfo {author} {\bibfnamefont {A.}~\bibnamefont {Laucht}},
		\bibinfo {author} {\bibfnamefont {C.~D.}\ \bibnamefont {Hill}},\ and\
		\bibinfo {author} {\bibfnamefont {A.}~\bibnamefont {Morello}},\ }\bibfield
	{title} {\bibinfo {title} {{Robust Two-Qubit Gates for Donors in Silicon
				Controlled by Hyperfine Interactions}},\ }\href
	{https://doi.org/10.1103/PhysRevX.4.021044} {\bibfield  {journal} {\bibinfo
			{journal} {Phys. Rev. X}\ }\textbf {\bibinfo {volume} {4}},\ \bibinfo {pages}
		{021044} (\bibinfo {year} {2014})}\BibitemShut {NoStop}%
	\bibitem [{\citenamefont {Stoneham}\ \emph {et~al.}(2003)\citenamefont
		{Stoneham}, \citenamefont {Fisher},\ and\ \citenamefont
		{Greenland}}]{Stoneham2003}%
	\BibitemOpen
	\bibfield  {author} {\bibinfo {author} {\bibfnamefont {A.~M.}\ \bibnamefont
			{Stoneham}}, \bibinfo {author} {\bibfnamefont {A.~J.}\ \bibnamefont
			{Fisher}},\ and\ \bibinfo {author} {\bibfnamefont {P.~T.}\ \bibnamefont
			{Greenland}},\ }\bibfield  {title} {\bibinfo {title} {{Optically driven
				silicon-based quantum gates with potential for high-temperature operation}},\
	}\href {https://doi.org/10.1088/0953-8984/15/27/102} {\bibfield  {journal}
		{\bibinfo  {journal} {J. Phys. Condens. Matter}\ }\textbf {\bibinfo {volume}
			{15}},\ \bibinfo {pages} {L447} (\bibinfo {year} {2003})}\BibitemShut
	{NoStop}%
	\bibitem [{\citenamefont {Tosi}\ \emph {et~al.}(2017)\citenamefont {Tosi},
		\citenamefont {Mohiyaddin}, \citenamefont {Schmitt}, \citenamefont {Tenberg},
		\citenamefont {Rahman}, \citenamefont {Klimeck},\ and\ \citenamefont
		{Morello}}]{Tosi2015}%
	\BibitemOpen
	\bibfield  {author} {\bibinfo {author} {\bibfnamefont {G.}~\bibnamefont
			{Tosi}}, \bibinfo {author} {\bibfnamefont {F.~A.}\ \bibnamefont
			{Mohiyaddin}}, \bibinfo {author} {\bibfnamefont {V.}~\bibnamefont {Schmitt}},
		\bibinfo {author} {\bibfnamefont {S.}~\bibnamefont {Tenberg}}, \bibinfo
		{author} {\bibfnamefont {R.}~\bibnamefont {Rahman}}, \bibinfo {author}
		{\bibfnamefont {G.}~\bibnamefont {Klimeck}},\ and\ \bibinfo {author}
		{\bibfnamefont {A.}~\bibnamefont {Morello}},\ }\bibfield  {title} {\bibinfo
		{title} {{Silicon quantum processor with robust long-distance qubit
				couplings}},\ }\href {https://doi.org/10.1038/s41467-017-00378-x} {\bibfield
		{journal} {\bibinfo  {journal} {Nature Communications}\ }\textbf {\bibinfo
			{volume} {8}},\ \bibinfo {pages} {450} (\bibinfo {year} {2017})}\BibitemShut
	{NoStop}%
	\bibitem [{\citenamefont {DiVincenzo}(2000)}]{DiVincenzo2000}%
	\BibitemOpen
	\bibfield  {author} {\bibinfo {author} {\bibfnamefont {D.~P.}\ \bibnamefont
			{DiVincenzo}},\ }\bibfield  {title} {\bibinfo {title} {{The Physical
				Implementation of Quantum Computation}},\ }\href
	{https://doi.org/10.1002/1521-3978(200009)48:9/11<771::AID-PROP771>3.0.CO;2-E}
	{\bibfield  {journal} {\bibinfo  {journal} {Fortschritte der Phys.}\ }\textbf
		{\bibinfo {volume} {48}},\ \bibinfo {pages} {771} (\bibinfo {year}
		{2000})}\BibitemShut {NoStop}%
	\bibitem [{\citenamefont {Stevens}(1952)}]{Stevens1952}%
	\BibitemOpen
	\bibfield  {author} {\bibinfo {author} {\bibfnamefont {K.~W.~H.}\
			\bibnamefont {Stevens}},\ }\bibfield  {title} {\bibinfo {title} {{Matrix
				Elements and Operator Equivalents Connected with the Magnetic Properties of
				Rare Earth Ions}},\ }\href {https://doi.org/10.1088/0370-1298/65/3/308}
	{\bibfield  {journal} {\bibinfo  {journal} {Proc. Phys. Soc. Sect. A}\
		}\textbf {\bibinfo {volume} {65}},\ \bibinfo {pages} {209} (\bibinfo {year}
		{1952})}\BibitemShut {NoStop}%
	\bibitem [{\citenamefont {Hutchings}(1964)}]{Hutchings1964}%
	\BibitemOpen
	\bibfield  {author} {\bibinfo {author} {\bibfnamefont {M.}~\bibnamefont
			{Hutchings}},\ }\bibfield  {title} {\bibinfo {title} {{Point-Charge
				Calculations of Energy Levels of Magnetic Ions in Crystalline Electric
				Fields}},\ }\href
	{https://doi.org/https://doi.org/10.1016/S0081-1947(08)60517-2} {\bibfield
		{journal} {\bibinfo  {journal} {Solid State Phys.}\ }\textbf {\bibinfo
			{volume} {16}},\ \bibinfo {pages} {227 } (\bibinfo {year}
		{1964})}\BibitemShut {NoStop}%
	\bibitem [{Note1()}]{Note1}%
	\BibitemOpen
	\bibinfo {note} {We note that the perfect degeneracy between the different
		passive qubit levels allows for resonant excitation hopping between RE ions,
		i.e.\ the transfer to a different RE of the $\mathinner {|{1}\delimiter
			"526930B }_\protect \mathrm {pas}$ state due to virtual transitions involving
		excited electronic states. The effective hopping scales as $J_\protect
		\mathrm {hop} \sim J_\protect \mathrm {dip} \left (A_J/E_\protect \mathrm {Z}
		\right )^2$ which amounts to about $J_\protect \mathrm {hop}/h \sim 10
		\protect \text {\protect \tmspace +\thinmuskip {.1667em}Hz}$ for a qubit
		spacing of $r=10 \protect \text {\protect \tmspace +\thinmuskip
			{.1667em}nm}$. This weak hopping is typically rendered off-resonant by small
		residual inhomogeneities among the RE sites.}\BibitemShut {Stop}%
	\bibitem [{\citenamefont {Lauritzen}\ \emph {et~al.}(2008)\citenamefont
		{Lauritzen}, \citenamefont {Hastings-Simon}, \citenamefont {de~Riedmatten},
		\citenamefont {Afzelius},\ and\ \citenamefont {Gisin}}]{Lauritzen2008}%
	\BibitemOpen
	\bibfield  {author} {\bibinfo {author} {\bibfnamefont {B.}~\bibnamefont
			{Lauritzen}}, \bibinfo {author} {\bibfnamefont {S.~R.}\ \bibnamefont
			{Hastings-Simon}}, \bibinfo {author} {\bibfnamefont {H.}~\bibnamefont
			{de~Riedmatten}}, \bibinfo {author} {\bibfnamefont {M.}~\bibnamefont
			{Afzelius}},\ and\ \bibinfo {author} {\bibfnamefont {N.}~\bibnamefont
			{Gisin}},\ }\bibfield  {title} {\bibinfo {title} {{State preparation by
				optical pumping in erbium-doped solids using stimulated emission and spin
				mixing}},\ }\href {https://doi.org/10.1103/PhysRevA.78.043402} {\bibfield
		{journal} {\bibinfo  {journal} {Phys. Rev. A}\ }\textbf {\bibinfo {volume}
			{78}},\ \bibinfo {pages} {043402} (\bibinfo {year} {2008})}\BibitemShut
	{NoStop}%
	\bibitem [{\citenamefont {Cruzeiro}\ \emph {et~al.}(2018)\citenamefont
		{Cruzeiro}, \citenamefont {Tiranov}, \citenamefont {Lavoie}, \citenamefont
		{Ferrier}, \citenamefont {Goldner}, \citenamefont {Gisin},\ and\
		\citenamefont {Afzelius}}]{Cruzeiro2018}%
	\BibitemOpen
	\bibfield  {author} {\bibinfo {author} {\bibfnamefont {E.~Z.}\ \bibnamefont
			{Cruzeiro}}, \bibinfo {author} {\bibfnamefont {A.}~\bibnamefont {Tiranov}},
		\bibinfo {author} {\bibfnamefont {J.}~\bibnamefont {Lavoie}}, \bibinfo
		{author} {\bibfnamefont {A.}~\bibnamefont {Ferrier}}, \bibinfo {author}
		{\bibfnamefont {P.}~\bibnamefont {Goldner}}, \bibinfo {author} {\bibfnamefont
			{N.}~\bibnamefont {Gisin}},\ and\ \bibinfo {author} {\bibfnamefont
			{M.}~\bibnamefont {Afzelius}},\ }\bibfield  {title} {\bibinfo {title}
		{{Efficient optical pumping using hyperfine levels in
				$^{145}$Nd$^{3+}$:Y$_2$SiO$_5$ and its application to optical storage}},\
	}\href {https://doi.org/10.1088/1367-2630/aabe3b} {\bibfield  {journal}
		{\bibinfo  {journal} {New J. Phys.}\ }\textbf {\bibinfo {volume} {20}},\
		\bibinfo {pages} {053013} (\bibinfo {year} {2018})}\BibitemShut {NoStop}%
	\bibitem [{Note2()}]{Note2}%
	\BibitemOpen
	\bibinfo {note} {The longitudinal part of the dipolar interaction is kept.
		These static dipolar fields can be taken into account as an effective static
		field that adds to the externally applied field.}\BibitemShut {Stop}%
	\bibitem [{Note3()}]{Note3}%
	\BibitemOpen
	\bibinfo {note} {The $g$-matrix governs how an external magnetic field
		$\protect \mathaccentV {vec}17E{B}$ couples to a doublet of CF states
		$H_\protect \mathrm {doublet} = \protect \frac {\mu _B}{2} \DOTSB \sum@
		\slimits@ _{\alpha ,\beta } B_\alpha g_{\alpha \beta } \sigma _\beta $. In
		many point symmetry groups the basis of the doublet can be chosen such that
		the $g$-matrix reduces to a diagonal matrix with a longitudinal and
		transverse component $g_\parallel , g_\perp $, i.e.\ \begin {equation*}
		H_\protect \mathrm {doublet} = \protect \frac {\mu _B}{2} [g_\parallel B_z
		\sigma _z+ g_\perp (B_x \sigma _x+ B_y \sigma _y)]. \end {equation*} For
		simplicity, we restrict to point symmetries where the g-matrix takes this
		form.}\BibitemShut {Stop}%
	\bibitem [{Note4()}]{Note4}%
	\BibitemOpen
	\bibinfo {note} {This symmetry consideration does not take into account the
		possibility of mixing two different, but energetically close CF levels by a
		static magnetic field. If two such states are connected by a finite $J_{x/y}$
		matrix element, a magnetic field hybridizes them to create two polarized
		states with a magnetic moment in the transverse direction, even for
		non-Kramers ions. However, significantly hybridized states with
		non-negligible transverse magnetization occur only for fine-tuned CF
		Hamiltonians.}\BibitemShut {Stop}%
	\bibitem [{\citenamefont {Ortu}\ \emph {et~al.}(2018)\citenamefont {Ortu},
		\citenamefont {Tiranov}, \citenamefont {Welinski}, \citenamefont
		{Fr{\"{o}}wis}, \citenamefont {Gisin}, \citenamefont {Ferrier}, \citenamefont
		{Goldner},\ and\ \citenamefont {Afzelius}}]{Ortu2018}%
	\BibitemOpen
	\bibfield  {author} {\bibinfo {author} {\bibfnamefont {A.}~\bibnamefont
			{Ortu}}, \bibinfo {author} {\bibfnamefont {A.}~\bibnamefont {Tiranov}},
		\bibinfo {author} {\bibfnamefont {S.}~\bibnamefont {Welinski}}, \bibinfo
		{author} {\bibfnamefont {F.}~\bibnamefont {Fr{\"{o}}wis}}, \bibinfo {author}
		{\bibfnamefont {N.}~\bibnamefont {Gisin}}, \bibinfo {author} {\bibfnamefont
			{A.}~\bibnamefont {Ferrier}}, \bibinfo {author} {\bibfnamefont
			{P.}~\bibnamefont {Goldner}},\ and\ \bibinfo {author} {\bibfnamefont
			{M.}~\bibnamefont {Afzelius}},\ }\bibfield  {title} {\bibinfo {title}
		{{Simultaneous coherence enhancement of optical and microwave transitions in
				solid-state electronic spins}},\ }\href
	{https://doi.org/10.1038/s41563-018-0138-x} {\bibfield  {journal} {\bibinfo
			{journal} {Nat. Mater.}\ }\textbf {\bibinfo {volume} {17}},\ \bibinfo {pages}
		{671} (\bibinfo {year} {2018})}\BibitemShut {NoStop}%
	\bibitem [{\citenamefont {B{\"{o}}ttger}\ \emph {et~al.}(2009)\citenamefont
		{B{\"{o}}ttger}, \citenamefont {Thiel}, \citenamefont {Cone},\ and\
		\citenamefont {Sun}}]{Bottger2009}%
	\BibitemOpen
	\bibfield  {author} {\bibinfo {author} {\bibfnamefont {T.}~\bibnamefont
			{B{\"{o}}ttger}}, \bibinfo {author} {\bibfnamefont {C.~W.}\ \bibnamefont
			{Thiel}}, \bibinfo {author} {\bibfnamefont {R.~L.}\ \bibnamefont {Cone}},\
		and\ \bibinfo {author} {\bibfnamefont {Y.}~\bibnamefont {Sun}},\ }\bibfield
	{title} {\bibinfo {title} {{Effects of magnetic field orientation on optical
				decoherence in Er$^{3+}$:Y$_2$SiO$_5$}},\ }\href
	{https://doi.org/10.1103/PhysRevB.79.115104} {\bibfield  {journal} {\bibinfo
			{journal} {Phys. Rev. B}\ }\textbf {\bibinfo {volume} {79}},\ \bibinfo
		{pages} {115104} (\bibinfo {year} {2009})}\BibitemShut {NoStop}%
	\bibitem [{Note5()}]{Note5}%
	\BibitemOpen
	\bibinfo {note} {For RE ions the photon emission rate is strongly suppressed,
		because the electric dipole operator only couples states with opposite
		parity.}\BibitemShut {Stop}%
	\bibitem [{\citenamefont {Hastings-Simon}\ \emph {et~al.}(2008)\citenamefont
		{Hastings-Simon}, \citenamefont {Lauritzen}, \citenamefont {Staudt},
		\citenamefont {{vanMechelen}}, \citenamefont {Simon}, \citenamefont
		{{deRiedmatten}}, \citenamefont {Afzelius},\ and\ \citenamefont
		{Gisin}}]{Hastings-Simon2008}%
	\BibitemOpen
	\bibfield  {author} {\bibinfo {author} {\bibfnamefont {S.~R.}\ \bibnamefont
			{Hastings-Simon}}, \bibinfo {author} {\bibfnamefont {B.}~\bibnamefont
			{Lauritzen}}, \bibinfo {author} {\bibfnamefont {M.~U.}\ \bibnamefont
			{Staudt}}, \bibinfo {author} {\bibfnamefont {J.}~\bibnamefont
			{{vanMechelen}}}, \bibinfo {author} {\bibfnamefont {C.}~\bibnamefont
			{Simon}}, \bibinfo {author} {\bibfnamefont {H.}~\bibnamefont
			{{deRiedmatten}}}, \bibinfo {author} {\bibfnamefont {M.}~\bibnamefont
			{Afzelius}},\ and\ \bibinfo {author} {\bibfnamefont {N.}~\bibnamefont
			{Gisin}},\ }\bibfield  {title} {\bibinfo {title} {{Zeeman-level lifetimes in
				Er$^{3+}$:Y$_2$SiO$_5$}},\ }\href
	{https://doi.org/10.1103/PhysRevB.78.085410} {\bibfield  {journal} {\bibinfo
			{journal} {Phys. Rev. B}\ }\textbf {\bibinfo {volume} {78}},\ \bibinfo
		{pages} {085410} (\bibinfo {year} {2008})}\BibitemShut {NoStop}%
	\bibitem [{Note6()}]{Note6}%
	\BibitemOpen
	\bibinfo {note} {Working instead in the regime where the magnetic field is of
		the order of the hyperfine interaction allows one to exploit states at
		avoided hyperfine crossings,where the relevant eigenstates are superpositions
		of both electronic polarization states, that are entangled with the nuclear
		spin. This structure enables fast transitions within the (passive) qubit due
		to purely electronic matrix elements~\cite
		{Morley2010,Morley2013,Wolfowicz2013}. This regime, however, hast the
		drawback that it is unclear how such qubits could be individually addressed
		and how magnetic dipolar interactions could be \IeC {\textquoteright
		}switched on\IeC {\textquoteright } in a fast manner so as to entangle
		qubits. Furthermore, the (passive) qubit life-times are drastically reduced
		compared to our proposed qubits, since two passive qubit can swap their
		states directly via the magnetic dipolar interaction.}\BibitemShut {Stop}%
	\bibitem [{Note7()}]{Note7}%
	\BibitemOpen
	\bibinfo {note} {In some cases it may nevertheless be favorable to choose an
		excited manifold which cannot be reached from the ground state by a magnetic
		dipole transition. In such cases the electric dipole transition from the
		ground state to the excited doublet can be used. The associated electric
		dipole moments are typically in the range $\mu _\protect \mathrm {e} =
		10^{-34}-10^{-32} \protect \tmspace +\thinmuskip {.1667em} \protect \text
		{C\protect \tmspace +\thinmuskip {.1667em}m}$ (values for LiYF$_4$:Ho$^{3+}$
		\cite {Matmon2016}). The speed-up as compared to direct nuclear spin
		transitions is still of the order of a factor $(c \mu _\protect \mathrm
		{e})/(3\mu _\protect \mathrm {N}) = 2-200$.}\BibitemShut {Stop}%
	\bibitem [{Note8()}]{Note8}%
	\BibitemOpen
	\bibinfo {note} {To minimize the single-qubit gate time, the overlap of the
		nuclear spin states in the passive and active qubits should be maximized. For
		$I=1/2$, the optimal angle between the polarization directions of the active
		and passive doublet is $\pi /2$, whereas for larger nuclear spins $I>1/2$ it
		is less than $\pi /2$.}\BibitemShut {Stop}%
	\bibitem [{\citenamefont {Dodd}\ \emph {et~al.}(2002)\citenamefont {Dodd},
		\citenamefont {Nielsen}, \citenamefont {Bremner},\ and\ \citenamefont
		{Thew}}]{Dodd2002}%
	\BibitemOpen
	\bibfield  {author} {\bibinfo {author} {\bibfnamefont {J.~L.}\ \bibnamefont
			{Dodd}}, \bibinfo {author} {\bibfnamefont {M.~A.}\ \bibnamefont {Nielsen}},
		\bibinfo {author} {\bibfnamefont {M.~J.}\ \bibnamefont {Bremner}},\ and\
		\bibinfo {author} {\bibfnamefont {R.~T.}\ \bibnamefont {Thew}},\ }\bibfield
	{title} {\bibinfo {title} {{Universal quantum computation and simulation
				using any entangling Hamiltonian and local unitaries}},\ }\href
	{https://doi.org/10.1103/PhysRevA.65.040301} {\bibfield  {journal} {\bibinfo
			{journal} {Phys. Rev. A}\ }\textbf {\bibinfo {volume} {65}},\ \bibinfo
		{pages} {040301(R)} (\bibinfo {year} {2002})}\BibitemShut {NoStop}%
	\bibitem [{\citenamefont {Vandersypen}\ and\ \citenamefont
		{Chuang}(2005)}]{Chuang2005}%
	\BibitemOpen
	\bibfield  {author} {\bibinfo {author} {\bibfnamefont {L.~M.~K.}\
			\bibnamefont {Vandersypen}}\ and\ \bibinfo {author} {\bibfnamefont {I.~L.}\
			\bibnamefont {Chuang}},\ }\bibfield  {title} {\bibinfo {title} {{NMR
				techniques for quantum control and computation}},\ }\href
	{https://doi.org/10.1103/RevModPhys.76.1037} {\bibfield  {journal} {\bibinfo
			{journal} {Rev. Mod. Phys.}\ }\textbf {\bibinfo {volume} {76}},\ \bibinfo
		{pages} {1037} (\bibinfo {year} {2005})}\BibitemShut {NoStop}%
	\bibitem [{Note9()}]{Note9}%
	\BibitemOpen
	\bibinfo {note} {Near field optical excitation could potentially go beyond
		this limit for samples with a thin, near-surface layer of rare-earth
		impurities}\BibitemShut {NoStop}%
	\bibitem [{\citenamefont {Ahlefeldt}\ \emph {et~al.}(2013)\citenamefont
		{Ahlefeldt}, \citenamefont {McAuslan}, \citenamefont {Longdell},
		\citenamefont {Manson},\ and\ \citenamefont {Sellars}}]{Ahlefeldt2013}%
	\BibitemOpen
	\bibfield  {author} {\bibinfo {author} {\bibfnamefont {R.~L.}\ \bibnamefont
			{Ahlefeldt}}, \bibinfo {author} {\bibfnamefont {D.~L.}\ \bibnamefont
			{McAuslan}}, \bibinfo {author} {\bibfnamefont {J.~J.}\ \bibnamefont
			{Longdell}}, \bibinfo {author} {\bibfnamefont {N.~B.}\ \bibnamefont
			{Manson}},\ and\ \bibinfo {author} {\bibfnamefont {M.~J.}\ \bibnamefont
			{Sellars}},\ }\bibfield  {title} {\bibinfo {title} {{Precision Measurement of
				Electronic Ion-Ion Interactions between Neighboring Eu$^{3+}$ Optical
				Centers}},\ }\href {https://doi.org/10.1103/PhysRevLett.111.240501}
	{\bibfield  {journal} {\bibinfo  {journal} {Phys. Rev. Lett.}\ }\textbf
		{\bibinfo {volume} {111}},\ \bibinfo {pages} {240501} (\bibinfo {year}
		{2013})}\BibitemShut {NoStop}%
	\bibitem [{\citenamefont {Longdell}\ \emph {et~al.}(2004)\citenamefont
		{Longdell}, \citenamefont {Sellars},\ and\ \citenamefont
		{Manson}}]{Longdell2004}%
	\BibitemOpen
	\bibfield  {author} {\bibinfo {author} {\bibfnamefont {J.~J.}\ \bibnamefont
			{Longdell}}, \bibinfo {author} {\bibfnamefont {M.~J.}\ \bibnamefont
			{Sellars}},\ and\ \bibinfo {author} {\bibfnamefont {N.~B.}\ \bibnamefont
			{Manson}},\ }\bibfield  {title} {\bibinfo {title} {{Demonstration of
				Conditional Quantum Phase Shift Between Ions in a Solid}},\ }\href
	{https://doi.org/10.1103/PhysRevLett.93.130503} {\bibfield  {journal}
		{\bibinfo  {journal} {Phys. Rev. Lett.}\ }\textbf {\bibinfo {volume} {93}},\
		\bibinfo {pages} {130503} (\bibinfo {year} {2004})}\BibitemShut {NoStop}%
	\bibitem [{\citenamefont {Ahlefeldt}\ \emph {et~al.}(2020)\citenamefont
		{Ahlefeldt}, \citenamefont {Pearce}, \citenamefont {Hush},\ and\
		\citenamefont {Sellars}}]{Ahlefeldt2020}%
	\BibitemOpen
	\bibfield  {author} {\bibinfo {author} {\bibfnamefont {R.~L.}\ \bibnamefont
			{Ahlefeldt}}, \bibinfo {author} {\bibfnamefont {M.~J.}\ \bibnamefont
			{Pearce}}, \bibinfo {author} {\bibfnamefont {M.~R.}\ \bibnamefont {Hush}},\
		and\ \bibinfo {author} {\bibfnamefont {M.~J.}\ \bibnamefont {Sellars}},\
	}\bibfield  {title} {\bibinfo {title} {{Quantum processing with ensembles of
				rare-earth ions in a stoichiometric crystal}},\ }\href
	{https://doi.org/10.1103/PhysRevA.101.012309} {\bibfield  {journal} {\bibinfo
			{journal} {Phys. Rev. A}\ }\textbf {\bibinfo {volume} {101}},\ \bibinfo
		{pages} {012309} (\bibinfo {year} {2020})}\BibitemShut {NoStop}%
	\bibitem [{Note10()}]{Note10}%
	\BibitemOpen
	\bibinfo {note} {The only exceptions occur in the groups $C_{3h}$, $D_{3h}$
		and $T_d$: For the double groups of $C_{3h}$ and $D_{3h}$ only the doublets
		of the $\protect \mathaccentV {bar}016{E}_3$ representation have an electric
		dipole moment (using the notation of Ref.~\cite {Bradley2010}). The group
		$T_d$ does not contain an inversion symmetry, but nevertheless does not allow
		for electric dipole moments in Kramers doublets.}\BibitemShut {Stop}%
	\bibitem [{\citenamefont {Raussendorf}\ and\ \citenamefont
		{Harrington}(2007)}]{Raussendorf2007}%
	\BibitemOpen
	\bibfield  {author} {\bibinfo {author} {\bibfnamefont {R.}~\bibnamefont
			{Raussendorf}}\ and\ \bibinfo {author} {\bibfnamefont {J.}~\bibnamefont
			{Harrington}},\ }\bibfield  {title} {\bibinfo {title} {{Fault-Tolerant
				Quantum Computation with High Threshold in Two Dimensions}},\ }\href
	{https://doi.org/10.1103/PhysRevLett.98.190504} {\bibfield  {journal}
		{\bibinfo  {journal} {Phys. Rev. Lett.}\ }\textbf {\bibinfo {volume} {98}},\
		\bibinfo {pages} {190504} (\bibinfo {year} {2007})}\BibitemShut {NoStop}%
	\bibitem [{\citenamefont {Rogin}\ and\ \citenamefont
		{Hulliger}(1997)}]{Rogin1997}%
	\BibitemOpen
	\bibfield  {author} {\bibinfo {author} {\bibfnamefont {P.}~\bibnamefont
			{Rogin}}\ and\ \bibinfo {author} {\bibfnamefont {J.}~\bibnamefont
			{Hulliger}},\ }\bibfield  {title} {\bibinfo {title} {{Liquid phase epitaxy of
				LiYF$_4$}},\ }\href {https://doi.org/10.1016/S0022-0248(97)00163-2}
	{\bibfield  {journal} {\bibinfo  {journal} {J. Cryst. Growth}\ }\textbf
		{\bibinfo {volume} {179}},\ \bibinfo {pages} {551} (\bibinfo {year}
		{1997})}\BibitemShut {NoStop}%
	\bibitem [{\citenamefont {Douysset-Bloch}\ \emph {et~al.}(1998)\citenamefont
		{Douysset-Bloch}, \citenamefont {Ferrand}, \citenamefont {Couchaud},
		\citenamefont {Fulbert}, \citenamefont {Joubert}, \citenamefont {Chadeyron},\
		and\ \citenamefont {Jacquier}}]{Douysset-Bloch1998}%
	\BibitemOpen
	\bibfield  {author} {\bibinfo {author} {\bibfnamefont {L.}~\bibnamefont
			{Douysset-Bloch}}, \bibinfo {author} {\bibfnamefont {B.}~\bibnamefont
			{Ferrand}}, \bibinfo {author} {\bibfnamefont {M.}~\bibnamefont {Couchaud}},
		\bibinfo {author} {\bibfnamefont {L.}~\bibnamefont {Fulbert}}, \bibinfo
		{author} {\bibfnamefont {M.}~\bibnamefont {Joubert}}, \bibinfo {author}
		{\bibfnamefont {G.}~\bibnamefont {Chadeyron}},\ and\ \bibinfo {author}
		{\bibfnamefont {B.}~\bibnamefont {Jacquier}},\ }\bibfield  {title} {\bibinfo
		{title} {{Growth by liquid phase epitaxy and characterization of Nd:YLiF4
				layers}},\ }\href
	{https://doi.org/https://doi.org/10.1016/S0925-8388(98)00275-8} {\bibfield
		{journal} {\bibinfo  {journal} {Journal of Alloys and Compounds}\ }\textbf
		{\bibinfo {volume} {275-277}},\ \bibinfo {pages} {67 } (\bibinfo {year}
		{1998})}\BibitemShut {NoStop}%
	\bibitem [{\citenamefont {Starecki}\ \emph {et~al.}(2013)\citenamefont
		{Starecki}, \citenamefont {Bola{\~{n}}os}, \citenamefont {Braud},
		\citenamefont {Doualan}, \citenamefont {Brasse}, \citenamefont {Benayad},
		\citenamefont {Nazabal}, \citenamefont {Xu}, \citenamefont {Moncorg{\'{e}}},\
		and\ \citenamefont {Camy}}]{Starecki2013}%
	\BibitemOpen
	\bibfield  {author} {\bibinfo {author} {\bibfnamefont {F.}~\bibnamefont
			{Starecki}}, \bibinfo {author} {\bibfnamefont {W.}~\bibnamefont
			{Bola{\~{n}}os}}, \bibinfo {author} {\bibfnamefont {A.}~\bibnamefont
			{Braud}}, \bibinfo {author} {\bibfnamefont {J.-L.}\ \bibnamefont {Doualan}},
		\bibinfo {author} {\bibfnamefont {G.}~\bibnamefont {Brasse}}, \bibinfo
		{author} {\bibfnamefont {A.}~\bibnamefont {Benayad}}, \bibinfo {author}
		{\bibfnamefont {V.}~\bibnamefont {Nazabal}}, \bibinfo {author} {\bibfnamefont
			{B.}~\bibnamefont {Xu}}, \bibinfo {author} {\bibfnamefont {R.}~\bibnamefont
			{Moncorg{\'{e}}}},\ and\ \bibinfo {author} {\bibfnamefont {P.}~\bibnamefont
			{Camy}},\ }\bibfield  {title} {\bibinfo {title} {{Red and orange
				Pr$^{3+}$:LiYF$_4$ planar waveguide laser}},\ }\href
	{https://doi.org/10.1364/OL.38.000455} {\bibfield  {journal} {\bibinfo
			{journal} {Opt. Lett.}\ }\textbf {\bibinfo {volume} {38}},\ \bibinfo {pages}
		{455} (\bibinfo {year} {2013})}\BibitemShut {NoStop}%
	\bibitem [{\citenamefont {Camposeo}\ \emph {et~al.}(2004)\citenamefont
		{Camposeo}, \citenamefont {Fuso}, \citenamefont {Arimondo}, \citenamefont
		{Toncelli},\ and\ \citenamefont {Tonelli}}]{Camposeo2004}%
	\BibitemOpen
	\bibfield  {author} {\bibinfo {author} {\bibfnamefont {A.}~\bibnamefont
			{Camposeo}}, \bibinfo {author} {\bibfnamefont {F.}~\bibnamefont {Fuso}},
		\bibinfo {author} {\bibfnamefont {E.}~\bibnamefont {Arimondo}}, \bibinfo
		{author} {\bibfnamefont {A.}~\bibnamefont {Toncelli}},\ and\ \bibinfo
		{author} {\bibfnamefont {M.}~\bibnamefont {Tonelli}},\ }\bibfield  {title}
	{\bibinfo {title} {{Er-LiYF$_4$ coating of Si-based substrates by pulsed
				laser deposition}},\ }\href {https://doi.org/10.1016/j.surfcoat.2003.10.103}
	{\bibfield  {journal} {\bibinfo  {journal} {Surf. Coatings Technol.}\
		}\textbf {\bibinfo {volume} {180-181}},\ \bibinfo {pages} {607} (\bibinfo
		{year} {2004})}\BibitemShut {NoStop}%
	\bibitem [{\citenamefont {{Anwar-ul-Haq}}\ \emph {et~al.}(2009)\citenamefont
		{{Anwar-ul-Haq}}, \citenamefont {{Barsanti}},\ and\ \citenamefont
		{{Bicchi}}}]{Anwar-ul-Haq2009}%
	\BibitemOpen
	\bibfield  {author} {\bibinfo {author} {\bibfnamefont {M.}~\bibnamefont
			{{Anwar-ul-Haq}}}, \bibinfo {author} {\bibfnamefont {S.}~\bibnamefont
			{{Barsanti}}},\ and\ \bibinfo {author} {\bibfnamefont {P.}~\bibnamefont
			{{Bicchi}}},\ }\bibfield  {title} {\bibinfo {title} {{Advances in the pulsed
				laser deposition of rare earth - Doped LiYF$_4$ thin films on LiYF$_4$
				substrates}},\ }in\ \href {https://ieeexplore.ieee.org/document/5394629}
	{\emph {\bibinfo {booktitle} {2009 9th IEEE Conference on Nanotechnology
				(IEEE-NANO)}}}\ (\bibinfo {year} {2009})\BibitemShut {NoStop}%
	\bibitem [{\citenamefont {Secu}\ \emph {et~al.}(2017)\citenamefont {Secu},
		\citenamefont {Secu}, \citenamefont {Stokker-Cheregi}, \citenamefont {Ion},
		\citenamefont {Brajnicov},\ and\ \citenamefont {Dinescu}}]{Secu2017}%
	\BibitemOpen
	\bibfield  {author} {\bibinfo {author} {\bibfnamefont {C.~E.}\ \bibnamefont
			{Secu}}, \bibinfo {author} {\bibfnamefont {M.}~\bibnamefont {Secu}}, \bibinfo
		{author} {\bibfnamefont {F.}~\bibnamefont {Stokker-Cheregi}}, \bibinfo
		{author} {\bibfnamefont {V.}~\bibnamefont {Ion}}, \bibinfo {author}
		{\bibfnamefont {S.}~\bibnamefont {Brajnicov}},\ and\ \bibinfo {author}
		{\bibfnamefont {M.}~\bibnamefont {Dinescu}},\ }\bibfield  {title} {\bibinfo
		{title} {{Laser processing of Yb$^{3+}$/Er$^{3+}$ co-doped LiYF$_4$ thin
				films with up-conversion properties}},\ }\href
	{https://doi.org/10.1016/j.tsf.2017.01.057} {\bibfield  {journal} {\bibinfo
			{journal} {Thin Solid Films}\ }\textbf {\bibinfo {volume} {625}},\ \bibinfo
		{pages} {6} (\bibinfo {year} {2017})}\BibitemShut {NoStop}%
	\bibitem [{\citenamefont {Zhong}\ and\ \citenamefont
		{Goldner}(2019)}]{Zhong2019}%
	\BibitemOpen
	\bibfield  {author} {\bibinfo {author} {\bibfnamefont {T.}~\bibnamefont
			{Zhong}}\ and\ \bibinfo {author} {\bibfnamefont {P.}~\bibnamefont
			{Goldner}},\ }\bibfield  {title} {\bibinfo {title} {{Emerging rare-earth
				doped material platforms for quantum nanophotonics}},\ }\href
	{https://doi.org/10.1515/nanoph-2019-0185} {\bibfield  {journal} {\bibinfo
			{journal} {Nanophotonics}\ }\textbf {\bibinfo {volume} {8}},\ \bibinfo
		{pages} {2003} (\bibinfo {year} {2019})}\BibitemShut {NoStop}%
	\bibitem [{\citenamefont {Tang}\ \emph {et~al.}(1989)\citenamefont {Tang},
		\citenamefont {Heasman}, \citenamefont {Gillin},\ and\ \citenamefont
		{Sealy}}]{Tang1989}%
	\BibitemOpen
	\bibfield  {author} {\bibinfo {author} {\bibfnamefont {Y.~S.}\ \bibnamefont
			{Tang}}, \bibinfo {author} {\bibfnamefont {K.~C.}\ \bibnamefont {Heasman}},
		\bibinfo {author} {\bibfnamefont {W.~P.}\ \bibnamefont {Gillin}},\ and\
		\bibinfo {author} {\bibfnamefont {B.~J.}\ \bibnamefont {Sealy}},\ }\bibfield
	{title} {\bibinfo {title} {{Characteristics of rare-earth element erbium
				implanted in silicon}},\ }\href {https://doi.org/10.1063/1.101888} {\bibfield
		{journal} {\bibinfo  {journal} {Appl. Phys. Lett.}\ }\textbf {\bibinfo
			{volume} {55}},\ \bibinfo {pages} {432} (\bibinfo {year} {1989})}\BibitemShut
	{NoStop}%
	\bibitem [{\citenamefont {Louren{\c{c}}o}\ \emph {et~al.}(2016)\citenamefont
		{Louren{\c{c}}o}, \citenamefont {Hughes}, \citenamefont {Lai}, \citenamefont
		{Sofi}, \citenamefont {Ludurczak}, \citenamefont {Wong}, \citenamefont
		{Gwilliam},\ and\ \citenamefont {Homewood}}]{Lourenco2016}%
	\BibitemOpen
	\bibfield  {author} {\bibinfo {author} {\bibfnamefont {M.~A.}\ \bibnamefont
			{Louren{\c{c}}o}}, \bibinfo {author} {\bibfnamefont {M.~A.}\ \bibnamefont
			{Hughes}}, \bibinfo {author} {\bibfnamefont {K.~T.}\ \bibnamefont {Lai}},
		\bibinfo {author} {\bibfnamefont {I.~M.}\ \bibnamefont {Sofi}}, \bibinfo
		{author} {\bibfnamefont {W.}~\bibnamefont {Ludurczak}}, \bibinfo {author}
		{\bibfnamefont {L.}~\bibnamefont {Wong}}, \bibinfo {author} {\bibfnamefont
			{R.~M.}\ \bibnamefont {Gwilliam}},\ and\ \bibinfo {author} {\bibfnamefont
			{K.~P.}\ \bibnamefont {Homewood}},\ }\bibfield  {title} {\bibinfo {title}
		{{Silicon-Modified Rare-Earth Transitions - A New Route to Near- and Mid-IR
				Photonics}},\ }\href {https://doi.org/10.1002/adfm.201504662} {\bibfield
		{journal} {\bibinfo  {journal} {Adv. Funct. Mater.}\ }\textbf {\bibinfo
			{volume} {26}},\ \bibinfo {pages} {1986} (\bibinfo {year}
		{2016})}\BibitemShut {NoStop}%
	\bibitem [{\citenamefont {Zhang}\ \emph {et~al.}(2019)\citenamefont {Zhang},
		\citenamefont {Hu}, \citenamefont {de~Boo}, \citenamefont {Rančić},
		\citenamefont {Johnson}, \citenamefont {McCallum}, \citenamefont {Du},
		\citenamefont {Sellars}, \citenamefont {Yin},\ and\ \citenamefont
		{Rogge}}]{Zhang2019}%
	\BibitemOpen
	\bibfield  {author} {\bibinfo {author} {\bibfnamefont {Q.}~\bibnamefont
			{Zhang}}, \bibinfo {author} {\bibfnamefont {G.}~\bibnamefont {Hu}}, \bibinfo
		{author} {\bibfnamefont {G.~G.}\ \bibnamefont {de~Boo}}, \bibinfo {author}
		{\bibfnamefont {M.}~\bibnamefont {Rančić}}, \bibinfo {author}
		{\bibfnamefont {B.~C.}\ \bibnamefont {Johnson}}, \bibinfo {author}
		{\bibfnamefont {J.~C.}\ \bibnamefont {McCallum}}, \bibinfo {author}
		{\bibfnamefont {J.}~\bibnamefont {Du}}, \bibinfo {author} {\bibfnamefont
			{M.~J.}\ \bibnamefont {Sellars}}, \bibinfo {author} {\bibfnamefont
			{C.}~\bibnamefont {Yin}},\ and\ \bibinfo {author} {\bibfnamefont
			{S.}~\bibnamefont {Rogge}},\ }\bibfield  {title} {\bibinfo {title} {{Single
				Rare-Earth Ions as Atomic-Scale Probes in Ultrascaled Transistors}},\ }\href
	{https://doi.org/10.1021/acs.nanolett.9b01281} {\bibfield  {journal}
		{\bibinfo  {journal} {Nano Lett.}\ }\textbf {\bibinfo {volume} {19}},\
		\bibinfo {pages} {5025} (\bibinfo {year} {2019})}\BibitemShut {NoStop}%
	\bibitem [{\citenamefont {England}\ \emph {et~al.}(2019)\citenamefont
		{England}, \citenamefont {Cox}, \citenamefont {Cassidy}, \citenamefont
		{Mirkhaydarov},\ and\ \citenamefont {Perez-Fadon}}]{England2019}%
	\BibitemOpen
	\bibfield  {author} {\bibinfo {author} {\bibfnamefont {J.}~\bibnamefont
			{England}}, \bibinfo {author} {\bibfnamefont {D.}~\bibnamefont {Cox}},
		\bibinfo {author} {\bibfnamefont {N.}~\bibnamefont {Cassidy}}, \bibinfo
		{author} {\bibfnamefont {B.}~\bibnamefont {Mirkhaydarov}},\ and\ \bibinfo
		{author} {\bibfnamefont {A.}~\bibnamefont {Perez-Fadon}},\ }\bibfield
	{title} {\bibinfo {title} {{Investigating the formation of isotopically pure
				layers for quantum computers using ion implantation and layer exchange}},\
	}\href {https://doi.org/https://doi.org/10.1016/j.nimb.2019.09.013}
	{\bibfield  {journal} {\bibinfo  {journal} {Nucl. Instrum. Methods Phys. Res.
				B}\ }\textbf {\bibinfo {volume} {461}},\ \bibinfo {pages} {30 } (\bibinfo
		{year} {2019})}\BibitemShut {NoStop}%
	\bibitem [{\citenamefont {Zwanenburg}\ \emph {et~al.}(2013)\citenamefont
		{Zwanenburg}, \citenamefont {Dzurak}, \citenamefont {Morello}, \citenamefont
		{Simmons}, \citenamefont {Hollenberg}, \citenamefont {Klimeck}, \citenamefont
		{Rogge}, \citenamefont {Coppersmith},\ and\ \citenamefont
		{Eriksson}}]{Zwanenburg2013}%
	\BibitemOpen
	\bibfield  {author} {\bibinfo {author} {\bibfnamefont {F.~A.}\ \bibnamefont
			{Zwanenburg}}, \bibinfo {author} {\bibfnamefont {A.~S.}\ \bibnamefont
			{Dzurak}}, \bibinfo {author} {\bibfnamefont {A.}~\bibnamefont {Morello}},
		\bibinfo {author} {\bibfnamefont {M.~Y.}\ \bibnamefont {Simmons}}, \bibinfo
		{author} {\bibfnamefont {L.~C.}\ \bibnamefont {Hollenberg}}, \bibinfo
		{author} {\bibfnamefont {G.}~\bibnamefont {Klimeck}}, \bibinfo {author}
		{\bibfnamefont {S.}~\bibnamefont {Rogge}}, \bibinfo {author} {\bibfnamefont
			{S.~N.}\ \bibnamefont {Coppersmith}},\ and\ \bibinfo {author} {\bibfnamefont
			{M.~A.}\ \bibnamefont {Eriksson}},\ }\bibfield  {title} {\bibinfo {title}
		{{Silicon quantum electronics}},\ }\href
	{https://doi.org/10.1103/RevModPhys.85.961} {\bibfield  {journal} {\bibinfo
			{journal} {Rev. Mod. Phys.}\ }\textbf {\bibinfo {volume} {85}},\ \bibinfo
		{pages} {961} (\bibinfo {year} {2013})}\BibitemShut {NoStop}%
	\bibitem [{\citenamefont {Sun}\ \emph {et~al.}(2008)\citenamefont {Sun},
		\citenamefont {B{\"{o}}ttger}, \citenamefont {Thiel},\ and\ \citenamefont
		{Cone}}]{Sun2008}%
	\BibitemOpen
	\bibfield  {author} {\bibinfo {author} {\bibfnamefont {Y.}~\bibnamefont
			{Sun}}, \bibinfo {author} {\bibfnamefont {T.}~\bibnamefont {B{\"{o}}ttger}},
		\bibinfo {author} {\bibfnamefont {C.~W.}\ \bibnamefont {Thiel}},\ and\
		\bibinfo {author} {\bibfnamefont {R.~L.}\ \bibnamefont {Cone}},\ }\bibfield
	{title} {\bibinfo {title} {{Magnetic $g$ tensors for the $^4$I$_{15/2}$ and
				$^4$I$_{13/2}$ states of Er$^{3+}$:Y$_2$SiO$_5$}},\ }\href
	{https://doi.org/10.1103/PhysRevB.77.085124} {\bibfield  {journal} {\bibinfo
			{journal} {Phys. Rev. B}\ }\textbf {\bibinfo {volume} {77}},\ \bibinfo
		{pages} {085124} (\bibinfo {year} {2008})}\BibitemShut {NoStop}%
	\bibitem [{\citenamefont {Wolfowicz}\ \emph {et~al.}(2015)\citenamefont
		{Wolfowicz}, \citenamefont {Maier-Flaig}, \citenamefont {Marino},
		\citenamefont {Ferrier}, \citenamefont {Vezin}, \citenamefont {Morton},\ and\
		\citenamefont {Goldner}}]{Wolfowicz2015}%
	\BibitemOpen
	\bibfield  {author} {\bibinfo {author} {\bibfnamefont {G.}~\bibnamefont
			{Wolfowicz}}, \bibinfo {author} {\bibfnamefont {H.}~\bibnamefont
			{Maier-Flaig}}, \bibinfo {author} {\bibfnamefont {R.}~\bibnamefont {Marino}},
		\bibinfo {author} {\bibfnamefont {A.}~\bibnamefont {Ferrier}}, \bibinfo
		{author} {\bibfnamefont {H.}~\bibnamefont {Vezin}}, \bibinfo {author}
		{\bibfnamefont {J.~J.}\ \bibnamefont {Morton}},\ and\ \bibinfo {author}
		{\bibfnamefont {P.}~\bibnamefont {Goldner}},\ }\bibfield  {title} {\bibinfo
		{title} {{Coherent Storage of Microwave Excitations in Rare-Earth Nuclear
				Spins}},\ }\href {https://doi.org/10.1103/PhysRevLett.114.170503} {\bibfield
		{journal} {\bibinfo  {journal} {Phys. Rev. Lett.}\ }\textbf {\bibinfo
			{volume} {114}},\ \bibinfo {pages} {170503} (\bibinfo {year}
		{2015})}\BibitemShut {NoStop}%
	\bibitem [{\citenamefont {Welinski}\ \emph {et~al.}(2016)\citenamefont
		{Welinski}, \citenamefont {Ferrier}, \citenamefont {Afzelius},\ and\
		\citenamefont {Goldner}}]{Welinski2016}%
	\BibitemOpen
	\bibfield  {author} {\bibinfo {author} {\bibfnamefont {S.}~\bibnamefont
			{Welinski}}, \bibinfo {author} {\bibfnamefont {A.}~\bibnamefont {Ferrier}},
		\bibinfo {author} {\bibfnamefont {M.}~\bibnamefont {Afzelius}},\ and\
		\bibinfo {author} {\bibfnamefont {P.}~\bibnamefont {Goldner}},\ }\bibfield
	{title} {\bibinfo {title} {{High-resolution optical spectroscopy and magnetic
				properties of Yb$^{3+}$ in Y$_2$SiO$_5$}},\ }\href
	{https://doi.org/10.1103/PhysRevB.94.155116} {\bibfield  {journal} {\bibinfo
			{journal} {Phys. Rev. B}\ }\textbf {\bibinfo {volume} {94}},\ \bibinfo
		{pages} {155116} (\bibinfo {year} {2016})}\BibitemShut {NoStop}%
	\bibitem [{\citenamefont {Yin}\ \emph {et~al.}(2013)\citenamefont {Yin},
		\citenamefont {Rancic}, \citenamefont {{De Boo}}, \citenamefont {Stavrias},
		\citenamefont {McCallum}, \citenamefont {Sellars},\ and\ \citenamefont
		{Rogge}}]{Yin2013}%
	\BibitemOpen
	\bibfield  {author} {\bibinfo {author} {\bibfnamefont {C.}~\bibnamefont
			{Yin}}, \bibinfo {author} {\bibfnamefont {M.}~\bibnamefont {Rancic}},
		\bibinfo {author} {\bibfnamefont {G.~G.}\ \bibnamefont {{De Boo}}}, \bibinfo
		{author} {\bibfnamefont {N.}~\bibnamefont {Stavrias}}, \bibinfo {author}
		{\bibfnamefont {J.~C.}\ \bibnamefont {McCallum}}, \bibinfo {author}
		{\bibfnamefont {M.~J.}\ \bibnamefont {Sellars}},\ and\ \bibinfo {author}
		{\bibfnamefont {S.}~\bibnamefont {Rogge}},\ }\bibfield  {title} {\bibinfo
		{title} {{Optical addressing of an individual erbium ion in silicon}},\
	}\href {https://doi.org/10.1038/nature12081} {\bibfield  {journal} {\bibinfo
			{journal} {Nature}\ }\textbf {\bibinfo {volume} {497}},\ \bibinfo {pages}
		{91} (\bibinfo {year} {2013})}\BibitemShut {NoStop}%
	\bibitem [{\citenamefont {Weiss}\ \emph {et~al.}(2020)\citenamefont {Weiss},
		\citenamefont {Gritsch}, \citenamefont {Merkel},\ and\ \citenamefont
		{Reiserer}}]{Weiss2020}%
	\BibitemOpen
	\bibfield  {author} {\bibinfo {author} {\bibfnamefont {L.}~\bibnamefont
			{Weiss}}, \bibinfo {author} {\bibfnamefont {A.}~\bibnamefont {Gritsch}},
		\bibinfo {author} {\bibfnamefont {B.}~\bibnamefont {Merkel}},\ and\ \bibinfo
		{author} {\bibfnamefont {A.}~\bibnamefont {Reiserer}},\ }\href@noop {}
	{\bibinfo {title} {{Erbium dopants in silicon nanophotonic waveguides}}}
	(\bibinfo {year} {2020}),\ \Eprint {https://arxiv.org/abs/2005.01775}
	{arXiv:2005.01775 [physics.app-ph]} \BibitemShut {NoStop}%
	\bibitem [{\citenamefont {Vasilev}\ and\ \citenamefont
		{Vitanov}(2004)}]{Vasilev2004}%
	\BibitemOpen
	\bibfield  {author} {\bibinfo {author} {\bibfnamefont {G.~S.}\ \bibnamefont
			{Vasilev}}\ and\ \bibinfo {author} {\bibfnamefont {N.~V.}\ \bibnamefont
			{Vitanov}},\ }\bibfield  {title} {\bibinfo {title} {{Coherent excitation of a
				two-state system by a Gaussian field}},\ }\href
	{https://doi.org/10.1103/PhysRevA.70.053407} {\bibfield  {journal} {\bibinfo
			{journal} {Phys. Rev. A}\ }\textbf {\bibinfo {volume} {70}},\ \bibinfo
		{pages} {053407} (\bibinfo {year} {2004})}\BibitemShut {NoStop}%
	\bibitem [{\citenamefont {Gilchrist}\ \emph {et~al.}(2005)\citenamefont
		{Gilchrist}, \citenamefont {Langford},\ and\ \citenamefont
		{Nielsen}}]{Gilchrist2005}%
	\BibitemOpen
	\bibfield  {author} {\bibinfo {author} {\bibfnamefont {A.}~\bibnamefont
			{Gilchrist}}, \bibinfo {author} {\bibfnamefont {N.~K.}\ \bibnamefont
			{Langford}},\ and\ \bibinfo {author} {\bibfnamefont {M.~A.}\ \bibnamefont
			{Nielsen}},\ }\bibfield  {title} {\bibinfo {title} {{Distance measures to
				compare real and ideal quantum processes}},\ }\href
	{https://doi.org/10.1103/PhysRevA.71.062310} {\bibfield  {journal} {\bibinfo
			{journal} {Phys. Rev. A}\ }\textbf {\bibinfo {volume} {71}},\ \bibinfo
		{pages} {062310} (\bibinfo {year} {2005})}\BibitemShut {NoStop}%
	\bibitem [{\citenamefont {Nielsen}\ and\ \citenamefont
		{Chuang}(2010)}]{Nielsen2010}%
	\BibitemOpen
	\bibfield  {author} {\bibinfo {author} {\bibfnamefont {M.~A.}\ \bibnamefont
			{Nielsen}}\ and\ \bibinfo {author} {\bibfnamefont {I.~L.}\ \bibnamefont
			{Chuang}},\ }\href {https://doi.org/10.1017/CBO9780511976667} {\emph
		{\bibinfo {title} {{Quantum Computation and Quantum Information}}}}\
	(\bibinfo  {publisher} {Cambridge University Press},\ \bibinfo {address}
	{Cambridge},\ \bibinfo {year} {2010})\BibitemShut {NoStop}%
	\bibitem [{\citenamefont {Drake}(2006)}]{Drake2006}%
	\BibitemOpen
	\bibfield  {author} {\bibinfo {author} {\bibfnamefont {G.~W.~F.}\
			\bibnamefont {Drake}},\ }\href
	{https://www.springer.com/de/book/9780387208022} {\emph {\bibinfo {title}
			{{Springer Handbook of Atomic, Molecular, and Optical Physics}}}},\ Springer
	Handbook of Atomic, Molecular, and Optical Physics\ (\bibinfo  {publisher}
	{Springer},\ \bibinfo {year} {2006})\BibitemShut {NoStop}%
	\bibitem [{Note11()}]{Note11}%
	\BibitemOpen
	\bibinfo {note} {The Rabi frequency associated with the activation of a qubit
		and a simultaneous spin-flip of a nearby (non-targeted) qubit is suppressed
		by a factor $J_\protect \mathrm {dip}/E_\protect \mathrm {Z}$ and the
		transition is detuned by the hyperfine energy as compared to the activation with no
		spin-flip. Since one already has to minimize activation errors due to
		undesired hyperfine transitions of the targeted qubit (which are detuned by the hyperfine
		energy as well) to achieve high single-qubit gate fidelities, simultaneous
		spin-flips of passive qubits are inherently suppressed. They scale with an
		additional factor $\sim \left (\protect \frac {J_\protect \mathrm
			{dip}}{E_\protect \mathrm {Z}}\right )^2 \sim \left ( \protect \frac {\mu _0
			\mu _B}{4 \pi r^3 B} \right )^2 \sim 10^{-12}$ for qubit spacings of
		$r=10\protect \text {\protect \tmspace +\thinmuskip {.1667em}nm}$ and
		magnetic fields $B=1 \protect \text {\protect \tmspace +\thinmuskip
			{.1667em}T}$, and are therefore negligible.}\BibitemShut {Stop}%
	\bibitem [{\citenamefont {Guillot-No{\"{e}}l}\ \emph
		{et~al.}(2006)\citenamefont {Guillot-No{\"{e}}l}, \citenamefont {Goldner},
		\citenamefont {Du}, \citenamefont {Baldit}, \citenamefont {Monnier},\ and\
		\citenamefont {Bencheikh}}]{Guillot-Noel2006}%
	\BibitemOpen
	\bibfield  {author} {\bibinfo {author} {\bibfnamefont {O.}~\bibnamefont
			{Guillot-No{\"{e}}l}}, \bibinfo {author} {\bibfnamefont {P.}~\bibnamefont
			{Goldner}}, \bibinfo {author} {\bibfnamefont {Y.~L.}\ \bibnamefont {Du}},
		\bibinfo {author} {\bibfnamefont {E.}~\bibnamefont {Baldit}}, \bibinfo
		{author} {\bibfnamefont {P.}~\bibnamefont {Monnier}},\ and\ \bibinfo {author}
		{\bibfnamefont {K.}~\bibnamefont {Bencheikh}},\ }\bibfield  {title} {\bibinfo
		{title} {{Hyperfine interaction of Er$^{3+}$ ions in Y$_2$SiO$_5$: An
				electron paramagnetic resonance spectroscopy study}},\ }\href
	{https://doi.org/10.1103/PhysRevB.74.214409} {\bibfield  {journal} {\bibinfo
			{journal} {Phys. Rev. B - Condens. Matter Mater. Phys.}\ }\textbf {\bibinfo
			{volume} {74}},\ \bibinfo {pages} {1} (\bibinfo {year} {2006})}\BibitemShut
	{NoStop}%
	\bibitem [{\citenamefont {Stone}(2005)}]{Stone2005}%
	\BibitemOpen
	\bibfield  {author} {\bibinfo {author} {\bibfnamefont {N.}~\bibnamefont
			{Stone}},\ }\bibfield  {title} {\bibinfo {title} {{Table of nuclear magnetic
				dipole and electric quadrupole moments}},\ }\href
	{https://doi.org/10.1016/j.adt.2005.04.001} {\bibfield  {journal} {\bibinfo
			{journal} {At. Data Nucl. Data Tables}\ }\textbf {\bibinfo {volume} {90}},\
		\bibinfo {pages} {75} (\bibinfo {year} {2005})}\BibitemShut {NoStop}%
	\bibitem [{\citenamefont {Thiel}\ \emph {et~al.}(2011)\citenamefont {Thiel},
		\citenamefont {B{\"{o}}ttger},\ and\ \citenamefont {Cone}}]{Thiel2011}%
	\BibitemOpen
	\bibfield  {author} {\bibinfo {author} {\bibfnamefont {C.}~\bibnamefont
			{Thiel}}, \bibinfo {author} {\bibfnamefont {T.}~\bibnamefont
			{B{\"{o}}ttger}},\ and\ \bibinfo {author} {\bibfnamefont {R.}~\bibnamefont
			{Cone}},\ }\bibfield  {title} {\bibinfo {title} {{Rare-earth-doped materials
				for applications in quantum information storage and signal processing}},\
	}\href {https://doi.org/10.1016/j.jlumin.2010.12.015} {\bibfield  {journal}
		{\bibinfo  {journal} {J. Lumin.}\ }\textbf {\bibinfo {volume} {131}},\
		\bibinfo {pages} {353} (\bibinfo {year} {2011})}\BibitemShut {NoStop}%
	\bibitem [{\citenamefont {Morley}\ \emph {et~al.}(2010)\citenamefont {Morley},
		\citenamefont {Warner}, \citenamefont {Stoneham}, \citenamefont {Greenland},
		\citenamefont {{Van Tol}}, \citenamefont {Kay},\ and\ \citenamefont
		{Aeppli}}]{Morley2010}%
	\BibitemOpen
	\bibfield  {author} {\bibinfo {author} {\bibfnamefont {G.~W.}\ \bibnamefont
			{Morley}}, \bibinfo {author} {\bibfnamefont {M.}~\bibnamefont {Warner}},
		\bibinfo {author} {\bibfnamefont {A.~M.}\ \bibnamefont {Stoneham}}, \bibinfo
		{author} {\bibfnamefont {P.~T.}\ \bibnamefont {Greenland}}, \bibinfo {author}
		{\bibfnamefont {J.}~\bibnamefont {{Van Tol}}}, \bibinfo {author}
		{\bibfnamefont {C.~W.}\ \bibnamefont {Kay}},\ and\ \bibinfo {author}
		{\bibfnamefont {G.}~\bibnamefont {Aeppli}},\ }\bibfield  {title} {\bibinfo
		{title} {{The initialization and manipulation of quantum information stored
				in silicon by bismuth dopants}},\ }\href {https://doi.org/10.1038/nmat2828}
	{\bibfield  {journal} {\bibinfo  {journal} {Nat. Mater.}\ }\textbf {\bibinfo
			{volume} {9}},\ \bibinfo {pages} {725} (\bibinfo {year} {2010})}\BibitemShut
	{NoStop}%
	\bibitem [{\citenamefont {Morley}\ \emph {et~al.}(2013)\citenamefont {Morley},
		\citenamefont {Lueders}, \citenamefont {{Hamed Mohammady}}, \citenamefont
		{Balian}, \citenamefont {Aeppli}, \citenamefont {Kay}, \citenamefont
		{Witzel}, \citenamefont {Jeschke},\ and\ \citenamefont
		{Monteiro}}]{Morley2013}%
	\BibitemOpen
	\bibfield  {author} {\bibinfo {author} {\bibfnamefont {G.~W.}\ \bibnamefont
			{Morley}}, \bibinfo {author} {\bibfnamefont {P.}~\bibnamefont {Lueders}},
		\bibinfo {author} {\bibfnamefont {M.}~\bibnamefont {{Hamed Mohammady}}},
		\bibinfo {author} {\bibfnamefont {S.~J.}\ \bibnamefont {Balian}}, \bibinfo
		{author} {\bibfnamefont {G.}~\bibnamefont {Aeppli}}, \bibinfo {author}
		{\bibfnamefont {C.~W.}\ \bibnamefont {Kay}}, \bibinfo {author} {\bibfnamefont
			{W.~M.}\ \bibnamefont {Witzel}}, \bibinfo {author} {\bibfnamefont
			{G.}~\bibnamefont {Jeschke}},\ and\ \bibinfo {author} {\bibfnamefont {T.~S.}\
			\bibnamefont {Monteiro}},\ }\bibfield  {title} {\bibinfo {title} {{Quantum
				control of hybrid nuclear-electronic qubits}},\ }\href
	{https://doi.org/10.1038/nmat3499} {\bibfield  {journal} {\bibinfo  {journal}
			{Nat. Mater.}\ }\textbf {\bibinfo {volume} {12}},\ \bibinfo {pages} {103}
		(\bibinfo {year} {2013})}\BibitemShut {NoStop}%
	\bibitem [{\citenamefont {Wolfowicz}\ \emph {et~al.}(2013)\citenamefont
		{Wolfowicz}, \citenamefont {Tyryshkin}, \citenamefont {George}, \citenamefont
		{Riemann}, \citenamefont {Abrosimov}, \citenamefont {Becker}, \citenamefont
		{Pohl}, \citenamefont {Thewalt}, \citenamefont {Lyon},\ and\ \citenamefont
		{Morton}}]{Wolfowicz2013}%
	\BibitemOpen
	\bibfield  {author} {\bibinfo {author} {\bibfnamefont {G.}~\bibnamefont
			{Wolfowicz}}, \bibinfo {author} {\bibfnamefont {A.~M.}\ \bibnamefont
			{Tyryshkin}}, \bibinfo {author} {\bibfnamefont {R.~E.}\ \bibnamefont
			{George}}, \bibinfo {author} {\bibfnamefont {H.}~\bibnamefont {Riemann}},
		\bibinfo {author} {\bibfnamefont {N.~V.}\ \bibnamefont {Abrosimov}}, \bibinfo
		{author} {\bibfnamefont {P.}~\bibnamefont {Becker}}, \bibinfo {author}
		{\bibfnamefont {H.~J.}\ \bibnamefont {Pohl}}, \bibinfo {author}
		{\bibfnamefont {M.~L.}\ \bibnamefont {Thewalt}}, \bibinfo {author}
		{\bibfnamefont {S.~A.}\ \bibnamefont {Lyon}},\ and\ \bibinfo {author}
		{\bibfnamefont {J.~J.}\ \bibnamefont {Morton}},\ }\bibfield  {title}
	{\bibinfo {title} {{Atomic clock transitions in silicon-based spin qubits}},\
	}\href {https://doi.org/10.1038/nnano.2013.117} {\bibfield  {journal}
		{\bibinfo  {journal} {Nat. Nanotechnol.}\ }\textbf {\bibinfo {volume} {8}},\
		\bibinfo {pages} {561} (\bibinfo {year} {2013})}\BibitemShut {NoStop}%
	\bibitem [{\citenamefont {Matmon}\ \emph {et~al.}(2016)\citenamefont {Matmon},
		\citenamefont {Lynch}, \citenamefont {Rosenbaum}, \citenamefont {Fisher},\
		and\ \citenamefont {Aeppli}}]{Matmon2016}%
	\BibitemOpen
	\bibfield  {author} {\bibinfo {author} {\bibfnamefont {G.}~\bibnamefont
			{Matmon}}, \bibinfo {author} {\bibfnamefont {S.~A.}\ \bibnamefont {Lynch}},
		\bibinfo {author} {\bibfnamefont {T.~F.}\ \bibnamefont {Rosenbaum}}, \bibinfo
		{author} {\bibfnamefont {A.~J.}\ \bibnamefont {Fisher}},\ and\ \bibinfo
		{author} {\bibfnamefont {G.}~\bibnamefont {Aeppli}},\ }\bibfield  {title}
	{\bibinfo {title} {{Optical response from terahertz to visible light of
				electronuclear transitions in LiYF$_4$:Ho$^{3+}$}},\ }\href
	{https://doi.org/10.1103/PhysRevB.94.205132} {\bibfield  {journal} {\bibinfo
			{journal} {Phys. Rev. B}\ }\textbf {\bibinfo {volume} {94}},\ \bibinfo
		{pages} {205132} (\bibinfo {year} {2016})}\BibitemShut {NoStop}%
	\bibitem [{\citenamefont {Bradley}\ and\ \citenamefont
		{Cracknell}(2010)}]{Bradley2010}%
	\BibitemOpen
	\bibfield  {author} {\bibinfo {author} {\bibfnamefont {C.}~\bibnamefont
			{Bradley}}\ and\ \bibinfo {author} {\bibfnamefont {A.}~\bibnamefont
			{Cracknell}},\ }\href
	{https://global.oup.com/academic/product/the-mathematical-theory-of-symmetry-in-solids-9780199582587?cc=ch&lang=en&}
	{\emph {\bibinfo {title} {{The Mathematical Theory of Symmetry in Solids:
					Representation Theory for Point Groups and Space Groups}}}}\ (\bibinfo
	{publisher} {OUP Oxford},\ \bibinfo {year} {2010})\BibitemShut {NoStop}%
\end{thebibliography}
\end{document}